\documentclass[onecolumn]{aa} 

\usepackage{graphicx}
\usepackage{ulem}
\usepackage[varg]{txfonts}
\usepackage{amsmath,amsfonts,amssymb}

\begin{document}

\title{Integrals of motion for non-axisymmetric potentials}


\author{O. Bienaym\'e\inst{1} }
\institute{Observatoire astronomique de Strasbourg, Universit\'e de Strasbourg, CNRS, 11 rue de l'Universit\'e, F-67000 Strasbourg, France }

   \date{  }

 \abstract{}{}{}{}{} 
 
 \abstract
   {The modelling of stationary galactic stellar populations can be performed using distribution functions.}  
%
%
  {This paper aims to write explicit  integrals of motion and distribution functions. }
   {We propose an analytic formulation of the integrals of motion with an explicit dependence on potential. This formulation applies to potentials with rotational symmetry or triaxial symmetry. It is exact for St\"ackel potentials and approximate for other potentials.}
  {Modelling a stationary stellar population using these  integrals of motion allows the force field to be found with satisfactory accuracy. On the other hand, the mass density distribution that generates the force field and the gravitational potential is  recovered with less accuracy due to lower precision in modelling  box-type orbits.

}
   {}


   \maketitle
%



\section{ Introduction }

It is well established that  orbits in galactic potentials  generally admit three integrals of motion \citep{con60,oll62} with the possible presence of ergodic orbits \citep{hen64}. 
The kinematic modelling of a stationary stellar population can therefore be done simply by using a distribution function that depends on three integrals of motion.
However, very few gravitational potentials allow such distribution functions to be constructed with analytic integrals. 
In addition to  rotationally symmetric systems with two analytic integrals, meaning the energy and the angular momentum, the most general case known concerns St\"ackel potentials with three analytic integrals. These integrals make  it possible to model certain potentials with triaxial symmetry\footnote{ We remark that the most studied St\"ackel potentials have analytic expressions of their three integrals with symmetries in confocal ellipsoidal coordinate systems  \citep{oll62,lyn62,dez85a}.
On the other hand, those with symmetries in a parabolic coordinate system have been little studied \citep{oll62,lyn62,tsi12} and left aside.}. 
Apart from these cases, known potentials with analytic integrals  are uncommon \cite[see for example][]{hie87}
 and have no obvious application for galactic dynamics.\\

To numerically compute integrals of motion, \cite{mcg90} and \cite{san14}
  \cite[see][for a review of different methods]{san16} 
 showed how the torus overlaid with  orbits in phase space could be modelled by  numerically determining  the angle-action  variables. Other related methods 
 \citep{bin12,san15}
 have been proposed with simpler numerical approaches that are sufficiently precise in many practical cases, either for axisymmetric  or triaxially symmetrical  potentials. These simpler methods apply for potentials for which the orbits are close to those of St\"ackel potentials.\\

In Section 2 of this article we detail  exact integrals of motion defined by  explicit analytic expressions. 
Our results are based on the same principles as those of \cite{san12} and \cite{san14}, save that the search for integrals is analytic and not numerical. These integrals of motion are exact for St\"ackel triaxial potentials and their expressions depend explicitly on the potential.
In Section 3, we test the efficiency of using these analytic expressions as approximate integrals for non-St\"ackel triaxial potentials.
We  note that the initial principle, that is the efficiency of representing each orbit by using an integral of a St\"ackel potential, has been shown by \cite{ken91}.
In Section 4, we detail a distribution function with a triaxial velocity ellipsoid  for a triaxial potential. We test the stationarity and accuracy of this distribution function and of  the approximate integrals using the Jeans equation.  Section 5 summarizes the results and proposes possible extensions of this work.\\



\section{Triaxial St\"ackel potentials and their integrals of motion }

Ellipsoidal coordinates and St\"ackel potentials have been detailed and discussed many times
\citep{whi1902,mor53,oll62,lyn62,dez85,dez85a,dez85b}.
St\"ackel potentials can be  written in ellipsoidal coordinates (see Appendix), depending on three arbitrary functions $\zeta(\lambda)$, $\eta(\mu),$ and $\theta(\nu)$:
 \begin{equation}
\Phi(\lambda,\mu,\nu)=
-4\,{\frac {\zeta  \left( \lambda \right)  \left( \lambda+\alpha
 \right)  \left( \lambda+\beta \right)  \left( \lambda+\gamma \right) 
}{ \left( \lambda-\mu \right)  \left( \lambda-\nu \right) }}-4\,{
\frac {\eta \left( \mu \right)  \left( \mu+\alpha \right)  \left( \mu+
\beta \right)  \left( \mu+\gamma \right) }{ \left( \mu-\nu \right) 
 \left( \mu-\lambda \right) }}-4\,{\frac {\theta \left( \nu \right) 
 \left( \nu+\alpha \right)  \left( \nu+\beta \right)  \left( \nu+
\gamma \right) }{ \left( \nu-\lambda \right)  \left( \nu-\mu \right) }
}.
\label{equ:appendix-phi1}
\end{equation}
With this coordinate system, the equations of motion separate and it can be shown that the St\"ackel potentials admit  three independent integrals. Here, we show that these integrals can be  written with an explicit dependence on positions and velocities and on the values of the potential at three different positions. This is obtained by noting that the usual free functions defining the potential can be  written as  functions of the potential along the  three axes of symmetry $x, y,$ and $z$ of the potential.
This alternative formulation can be  applied to obtain approximate integrals of motion for any potential that is close to a St\"ackel potential and for the related orbits.  Thus, possible fitting of  a given potential $V$ with a St\"ackel potential $\Phi$ consists in equating both potentials along these three  different axes. Other possibilities are also presented.
 \\
 
If $\Phi$ is a continuous function at the origin, the formulation in Eq.\ref{equ:appendix-phi1}
implies that  $\Phi(-\alpha,-\beta,-\gamma)=0$.
Evaluating $\Phi$ at positions $(\lambda,\mu=-\beta,\nu=-\gamma)$, we  deduce $\zeta(\lambda)$:
\begin{equation}
\zeta \left( \lambda \right) =-1/4\,{\frac {\Phi \left( \lambda,-\beta,-
\gamma \right) }{\lambda+\alpha}}. 
\end{equation}
Similarly, $\eta(\mu)$ and $\theta(\nu)$  are obtained  evaluating     $\Phi \left( -\alpha,-\beta,\nu \right)$  and $\Phi \left( -\alpha,\mu,-\gamma \right)$:
\begin{equation}
\eta \left( \mu \right) =-1/4\,{\frac {\Phi \left( -\alpha, \mu,-
\gamma \right) }{\mu+\beta}} ,
\end{equation}
\begin{equation}
\theta \left( \nu \right) =-1/4\,{\frac {\Phi \left( -\alpha,-\beta,
\nu \right) }{\nu+\gamma}} .
\end{equation}
Then, substituting these expressions of $\zeta,\eta,$ and $\theta$ within the expression of  $\Phi$ (Eq.\ref{equ:appendix-phi1}),
we obtain
%
%
\begin{equation}
\Phi(\lambda,\mu,\nu)=
{\frac { \left( \lambda+\beta \right)  \left( \lambda+\gamma \right) 
}{ \left( \lambda-\mu \right)  \left( \lambda-\nu \right) }}
\, \Phi \left( \lambda,-\beta,-\gamma \right) 
+
{\frac { \left( \mu+\alpha \right)  \left( \mu+\gamma \right) }{
 \left( \mu-\nu \right)  \left( \mu-\lambda \right) }}
\, \Phi \left( -\alpha,\mu,-\gamma \right) 
+
{\frac { \left( \nu+\alpha \right)  \left( \nu+\beta \right) }{
 \left( \nu-\lambda \right)  \left( \nu-\mu \right) }}
\, \Phi \left( -\alpha,-\beta,\nu \right)  ,
\end{equation}
a valid expression if $\Phi(-\alpha,-\beta,-\gamma)=0$.\\

The expressions of the three integrals of motion,  the  Hamiltonian $H,$ and   two integrals $J$ and $K$ are given by \citet[see equations 7-9]{dez85}. They also detail the integrals $I_2$ and $I_3$ that have useful properties particularly  for building  stellar distribution functions for stellar discs or halos.
We use the notations and definitions adopted by \citet[][equations 7 and 14]{dez85}, and for convenience write the exact integrals in compact form as
\begin{equation}H=\Phi(\lambda,\mu,\nu) +1/2\,{{ v_x}}^{2}+1/2\,{{ v_y}}^{2}+1/2\,{{ v_z}}^{2}\end{equation}
\begin{equation}J=\Psi(\lambda,\mu,\nu) +f_J(L^2,v_x,v_y,v_z)\end{equation}
\begin{equation}K=\Xi(\lambda,\mu,\nu) +f_K(L_x^2,L_y^2,L_z^2,v_x,v_y,v_z)\end{equation}
and 
\begin{equation}I_2=\frac{\alpha^2 H +\alpha J +L}{\gamma-\alpha}
=\psi(\lambda,\mu,\nu) 
+1/2\,{ \left( \alpha-\beta \right) { v_x}}^{2}+1/2\,{\frac { \left( 
\alpha-\beta \right) }{\left( \alpha-\gamma \right)}}{{\it L_y}}^{2}+1/2\,{{\it L_z}}^{
2}
\label{equ:I2}
\end{equation}
\begin{equation}I_3=\frac{\gamma^2 H +\gamma J +L}{\alpha-\gamma}
=\xi(\lambda,\mu,\nu) +
1/2\, \left( \gamma-\beta \right) {{ v_z}}^{2}+1/2\,{{\it L_x}}^{2}
+1/2\,{ \frac { \left(\gamma-\beta \right) } {\left( \gamma-\alpha \right)}}{{\it L_y}}^{2} 
\label{equ:I3}
\end{equation}

 According to \citet{dez85}, who give a complete discussion, "$I_2$ and $I_3$ can be considered as generalizations of the energy integrals  that exist in a potential separable in Cartesian coordinates, or generalization of angular momentum integrals in axisymmetric or spherical potentials".
For instance, in the limit of an axisymmetric potential ($\alpha=\beta$) with the $z$ axis for the rotational symmetry,
$I_2=1/2\, L_z^2$.
We note that the third integral is null, $I_3=0$ for motion within the $(x,y)$ plane (i.e. $v_z=0$).

After substitution of $\zeta,  \eta,$ and $\theta$,    we obtain the following  expressions  depending on the potential at three distinct positions,  one on each of the  three axes of symmetry of the potential (these expressions are valid only if $\Phi(-\alpha,-\beta,-\gamma)=0$):
\begin{equation}
\psi=
{\frac { \left( \mu+\alpha \right)  \left( \nu+\alpha \right) }{\alpha
-\gamma}}
\left[
{\frac { \left( \lambda+\beta \right)  \left( \lambda+\gamma \right) 
}{ \left( \lambda-\mu \right)  \left( \lambda-\nu \right) }}
\,
\Phi \left( \lambda,-\beta,-\gamma \right)
+
{\frac { \left( \mu+\gamma \right)  \left( \lambda+\alpha \right) }{
 \left( \mu-\nu \right)  \left( \mu-\lambda \right) }}
\,
\Phi \left( -\alpha,\mu,-\gamma \right) 
+
{\frac { \left( \nu+\beta \right)  \left( \lambda+\alpha \right) }{
 \left( \nu-\lambda \right)  \left( \nu-\mu \right) }}
\,
\Phi \left( -\alpha,-\beta,\nu \right) 
\right] ,
\end{equation}
\begin{equation}
\xi=
{\frac { \left( \mu+\gamma \right)  \left( \lambda+\gamma \right) }{\gamma-\alpha}}
\,
\left[
{\frac { \left( \lambda+\beta \right)  \left( \nu+\gamma \right) }{
 \left( \lambda-\mu \right)  \left( \lambda-\nu \right) }}
\,
\Phi \left( \lambda,-\beta,-\gamma \right)
+
{\frac { \left( \mu+\alpha \right)  \left( \nu+\gamma \right) }{
 \left( \mu-\nu \right)  \left( \mu-\lambda \right) }}
\,
\Phi \left( -\alpha,\mu,-\gamma \right) 
+
{\frac { \left( \nu+\alpha \right)  \left( \nu+\beta \right) }{
 \left( \nu-\lambda \right)  \left( \nu-\mu \right) }}
\,
\Phi \left( -\alpha,-\beta,\nu \right) 
\right] .
\label{equ:appendix-xi}
\end{equation}
\\
Other choices of the three positions are possible but  require more algebra.
However, a simple case can be  deduced, again evaluating $\Phi$ at three positions; determining $\zeta$, $\eta$, and $\theta$; and substituting in   $\Phi$, $\psi,$ and $\xi$ expressions, now  being defined with  the potential at three positions on  two planes and one axis of symmetry  (for all the expressions below, it is no longer  necessary that the potential be zero at the origin):
\begin{equation}
\Phi(\lambda, \mu, \nu)= {\frac {   \left( \mu+\gamma \right) }{
 \left( \mu-\nu \right)  }}
 \,\Phi \left( \lambda,\mu,-\gamma \right)
 +{\frac {   \left( \nu+\beta \right) }{
   \left( \nu-\mu \right) }}
\,\Phi \left( \lambda,-\beta,\nu \right)
+ {\frac {  \left(\beta-\gamma\right)}{  \left( \mu-\nu  \right) } }
\,\Phi \left( \lambda,-\beta,-\gamma \right) ,
\label{equ:appendix-phi2}
\end{equation}
\begin{multline}
\psi (\lambda,\mu,\nu)=
{\frac { \left( \lambda+\alpha \right)  \left( \mu+\gamma \right) 
 \left( \nu+\alpha \right) }{ \left( \mu-\nu \right)  \left( \alpha-
\gamma \right) }}
\Phi \left( \lambda,\mu,-\gamma \right) 
-{\frac { \left( \lambda+\alpha \right)   \left( \mu+\alpha \right) \left( \nu+\beta \right) 
 }{ \left( \mu-\nu \right)  \left( \alpha-
\gamma \right) }}
\Phi \left( \lambda,-\beta,\nu \right) 
\\
-\left[
{\frac { \left( \lambda+\alpha \right)   \left( \lambda+\beta \right) \left( \mu+\gamma \right) 
 \left( \nu+\alpha \right)  }{  \left( \lambda-\mu \right)  \left( \mu
-\nu \right)  \left( \alpha-\gamma
 \right) }}
 +{\frac { \left( \lambda+\alpha \right)   \left( \lambda+\gamma \right) \left( \mu+\alpha
 \right)  \left( \nu+\beta \right)  }{
 \left( \nu-\lambda \right)  \left( \mu-\nu \right)   \left( \alpha-
\gamma \right) }}
+{\frac { \left( \lambda+\beta \right)  \left( 
\lambda+\gamma \right)  \left( \mu+\alpha \right)  \left( \nu+\alpha
 \right) }{ \left( \lambda-\mu \right)  \left( \nu-\lambda \right) 
 \left( \alpha-\gamma \right) }}
\right]
\Phi \left( \lambda,-\beta,-\gamma \right)  ,
\end{multline}
\begin{equation}
\xi (\lambda,\mu,\nu)= \,
\frac {  \left( \lambda+\gamma \right)  \left( \mu+\gamma \right) }{
 \left( \mu-\nu \right)  \left( \gamma-\alpha \right) }
\left[
( \nu+\gamma ) \Phi ( \lambda,\mu,-\gamma ) -
 ( \nu+\beta ) \Phi ( \lambda,-\beta,\nu \right) -
 (\gamma -\beta ) \Phi ( \lambda,-\beta,-\gamma) ] .
\end{equation}
\\
A simpler combination is obtained after   substitution of $\Phi(\lambda,-\beta,\nu)$ deduced from  Eq. \ref{equ:appendix-phi2}:
\begin{equation}
 {\psi} \left( \lambda,\mu,\nu \right) 
=
{\frac {  \left( \lambda+\alpha \right) \left( \mu+\alpha \right) }{
\alpha-\gamma}}
\Phi \left( \lambda,\mu,\nu \right) 
-{\frac {\left( \lambda+\alpha \right)  \left( \mu+\gamma \right)  }{
\alpha-\gamma}}
\Phi \left( \lambda,\mu,-\gamma \right) 
- \left( \lambda+\beta \right) 
\Phi \left( \lambda,-\beta,-\gamma
 \right)  ,
 \label{equ:psi}
\end{equation}
\begin{equation}
\xi \left( \lambda,\mu,\nu \right) =
{\frac {  \left( \lambda+\gamma \right) \left( \mu+\gamma \right)  }{\gamma-
\alpha}} \,
[ \Phi \left( \lambda,\mu,\nu
 \right) 
 -\Phi \left( \lambda,\mu,-\gamma \right)  ] \, .
 \label{equ:xi}
\end{equation}
We note that  within the plane $z=0$, $\xi \left( \lambda,\mu,\nu=-\gamma \right)$ is null.
Moreover, along the $x$ axis, 
$  \psi  \left( \lambda,-\beta,-\gamma \right) =\left( -\beta+\alpha \right) \Phi \left( \lambda,-\beta,-\gamma \right), $
   is null in the case of  an axisymmetric potential with $\alpha=\beta$.
\\
Other expressions of the three integrals depending on $\Phi$ at three distinct positions can be obtained, we suppose under 
the necessary condition  that the three chosen positions are on    different  sheets defined by the ellipsoidal coordinate system and  including the  $(\lambda,\mu,\nu)$ position.\\

 In the case of an axisymmetric potential with $\alpha=\beta$, from Eq. \ref{equ:xi} we recover the formula
(exact for a St\"ackel potential)  used for approximate integrals for orbits within the potential of the Besan\c{c}on Galaxy Model  \citep[][equation A7]{bie15}.
\begin{equation}
\xi \left( \lambda,\nu \right) =
 \left( \lambda+\gamma \right)  \,
[ \Phi \left( \lambda,\nu
 \right) 
 -\Phi \left( \lambda,-\gamma \right)  ] \, .
 \label{equ:appendix-xi2}
\end{equation}
For an axisymmetric potential with setting  $\alpha=\beta$,  we can, for instance, define another  expression from Eq.\,\ref{equ:appendix-xi}, 
and express the third integral with the potential at two positions, one  on the $z$ axis  and the other one on the $z=0$ plane. It gives us
\begin{equation}
\xi(\lambda,\nu)=
\,
\frac { \left( \lambda+\gamma \right) } {(\lambda-\nu)}
\left[
  \left( \nu+\gamma \right)   
\,
\Phi \left( \lambda,-\gamma \right)
-
{ \left( \nu+\alpha \right)  }
\,
\Phi \left( -\alpha,\nu \right) 
\right]
\end{equation}
This may be compared with the different variants of the St\"ackel fudge for axisymmetric potentials proposed by \citet{vas19} and based on different sets  of positions. Finally, we note that it is possible to write expressions depending on more than three positions.\\



\section{Tests}

We chose a  St\"ackel potential given by
\begin{equation}
\Phi= (\beta + \gamma) \,x^2 + (\gamma+\alpha ) \,y^2 + (\alpha + \beta) \,z^2
+(-\alpha-\beta-\gamma +x^2 +y^2 +z^2)^2  
\label{equ:tests-1}
\end{equation}
to verify the accuracy of our programs, as well as the validity of the various analytic expressions of the  integrals of motion presented above. 
With $\alpha=-2$, $\beta=-1$ and $\gamma=0,$ this potential has a unique minimum at $x=y=z=0$ and all orbits are bound.
With these values of the potential parameters, 
we  examined the accuracy with which the integrals of motion $I_2$ (Eqs. \ref{equ:I2}  and \ref{equ:psi}) and $I_3$ (Eqs. \ref{equ:I3}  and \ref{equ:xi}) are numerically preserved.
Over a period of 300 rotations around the centre, the energy of the orbits with an initial value of $E=$1 is maintained to within $10^{-14}$. This accuracy is limited by rounding errors due to double-precision calculation. The other integrals $I_2$ and $I_3$ have values between about zero and one and each keep their initial value with a variation of less than $10^{-8}$, but greater than the accuracy that would result only from rounding errors.\\

We summarize our results and consider a potential more likely to represent a galaxy and similar to those chosen by \citet{san14} and \citet{san15}. The latter tested the conservation of quasi-integrals, quasi-actions, with an approximation also based on a St\"ackel triaxial potential. This allows us to compare their results with ours. 

We present our results obtained for the  logarithmic potential: 
\begin{equation}
  \Phi(x,y,z) = \frac{1}{2} \log \left(  m_0^2 + m^2 \right) 
\label{equ:tests-2}
\end{equation}
where $$ m^2 = x^2 + \frac{y^2}{q_y^2}+ \frac{z^2}{q_z^2} $$

with $m_0=0.3$, $q_y=0.95$, and $q_z=0.85$. These flattening values $q_y$ and $q_z$ correspond to those used in the model of \citet{san15}. Note that these two values of the potential flattening $q_\Phi$ correspond to density flattening values of $q_\rho\sim0.86$ and 0.56 respectively.
The commonly published values for flattening the dark halo of our Galaxy do not always agree with each other but still indicate that it is rather spherical. Thus, \citet{pos19} obtained $q_\rho=1.22 \pm $0.23 from the analysis of globular cluster orbits while \citet{mal18b} found a flattening of the density $q_\rho=0.86$ to explain the shape and velocity of the giant stellar current GD-1. 
In addition, \citet{law10}  proposed a triaxial halo to explain the Sagittarius Galaxy stream.

From the potential of the Eq. \ref{equ:tests-2}, we calculated 5000 orbits of energy $E=1$ with a Runge-Kutta of order 8 \citep{fel68} over a period of about three hundred rotations around the centre. 
 The initial position of the orbits is randomly selected on the sphere of radius $r=1$ and the modulus of the initial velocity is then of the order of 1. The direction of the velocity is also randomly drawn. 
 This choice makes it possible to cover many different types of orbits. 
 We note that in the case of a logarithmic potential with $m_0=0$ (i.e. a scale free potential) this choice would cover all the different types of orbit.
 Here with $m_0=0.3$ we excluded the very inner orbits but still include many eccentric orbits, the most difficult to model accurately with our approximate integrals.

Once the orbital trajectories are determined, we  adjusted the free parameters $\Delta_1 =\sqrt{\beta-\alpha}$, $\Delta_2=\sqrt{\gamma-\beta}$ and $\Delta_3=\sqrt{\gamma-\alpha}$ (positions of the foci of the ellipsoidal coordinate system) to obtain the best conservation of the quasi-integral $I_2$ and $I_3$ along all orbits. We obtained essentially the same adjusted focal points positions for this operation if we consider all orbits or if we simply limit ourselves to the three main closed orbits located in the three planes of symmetry.

\begin{figure}[!htbp]
\begin{center}
\centerline{\includegraphics[scale=0.37]{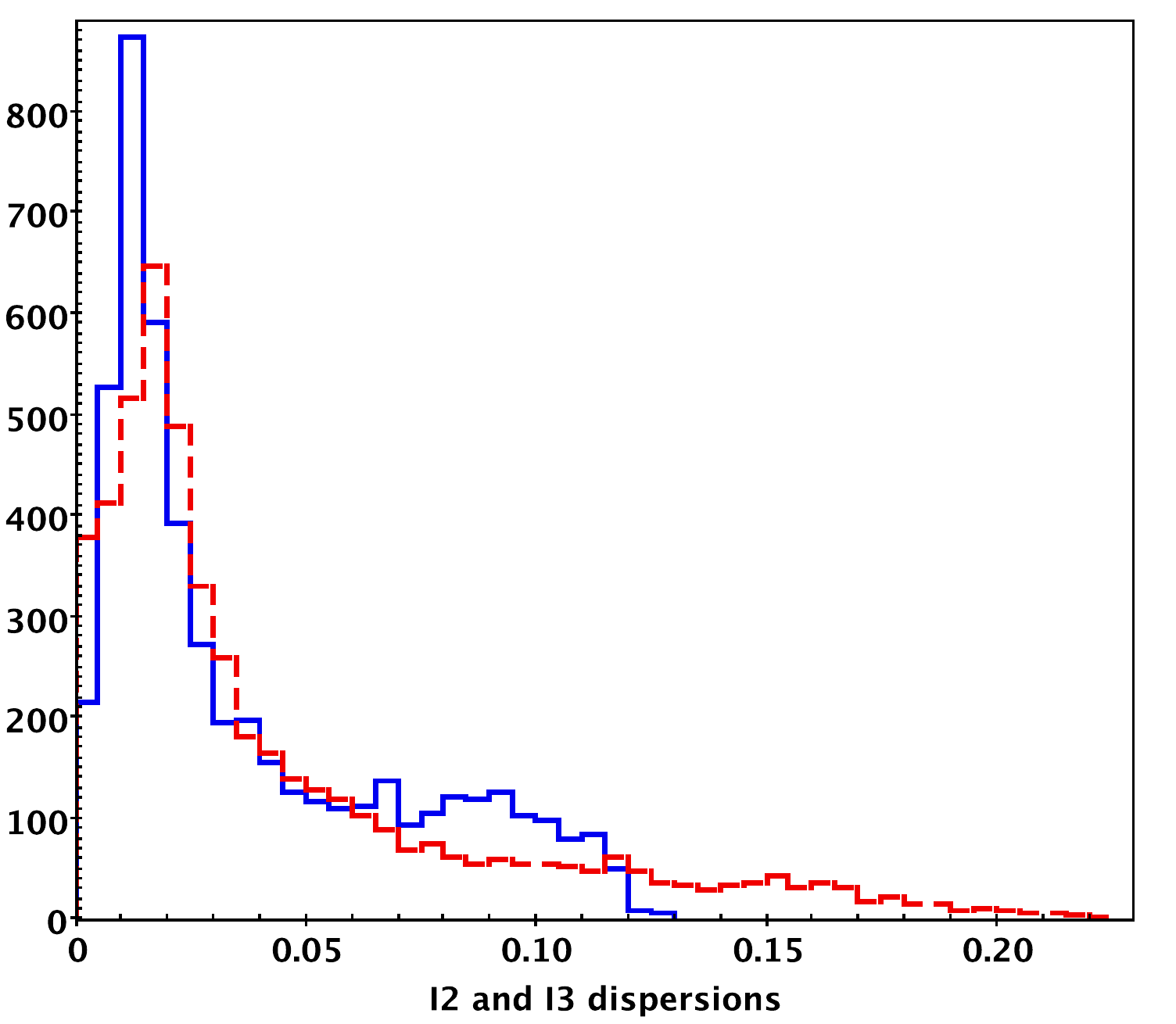}}
\caption{ Histograms for dispersions, $\sigma_{I2}$ and   $\sigma_{I3}$,  of  quasi integrals for 5000 orbits with $E=1$. 
The blue continuous line indicates dispersion for the integral $I_2$.
The red dashed  line indicates dispersion for the integral $I_3$. 
}
\label{fig:f1}
\end{center}
\end{figure}

We obtained the values $\Delta_1=0.49$ and $\Delta_3=0.33,$ which give the quasi-integrals with the smallest fluctuations.
The integrals of motion $I_2$ and $I_3$  then take values within the intervals [-0.28, 0.24], and [0.0.578], respectively. The relative variation, $\sigma_{I2}$ or $\sigma_{I3}$, of integrals (i.e. the variation divided by the amplitude of these intervals that is respectively 0.520 and 0.578 for $I_2$ and $I_3$) are shown in  Fig. \ref{fig:f1} in histogram form. 
The amplitude of this relative variations tells us the possibility of deciding when two orbits are distinct.
 For example, a relative variation of two per cent for $I_2$ and $I_3$ implies that the measurements of $I_2$ and $I_3$ make it possible to distinguish $50 \times 50$ orbits with the same energy. In addition, if we also have a two per cent accuracy on the energy, this would allow us to distinguish  $125,000$ separate orbits just by using  kinematic criteria.

\begin{figure}[!htbp]
\begin{center}
\centerline{\includegraphics[scale=0.4]{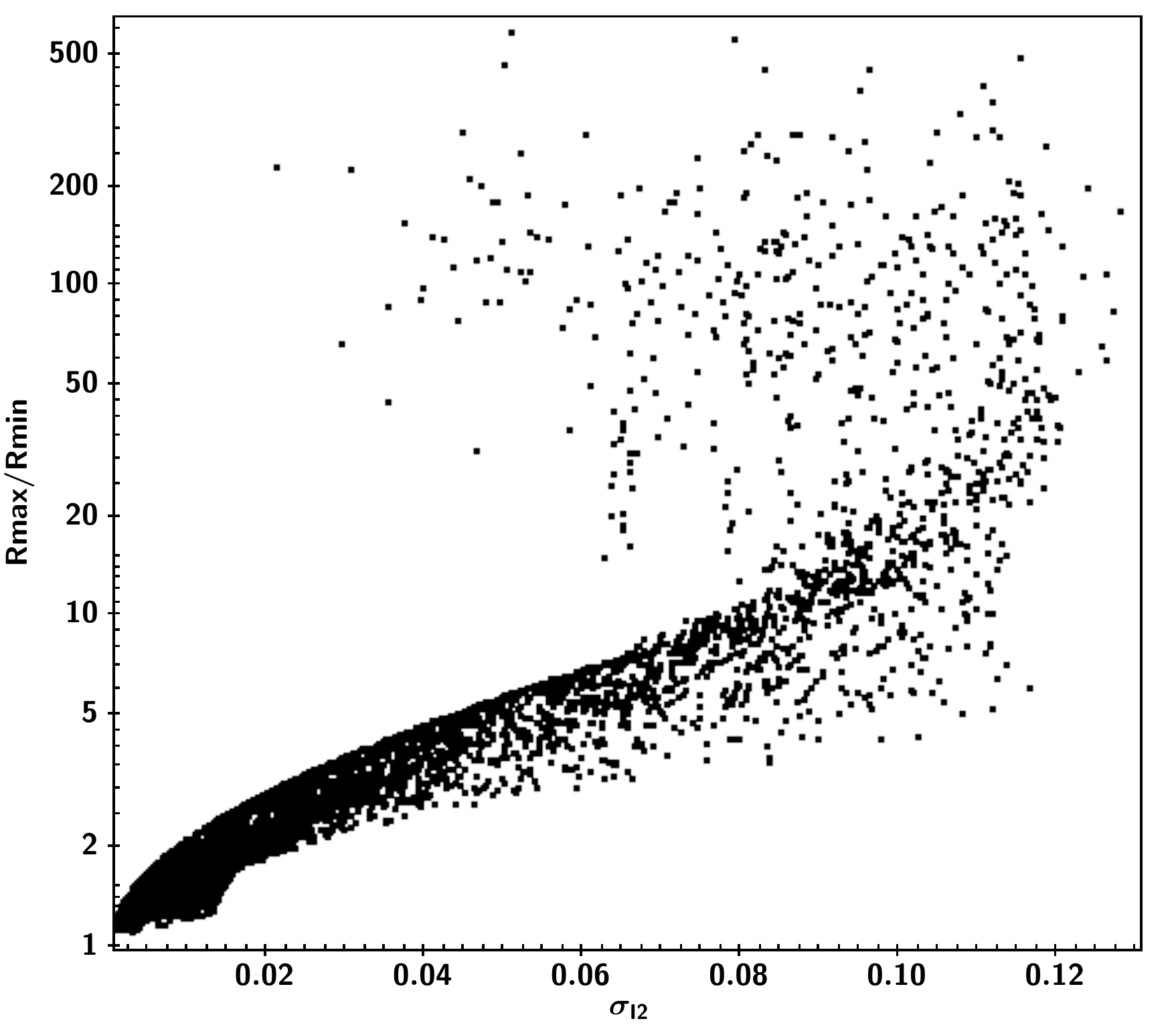},
{\includegraphics[scale=0.4]{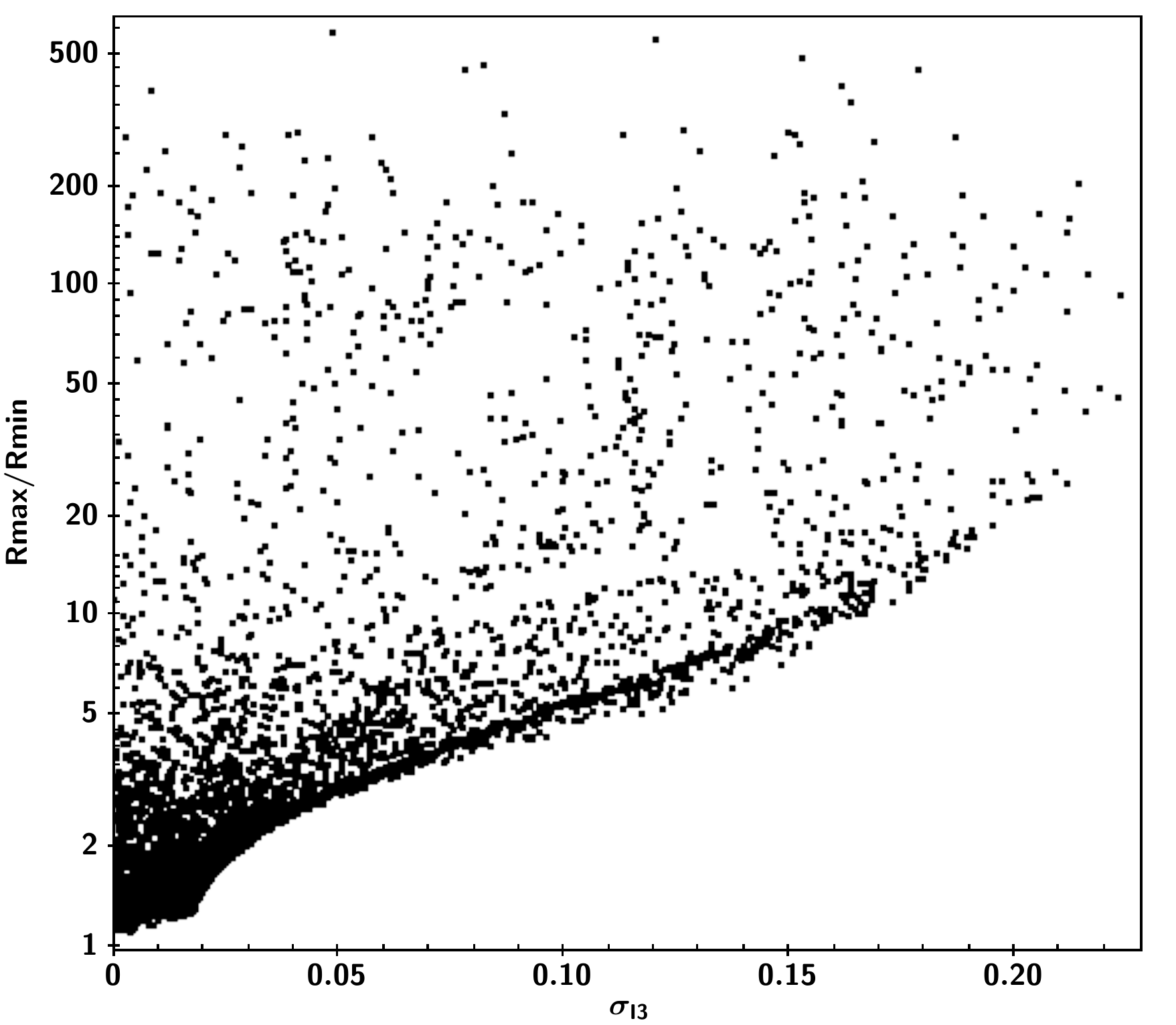}}
}
\caption{Left: $\sigma_{I2}$ uncertainties on the $I_2$ determinations for 5000 orbits with $E=1$ versus their apocentre/pericentre ratio.
Right: $\sigma_{I3}$  uncertainties on the $I_3$ determinations for 5000 orbits with $E=1$ versus their apocentre/pericentre ratio. }
\label{fig:f2}
\end{center}
\end{figure}

Figure \ref{fig:f1}  shows that 45 per cent  of the orbits  have integrals with an accuracy better than two per cent, 68 per cent better than four per cent, and 91 per cent better than ten per cent.
For the smallest values of $\sigma_{I2}$ and $\sigma_{I3}$, these dispersions are very strongly correlated. 
Figure\,\ref{fig:f2} shows a  correlation between the apocentre to pericentre  ratio $r_{max}/r_{min}$  of the orbit, and the accuracy of $I_2$ or $I_3$. 
This shows that the tube orbits, which are the least eccentric orbits, have the most accurate quasi-integrals
 Thus, orbits with apocenter to pericenter ratios of less than three have integrals with dispersions of less than three per cent. When we consider $\sigma_{I3}$, this correlation mainly concerns the tube orbits (Fig. \ref{fig:f2}).

\begin{figure}[!htbp]
\begin{center}
\centerline{\includegraphics[scale=0.40]{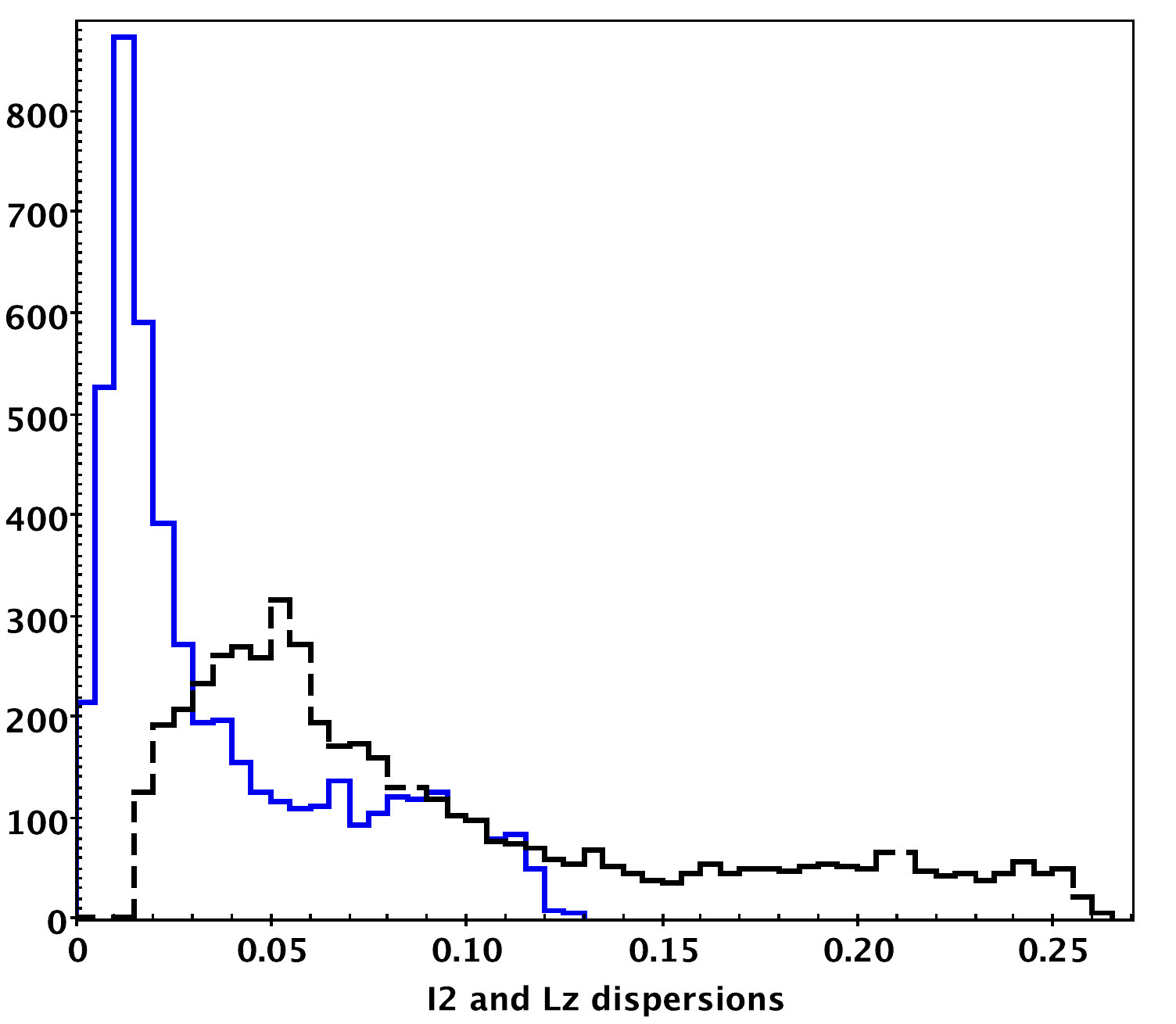},\includegraphics[scale=0.40]{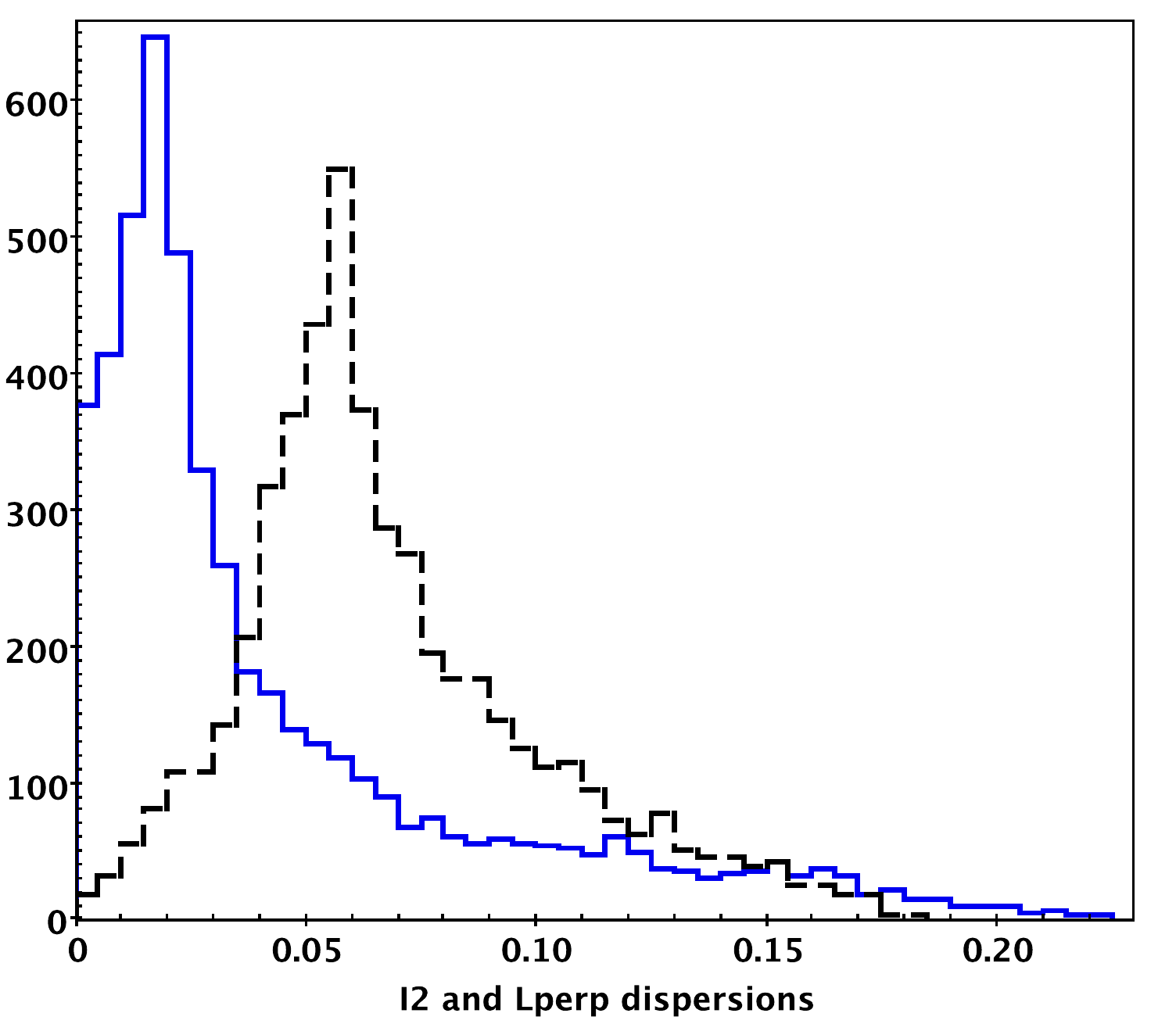}}
\caption{Comparison of accuracies achieved using St\"ackel (blue continuous line) or angular momentum (black dashed line) approximations.
 Left: histograms of $\sigma_{I2}$ and $ \sigma_{Lz}$.
  Right: histograms of $\sigma_{I3}$ and $ \sigma_{L\perp}$.
 }
\label{fig:f4}
\end{center}
\end{figure}

It is interesting to compare these integrals with the angular momenta $|L_z|$ and $L_\perp =\sqrt{L_x^2+L_y^2}$ which are frequently used as quasi-integrals \citep[for example, see][]{hel17}. These last quantities are easily computed and do not require the integration of orbits. These angular momenta are approximate integrals that remain tolerable in the case of an almost spherical potential. For the potential considered here, which is  close to spherical, they are however  significantly less accurate than St\"ackel approximations.
Figure\,\ref{fig:f4} shows a comparison of the histograms of the relative accuracies of these different quasi-integrals. The difference is most noticeable for the least eccentric orbits, which are also the most accurate. For these orbits, we note an improvement by a  factor of five ($I_2$) and of three ($I_3$) for the accuracy gain. The same gain is obtained  on the position of the peak of the $I_2$ and  $I_3$ histograms, also  in favour of the St\"ackel approximations.    

\begin{figure}[!htbp]
\begin{center}
\centerline{\includegraphics[scale=0.4]{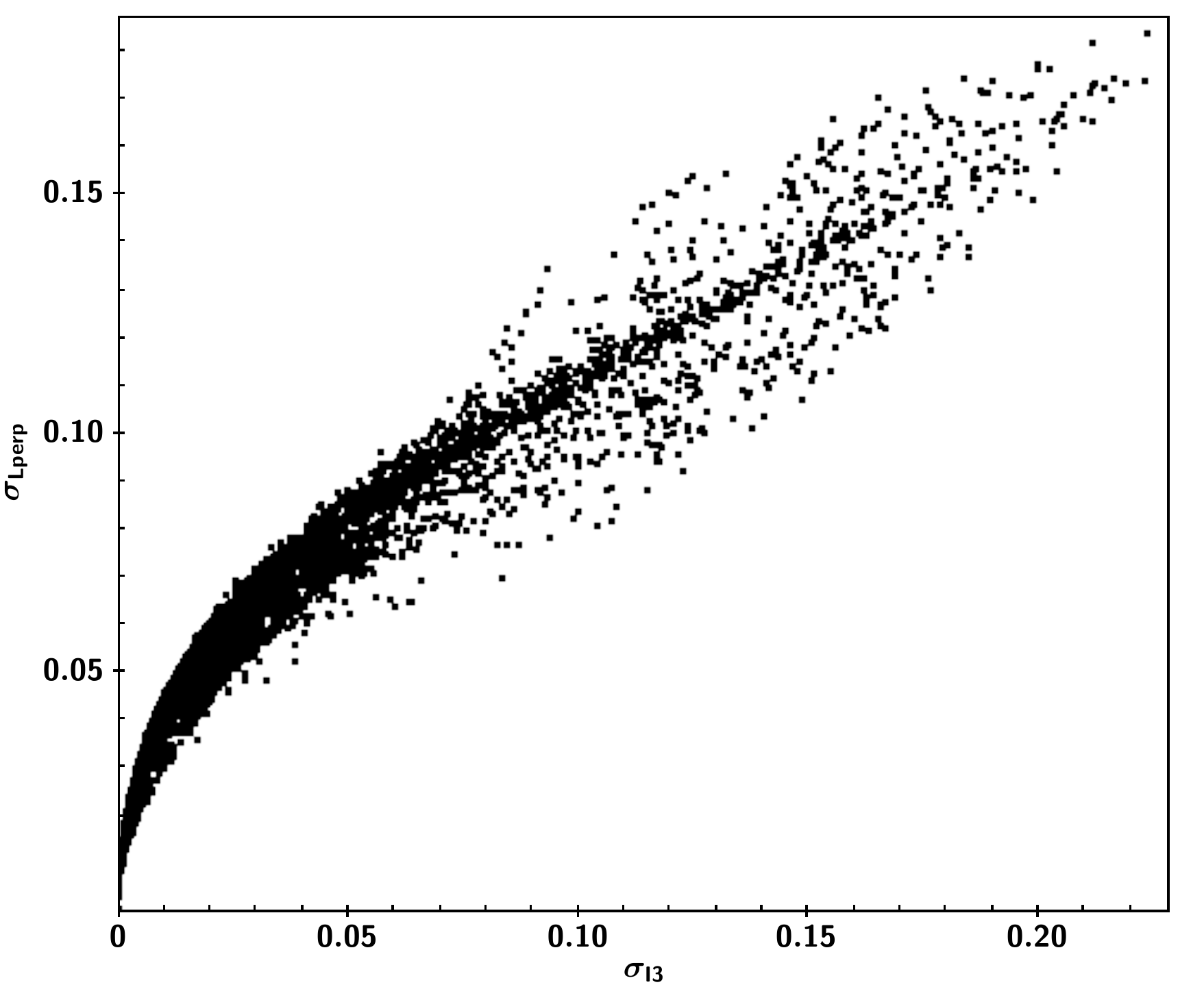}
, \includegraphics[scale=0.4]{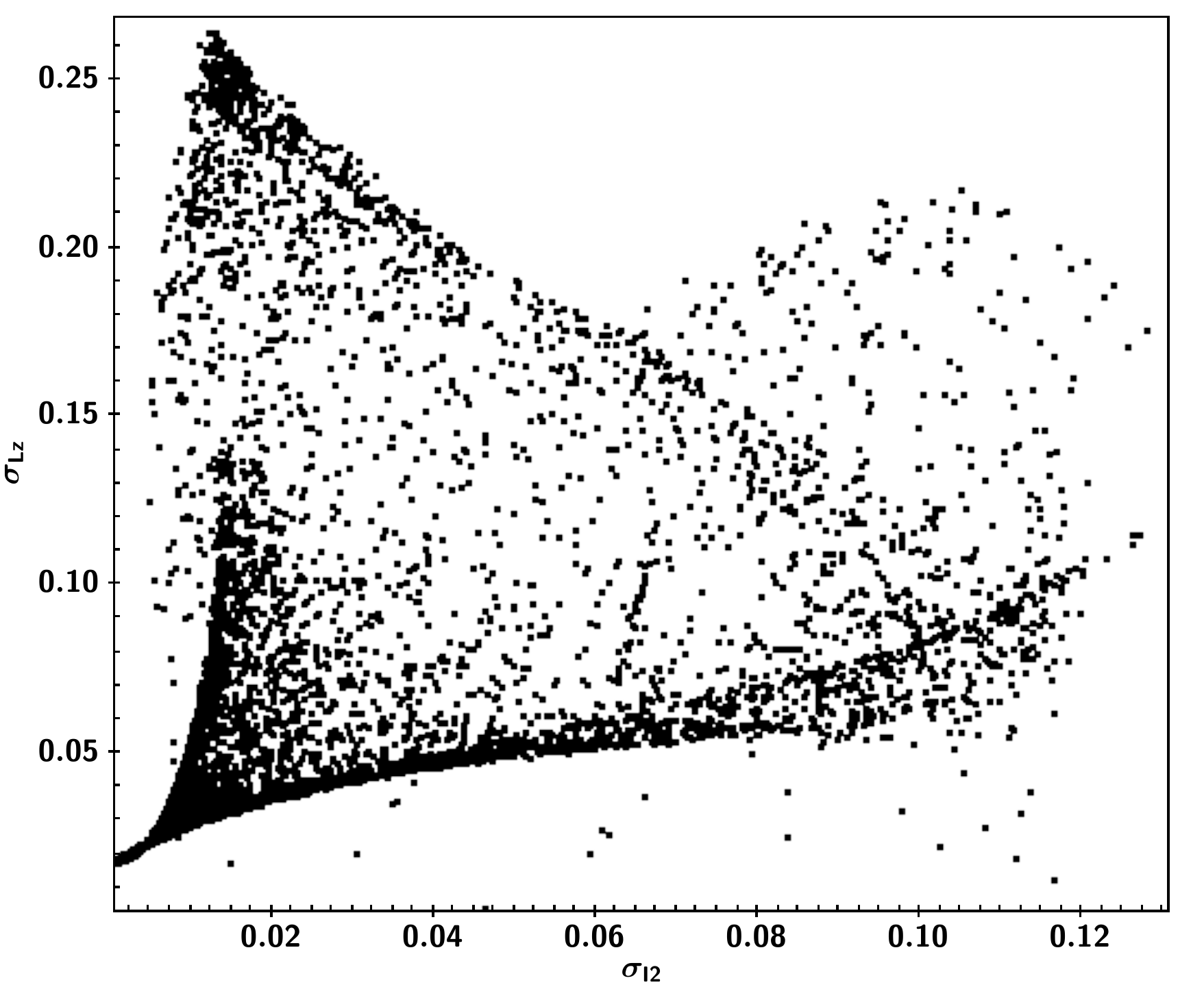}
}
\caption{Comparison of accuracies. Left:  $\sigma_{L_\perp}$ versus $\sigma_{I3}$ for 5000 orbits.
Right: $\sigma_{Lz}$ versus $\sigma_{I2}$ for 5000 orbits.
}
\label{fig:f5}
\end{center}
\end{figure}

\begin{figure}[!htbp]
\begin{center}
\centerline{\includegraphics[scale=0.37]{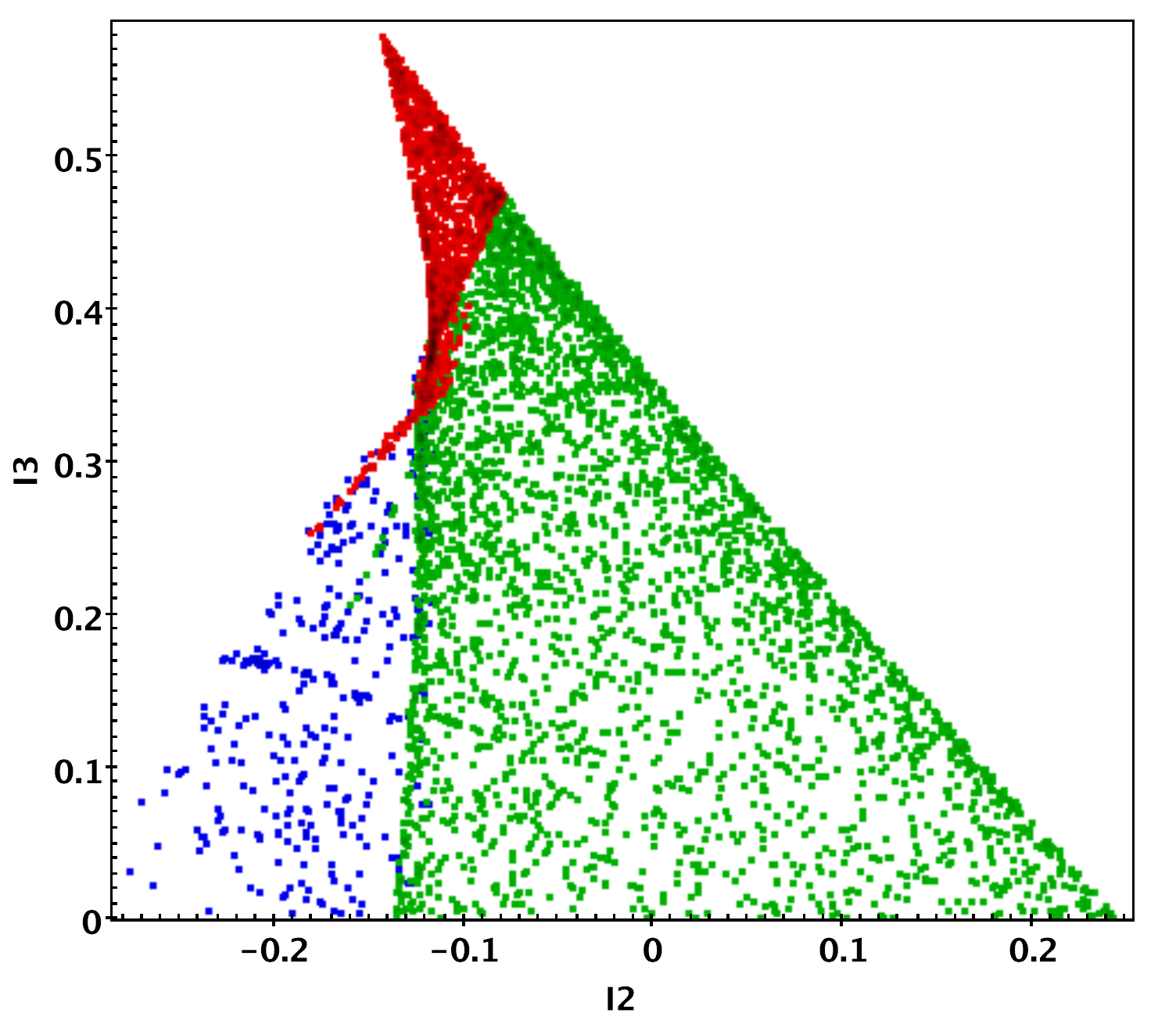}\space
\includegraphics[scale=0.37]{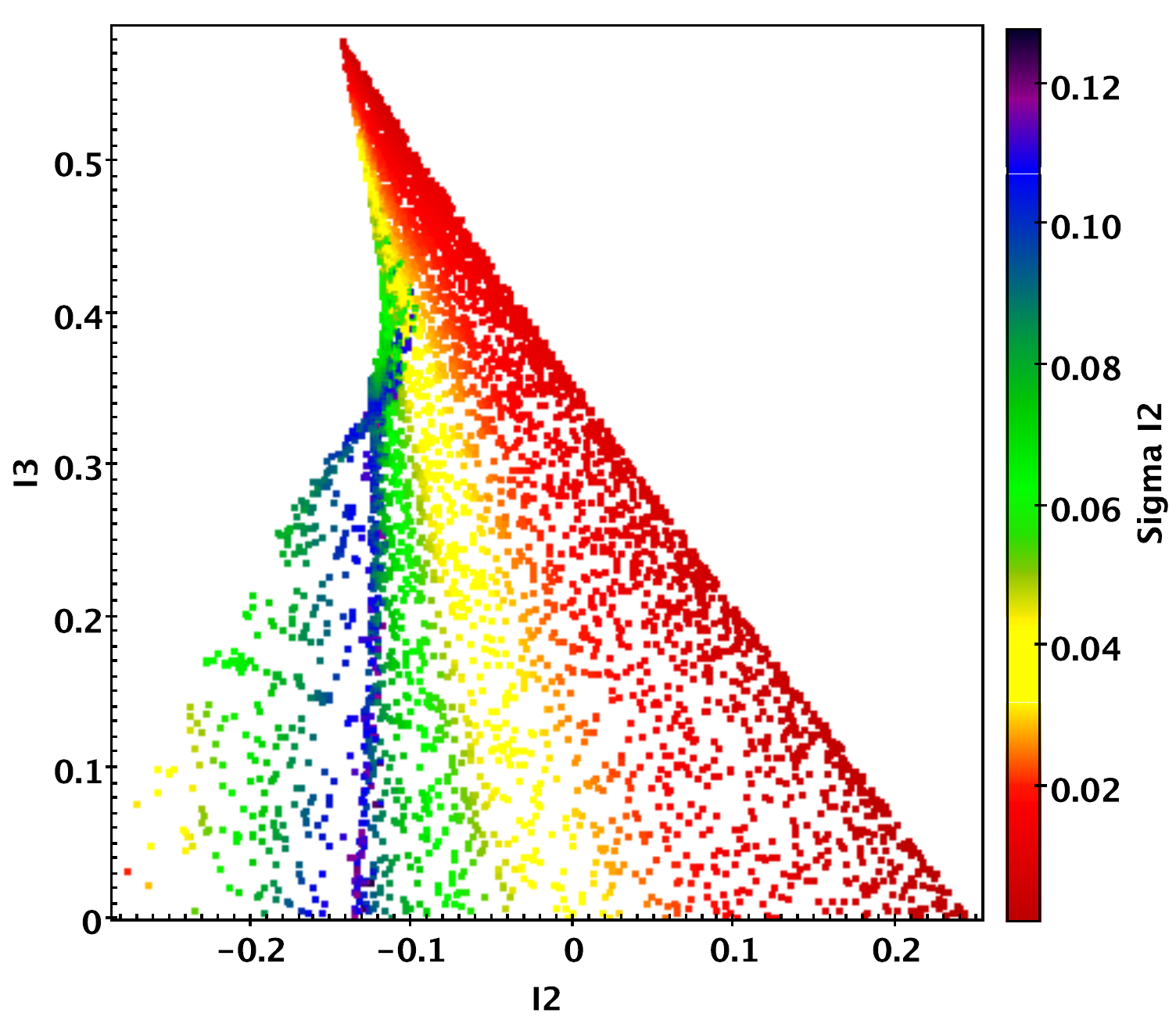}\space
\includegraphics[scale=0.37]{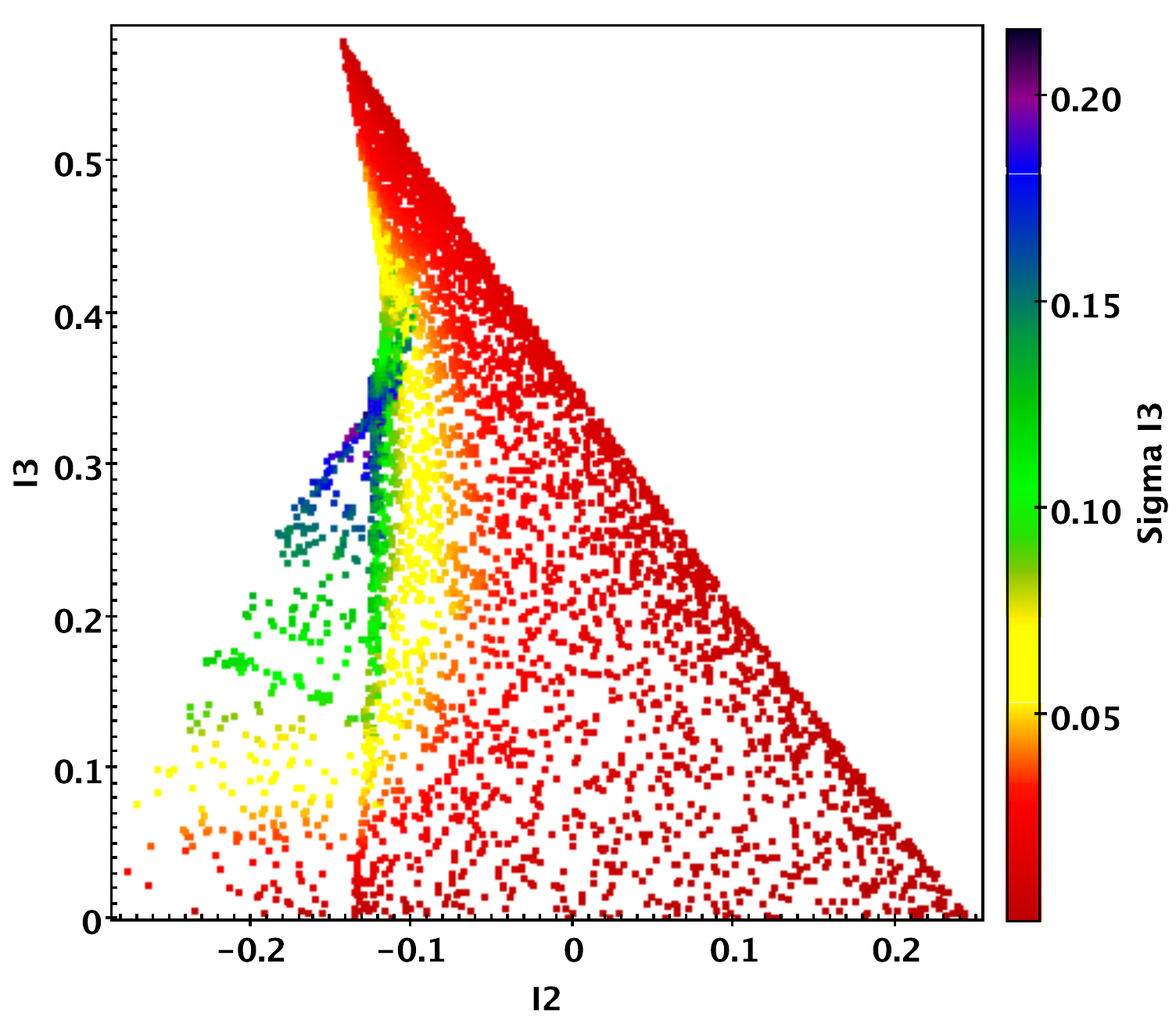}}
\caption{Two-dimensional classification of orbits. Each point $(i_2, i_3)$ corresponds to an orbit. Left: Different types of orbits are the box orbits (blue), the outer long axis tube orbits (red), and the short axis tube orbits (green).
Centre: The $\sigma_{I2}$ dispersion is colour-coded.
Right: The $\sigma_{I3}$ dispersion is colour-coded.}
\label{fig:f7}
\end{center}
\end{figure}

\begin{figure}[!htbp]
\begin{center}
\centerline{\includegraphics[scale=0.25]{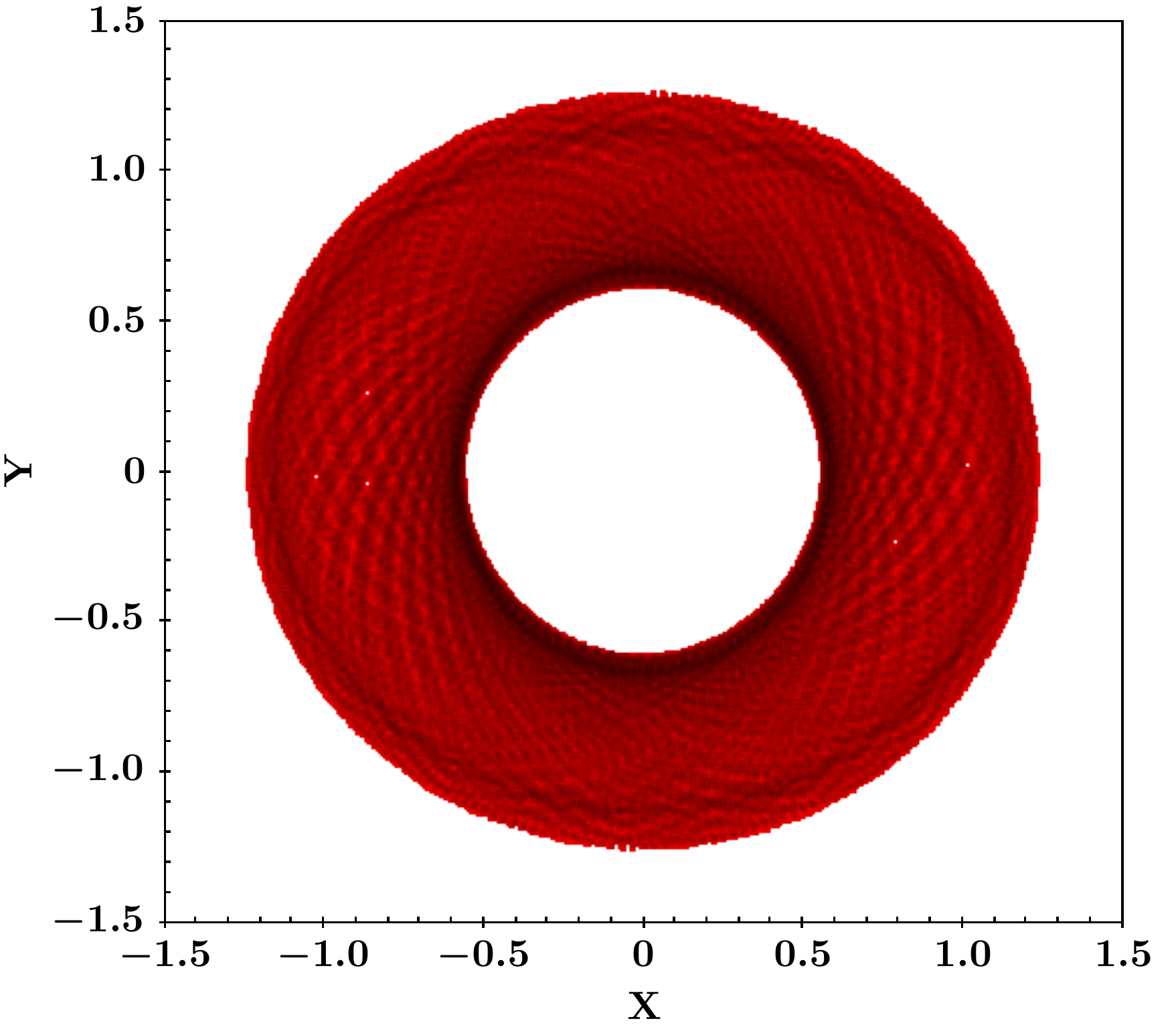},\includegraphics[scale=0.25]{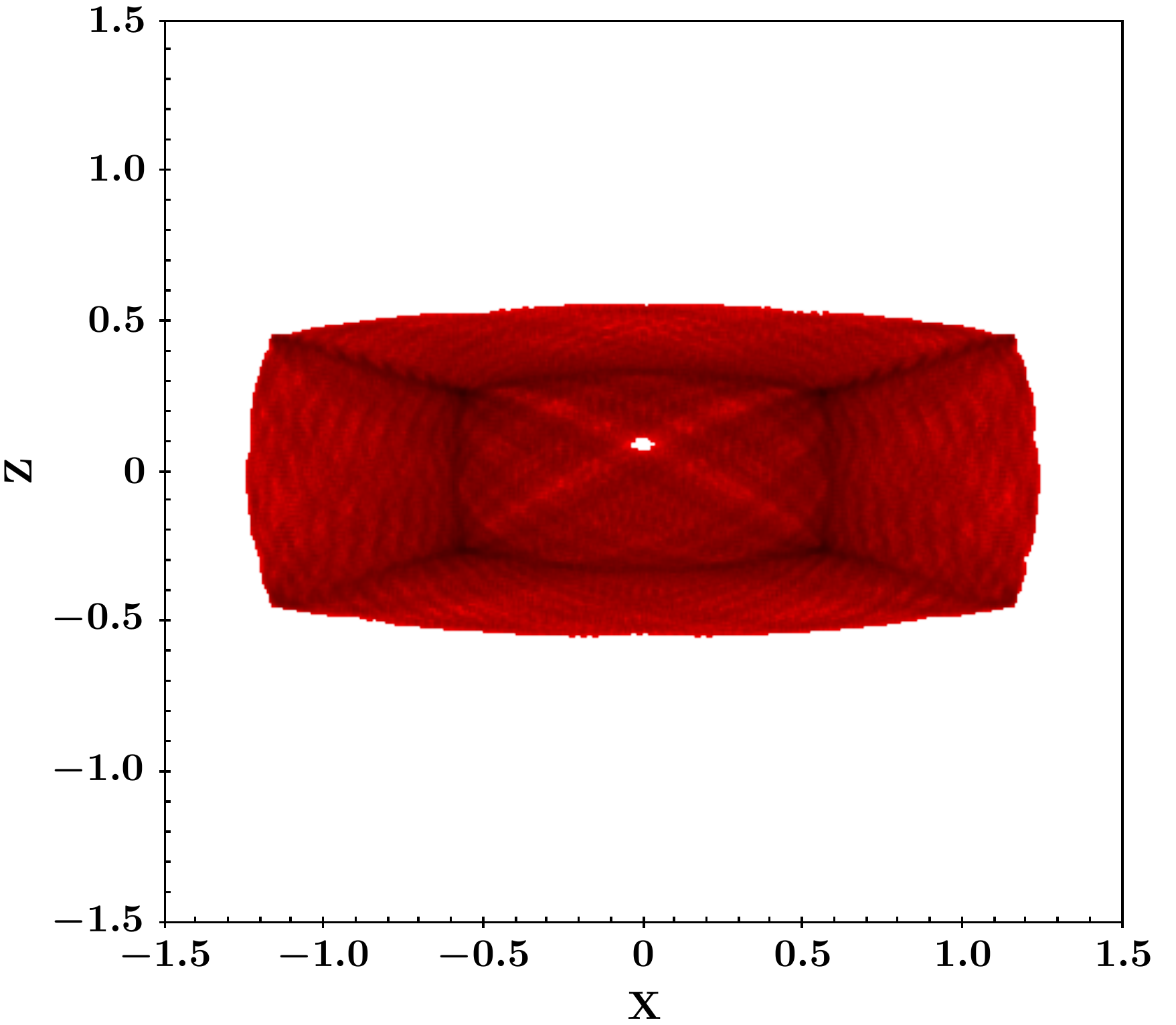},\includegraphics[scale=0.25]{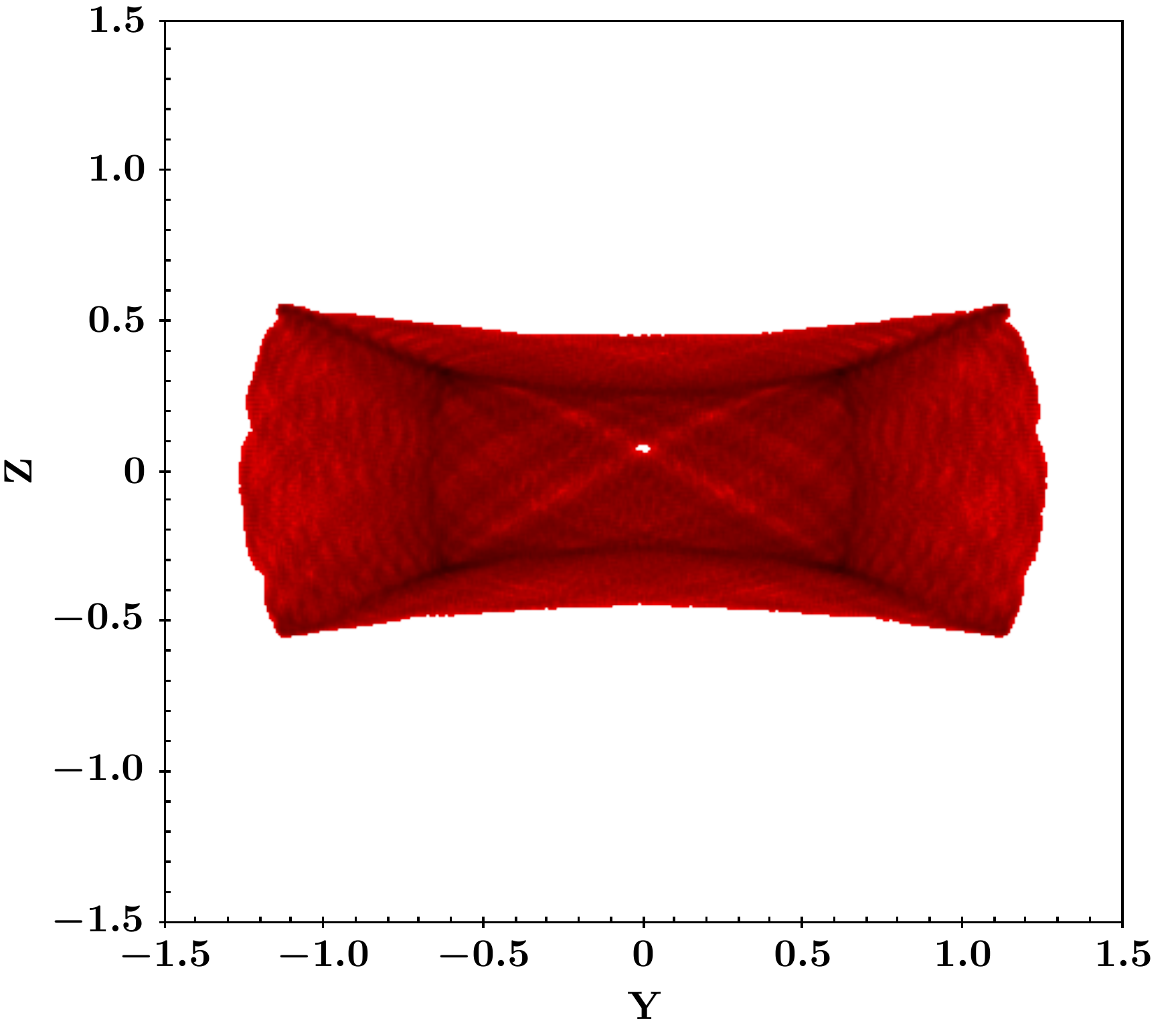}}
\centerline{\includegraphics[scale=0.25]{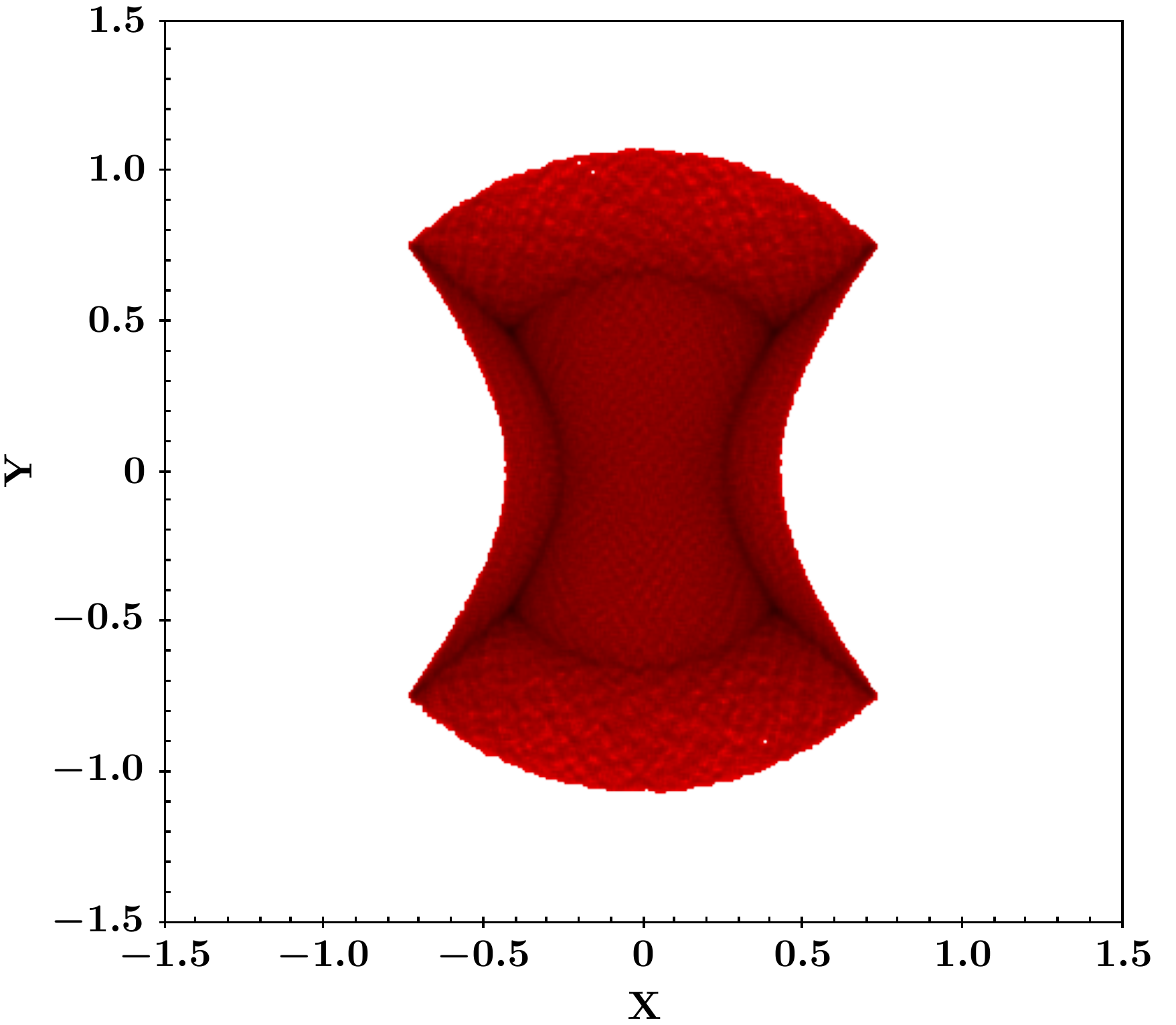},\includegraphics[scale=0.25]{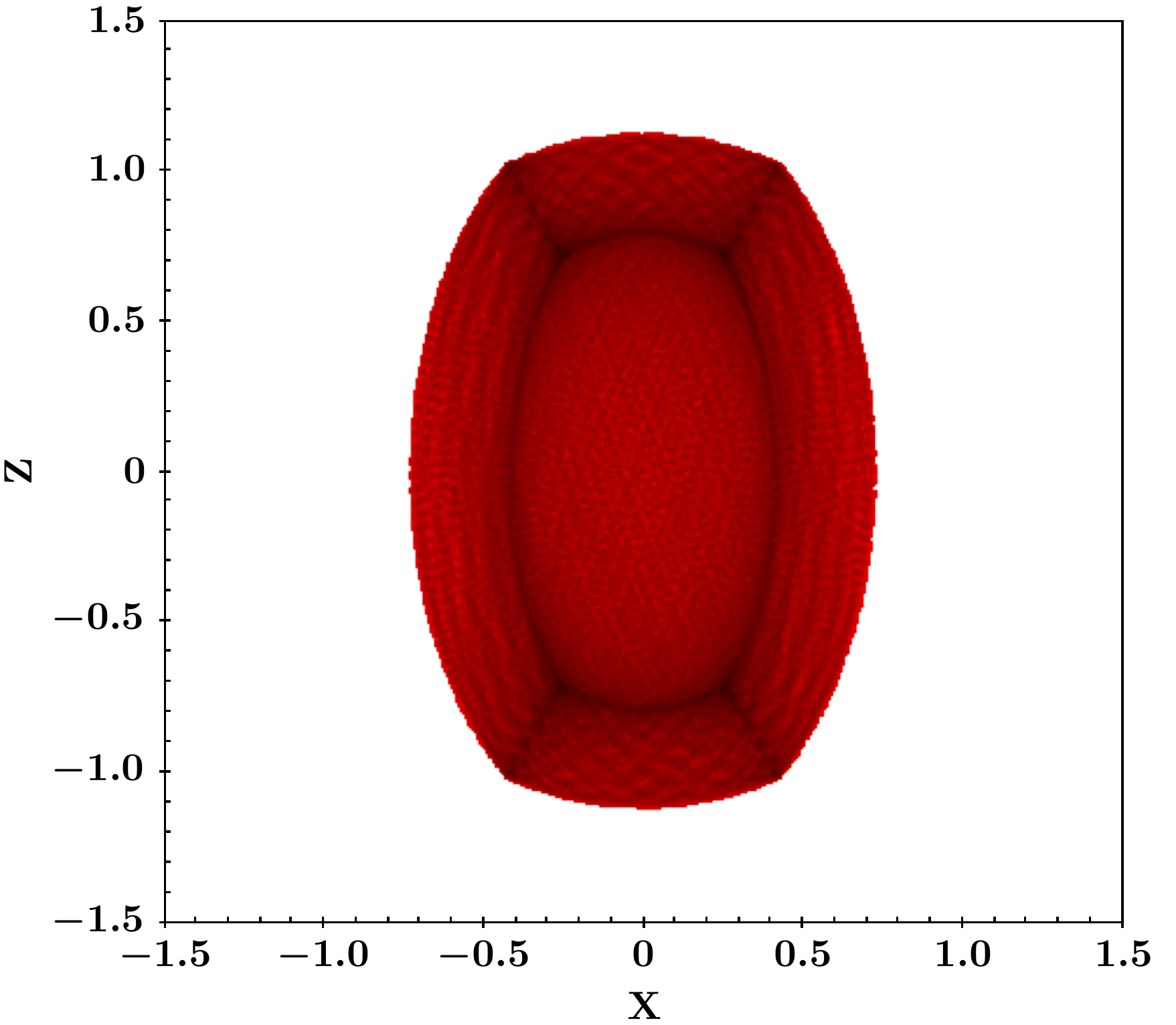},\includegraphics[scale=0.25]{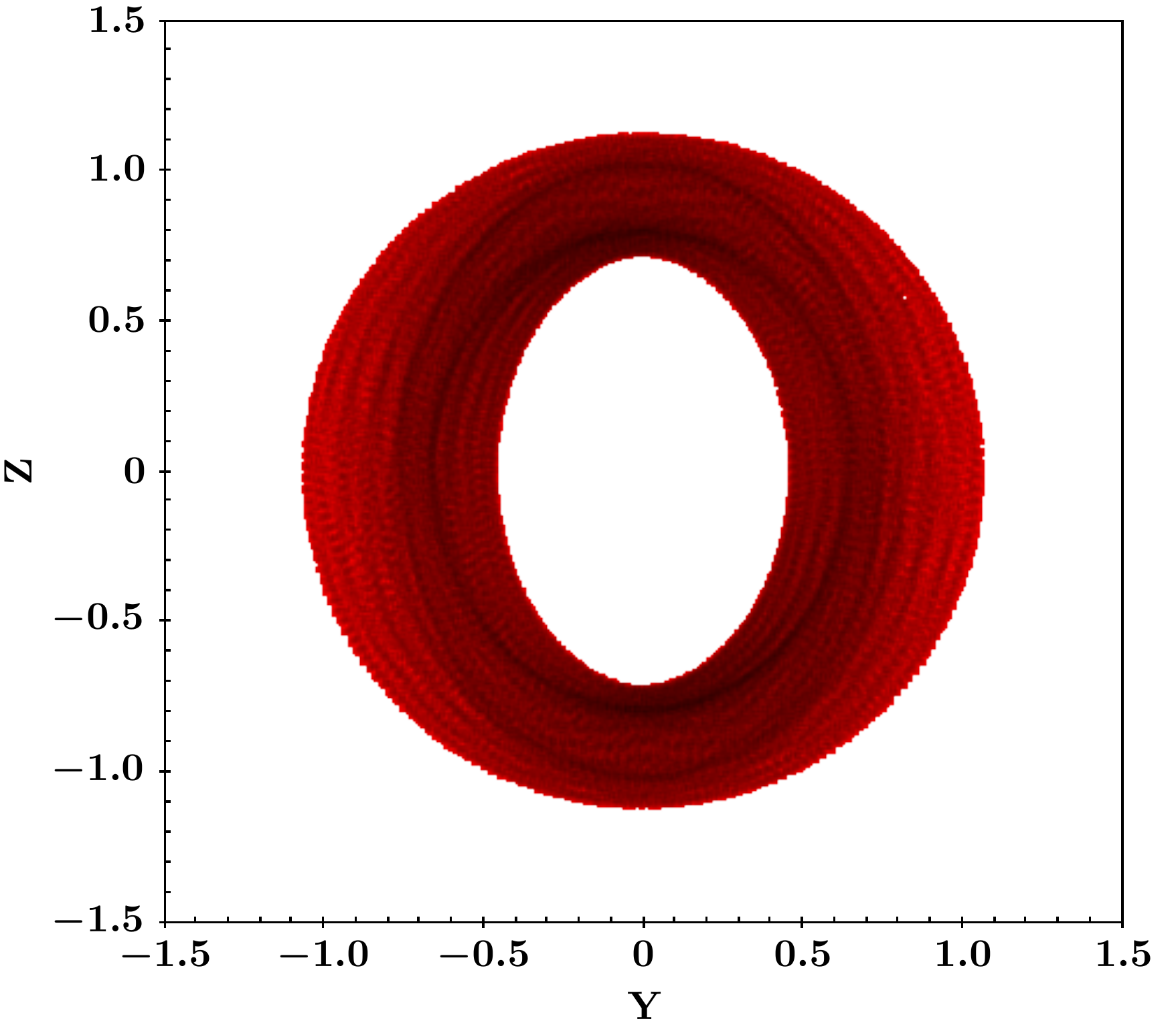}}
\centerline{\includegraphics[scale=0.25]{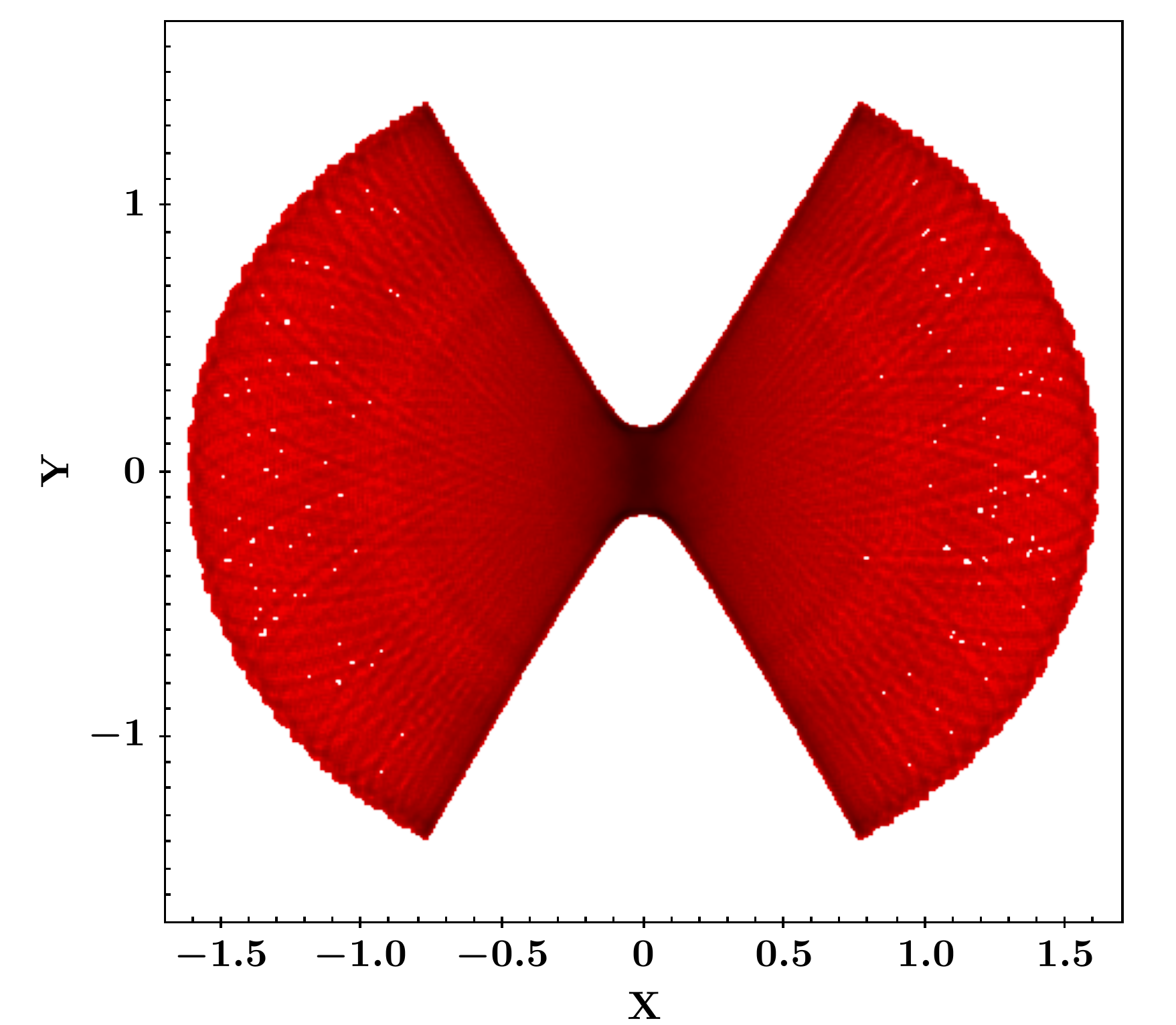},\includegraphics[scale=0.25]{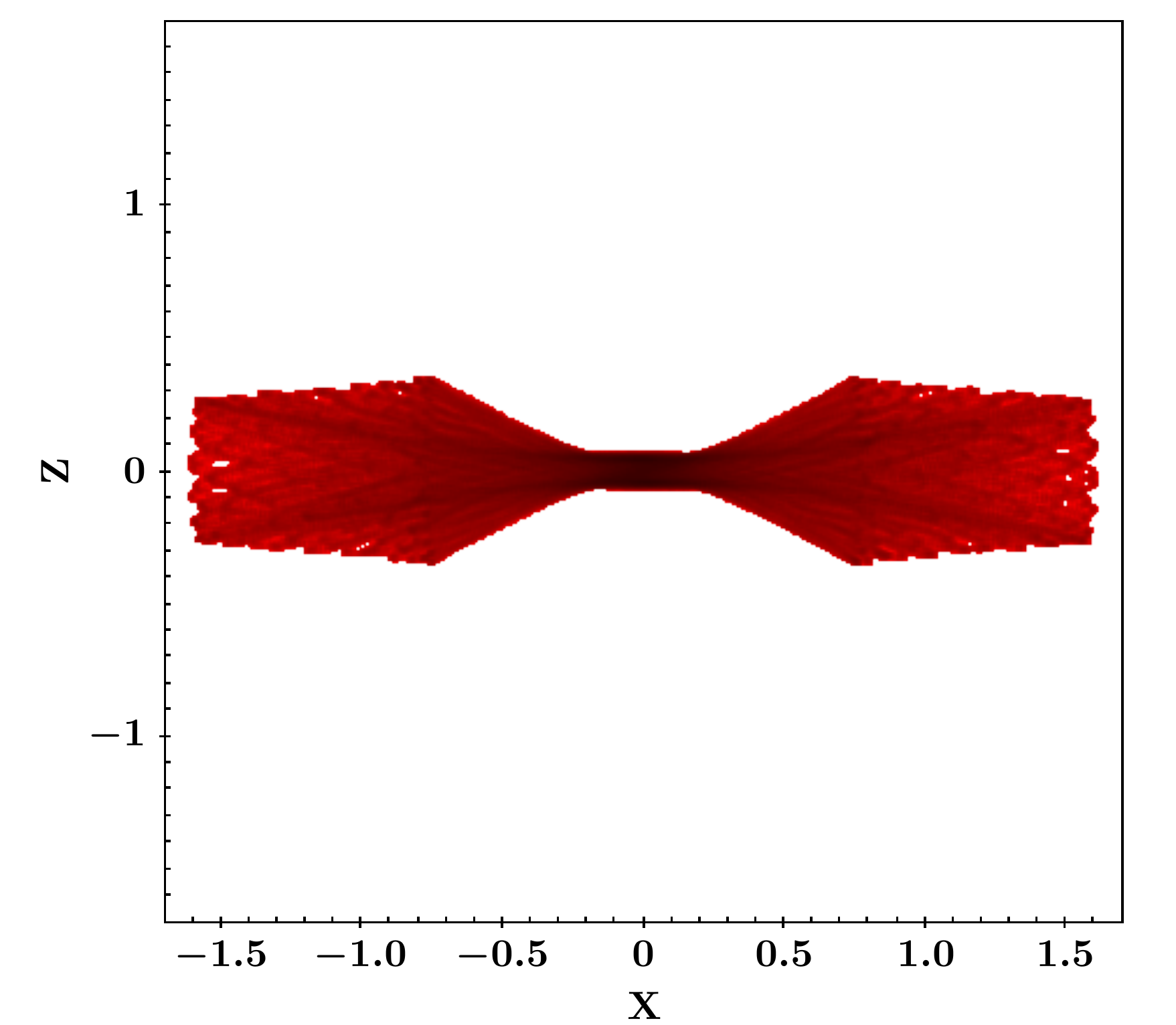},
\includegraphics[scale=0.25]{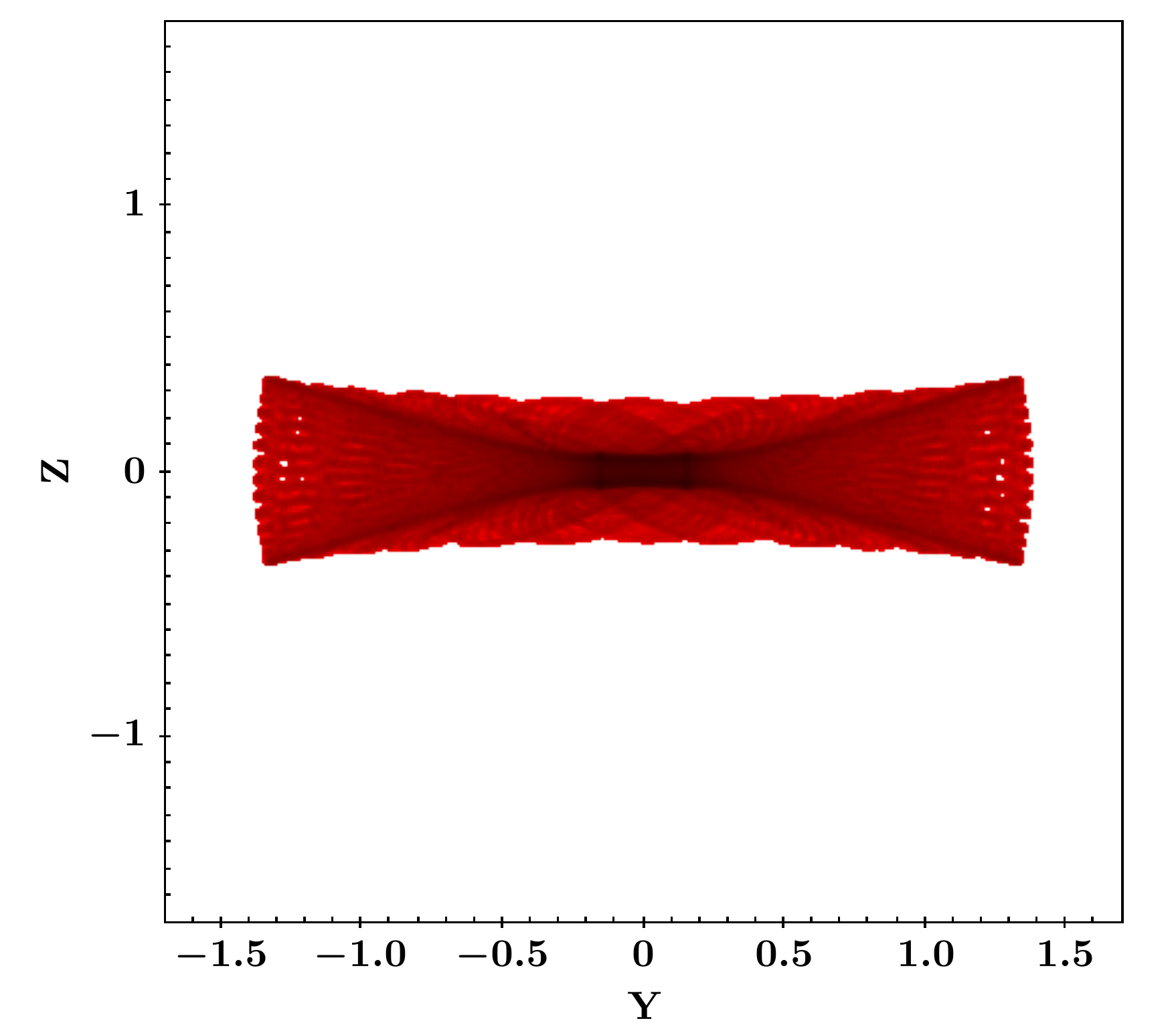}}
\caption{From top to bottom: a short-axis tube orbit, a long-axis tube orbit, and a box orbit } 
\label{fig:f8}
\end{center}
\end{figure}

Figure\,\ref{fig:f5} shows the correlation between the accuracies $\sigma_{I_3}$ and $\sigma_{L_{\perp}}$.
  For $\sigma_{I2}$ and $\sigma_{L_z}$ the correlation also exists  but mainly for tube orbits while box orbits have lower accuracies (Fig. \ref{fig:f5}). 
Figure\,\ref{fig:f7}a shows the distribution of the three main types of orbits  in the $(I_2, I_3)$ plane:
 box orbits and tube orbits, these latter ones classified according to their orientation along the major axis or the minor axis of symmetry of the potential. Figure \ref{fig:f7}a can be compared to Figs. 17 and 18 in \citet{dez85a}, which classifies the different families of orbits (see Fig. \ref{fig:f8}).
  His figures show a three-dimensional classification of orbits in terms of $E$, $I_2$, and $I_3$ for bound orbits of a triaxial St\"ackel potential and a cross-section for a given energy $E$.
Figures \ref{fig:f7}b and \ref{fig:f7}c show the distribution of orbits  with their accuracy colour-coded.

\begin{figure}[!htbp]
\begin{center}
\centerline{\includegraphics[scale=0.38]{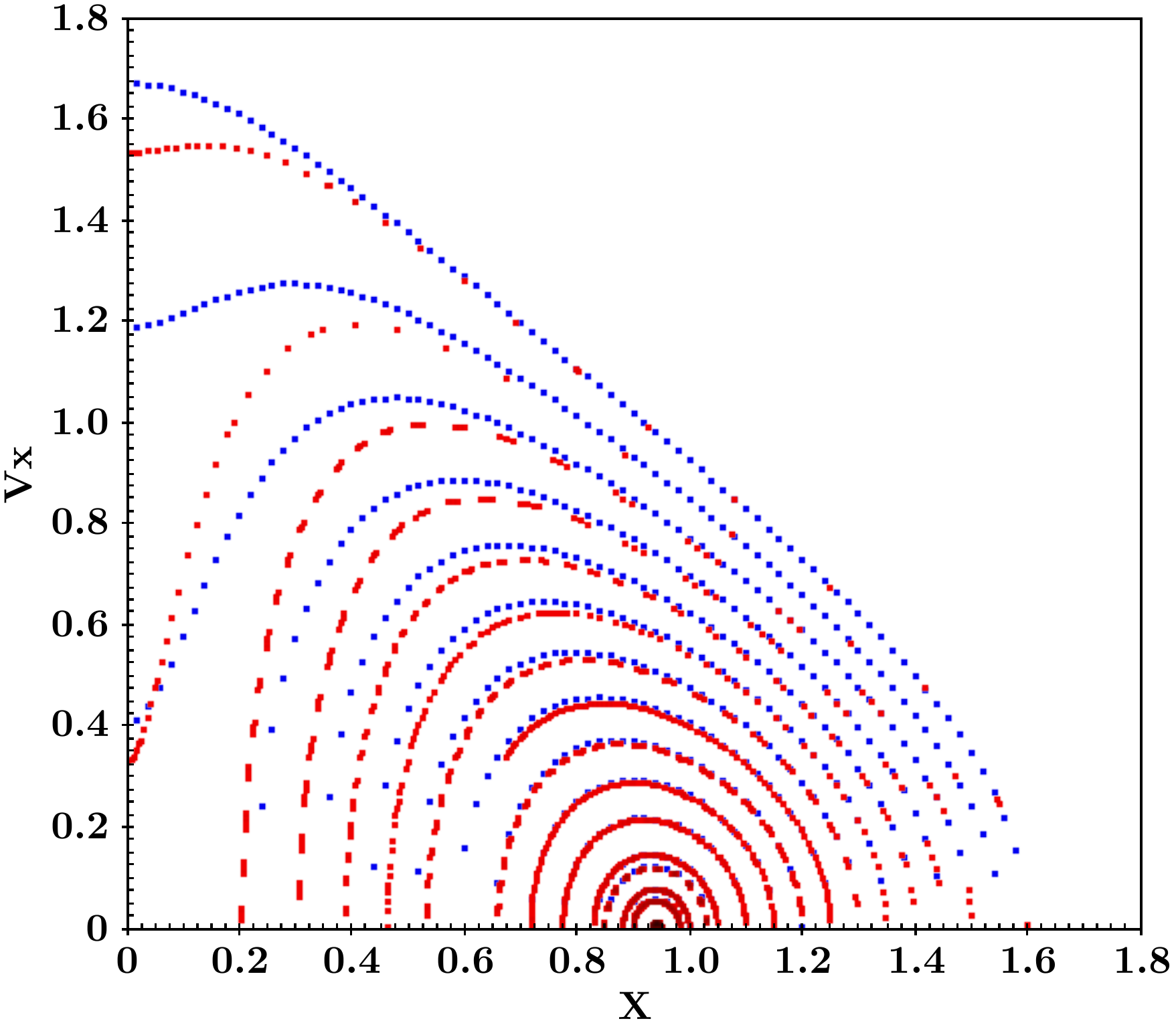},\includegraphics[scale=0.38]{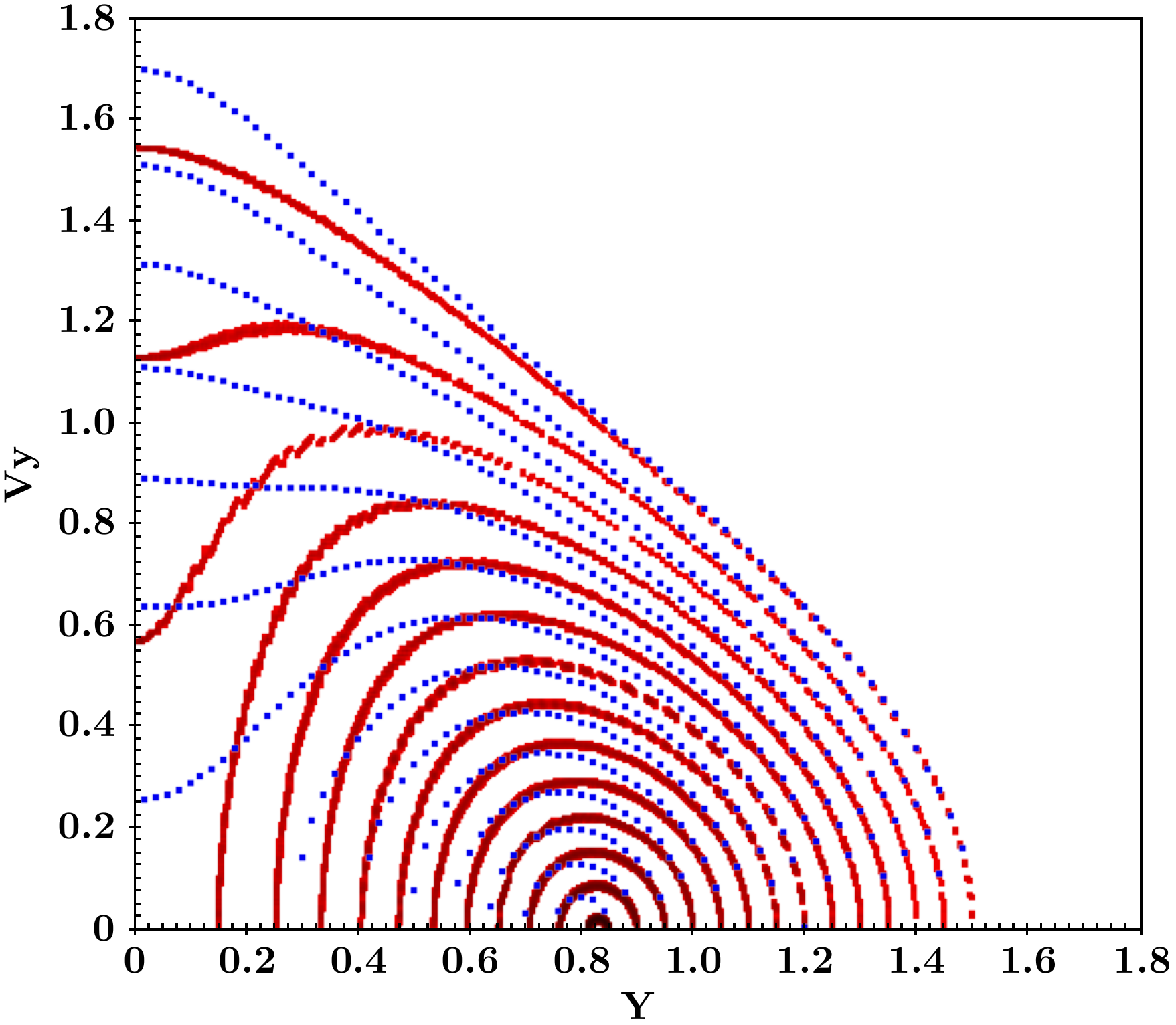}}
\caption{Poincar\'e sections. Red: orbit computation. Blue: from the analytic quasi-integrals. 
Left: $(x,v_x)$ section for orbits within the $(x,y)$ plane, ($z=\dot{z}=0$).
Right: $(y,v_y)$ section for orbits within the $(y,z)$ plane, $x=\dot{x}=0$.
 }
\label{fig:f11}
\end{center}
\end{figure}

Figure \ref{fig:f11} shows a section of the phase space for two series of orbits, each figure corresponding to orbits in the $(x,y)$ and $(y,z)$ planes of symmetry of the potential.  Contours in sections are calculated numerically from the orbital trajectories on the one hand, and on the other hand they are obtained using our analytic expressions of the approximate integrals.
The tube orbits close to the central periodic  tube orbit are best represented by the analytic expression  since the parameters $\Delta_1$ and $\Delta_3$ of the St\"ackel potentials have been adjusted on these orbits 

\begin{figure}[!htbp]
\begin{center}
\centerline{\includegraphics[scale=0.4]{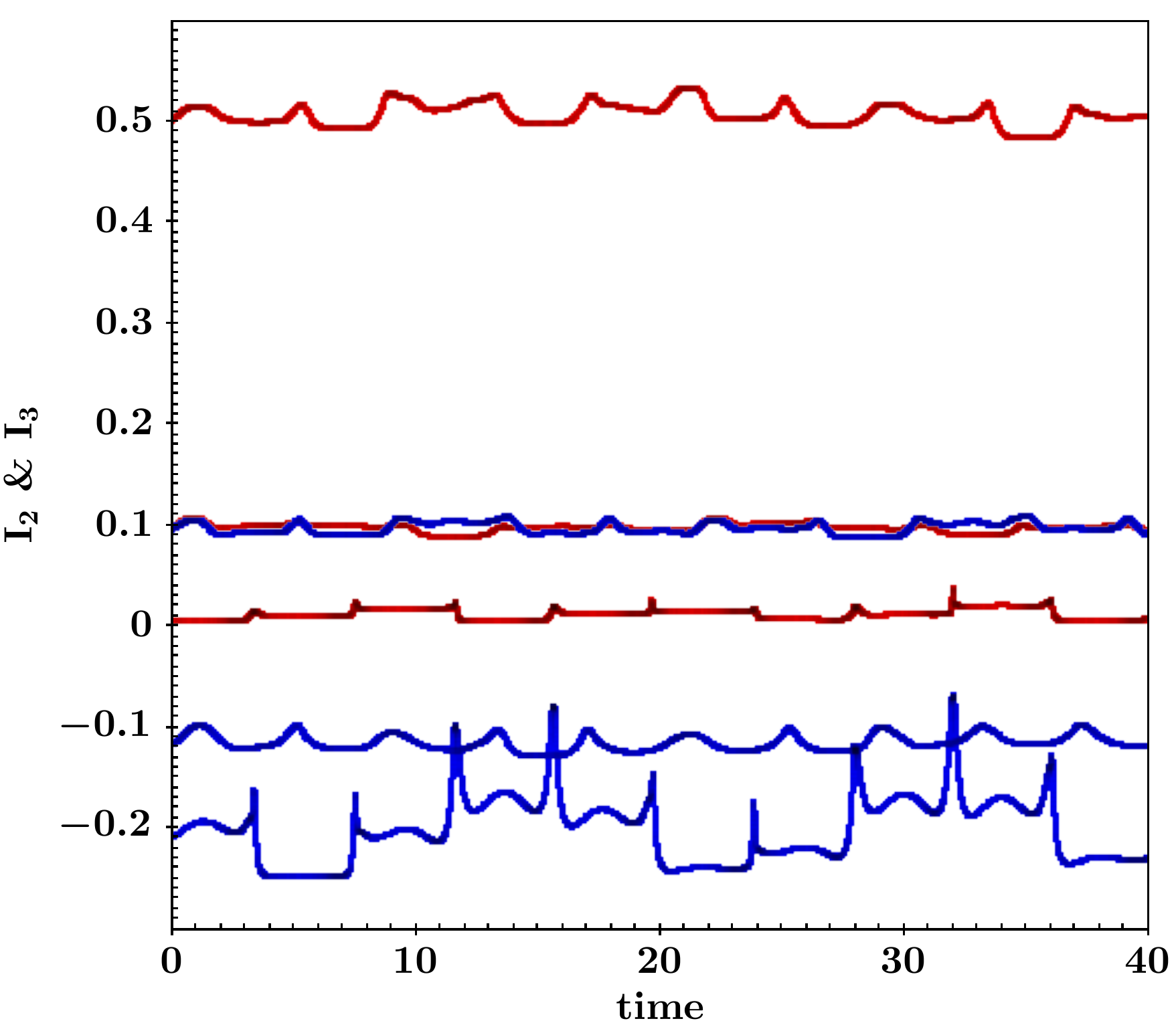}}
\caption{
Variation of $I_2$ (blue) and $I_3$ (red) for three orbits. Over the short time-length of the order of an average rotation period $(\Delta t=3),$ the variations of the quasi-integrals are significantly smaller compared to the variations over long periods.
 The largest variations of the quasi-integrals are observed for $I_3$($\sim-0.2$) (bottom curve) of the box orbit ($I_2\sim0$).
The sharp variations of $I_3$ are related to the passage at the pericentre.
 The short-axis tube ($I_2\sim I_3\sim0.1$) and the long-axis tube ($I_2\sim-0.1$ and $I_3\sim0.5$) orbits have smoother variations of the quasi-integrals.
 }
\label{fig:f12}
\end{center}
\end{figure}

It must be noted that we calculated the variations of $I_2$ and $I_3$  over a  long time interval, several thousand rotations around the centre. If we were interested in studying Galactic stellar streams, for example, these would have  short extension along their orbit, and  the dispersions of $I_2$ or $I_3$ along that extension would be significantly smaller. Figure \ref{fig:f12} shows that abrupt variations of the quasi-integrals occur during the transit close to the pericentre where the  potential variation is the most rapid. This is very sensitive for box orbits that show the greatest variations, but between two transits, away from the pericentre,  the quasi-integrals  are significantly more stationary.

In Fig.\,\ref{fig:f11} we note the lack of visible resonant orbits: either they are actually absent or they occupy a small space. To conclude, it can be expected that some simple modifications of the analytic expressions of $I_2$ and $I_3$ should lead to better agreement between the analytic and numerical phase space sections (Fig. \ref{fig:f11}). A difficulty to obtaining better analytic  quasi-integrals is to extend this modification to the entire volume of the phase space and not just to a few sections of the phase space. Such a modification with additional parameters must however exist as it has been obtained in \citet{bie13}, but in this latter case using a considerable number of free parameters to build approximate integrals.
\\


\section{Distribution functions and Jeans equations}

The developments described above to model the gravitational dynamics can have several practical applications.
An initial application is the search for stellar streams in the Galactic halo, which can be carried out by identifying the membership of stars in nearby orbits \cite{mal18a}  or by identifying 
  concentrations in the space of the integrals of motion. The quasi-integrals developed here allow us to take into consideration the probable triaxiality of the Galactic halo \citep{law10}. Conversely and simultaneously, this makes it possible to determine the Galactic potential  \citep{mal18b}.
Another important application is the construction of distribution functions dependent on three integrals of motion. Such distribution functions make it possible to calculate the velocity moments and density of a stationary stellar population. In addition to describing stellar populations, they can also be used to measure the gravitational potential using the Jeans equations in the case of triaxial potentials.

We test here the accuracy of a triaxial distribution function by applying the Jeans equations to find the potential used to build this distribution function. This test has already been used to validate the stellar disc distribution functions
\cite[generalisation of Shu distribution,][]{shu69}
for the Besan\c{c}on Galaxy model \citep{bie15}. This last model is axisymmetric and in addition to the energy and kinetic moment, we used a third approximate integral of the St\"ackel type, a particular case of the study presented here.
\cite{san14} also used the Jeans equations as tests of their approximate integrals for axisymmetric models  for non-axisymmetric models.

A second possible application of the two integrals $I_2$ and $I_3$ is the generalisation of Shu's distribution functions for stellar discs with slight ellipticity.  In this case, 
in Shu's three-dimensional axisymmetric version \citep[Eqs. 8-9 in ][]{bie15}, it is sufficient to replace the $L_z$ angular momentum with its generalisation $I_2,$ and to replace the $E_{circ}$ energy of the circular orbit having the $L_z$ angular momentum with the energy of the  main closed orbit having the value $I_2$ as a second integral.

Another application, the one we develop here, is the use of integrals to build a triaxial distribution function that models the halo stellar distribution.

\subsection{Prescribed  halo distribution functions}

\subsubsection{$f(E,L^2$) }

In order to test the relevance of the quasi-integrals  through the Jeans equations, we first constructed a distribution function for the stellar halo in the particular case of an axisymmetric logarithmic potential.
This distribution function for a spherical potential is inspired by the stationary distribution: 
\begin{equation}
f(E,L)= \exp \left[-(E+  L^2/2\alpha^2) \right] \, .
\label{equ:}
\end{equation}
\cite{osi79} and \cite{mer85} showed that for a spherical potential, any distribution function of the form $f(E+ L^2/2 \alpha^2)$ 
 has an isotropic velocity distribution at the centre and is radially anisotropic at a large distance from the centre, 
 ($\sigma_r$ does not depend on $r$
and $\alpha$ is the transition radius from the inner isotropic core to the outer regions) by following the relationship
\begin{equation}
\frac{\sigma_r^2}{\sigma_t^2}=\frac{\sigma_r^2}{ \frac{1}{2} <v_t^2> }  = 1 + \frac{r^2}{\alpha^2} \, .
\label{equ:}
\end{equation}
The ratio of radial-to-tangential velocity dispersions changes continuously with  distance to the centre, and the parameter $\alpha$ defines the transition radius.
In the specific case of a logarithmic potential, this scale factor $\alpha$ can simply be made ineffective by replacing it in the distribution function by an amount that is both energy dependent and proportional to the $r$ radius.
Thus, with the potential
\begin{equation}
\Phi(r)= v_0^2 \log r \, ,
\label{equ:}
\end{equation}
  the radius of the circular orbit $r_{circ}$ of energy E is given by
\begin{equation}
E=v_0^2 \, \log(r_{circ}) + 1/2 \,v_0^2
\label{equ:}
,\end{equation}
and we deduced
 the following quantity related to it
\begin{equation}
 r_c(E)=  \exp [ E/v_0^2 -1/2 ] =
r \,\, \exp \left[  \frac{1}{2} \frac{v_r^2+v_t^2}{v_0^2} -\frac{1}{2} \right] \,.
\label{equ:}
\end{equation}
 
We then set for the distribution function:
\begin{equation}
f(E,L) = \exp \left[-  \frac{n}{v_0^2} \left( {E + a \, \, \frac{L^2} {r_c(E)^2} } \right) \right] \, , 
\label{equ:dfSph}
\end{equation}

with only  two free parameters, $n$ and $a$. We can see that with the potential being logarithmic, the term on the right-hand side separates into two terms. A first term contains only the variable $r,$  which gives for the associated density law: $\nu(r) \propto r^{-n}$. The second term contains only the velocity components, and therefore the velocity moments  are independent of the $r$ radius.
It should be noted that in the case of the isothermal sphere, $a=0$, the velocity dispersion is isotropic and the radial and tangential velocity dispersions are $\sigma_r=\sigma_t/\sqrt{2}=v_0/\sqrt{n}$.

\begin{figure}[!htbp]
\begin{center}
\centerline{\includegraphics[scale=0.3]{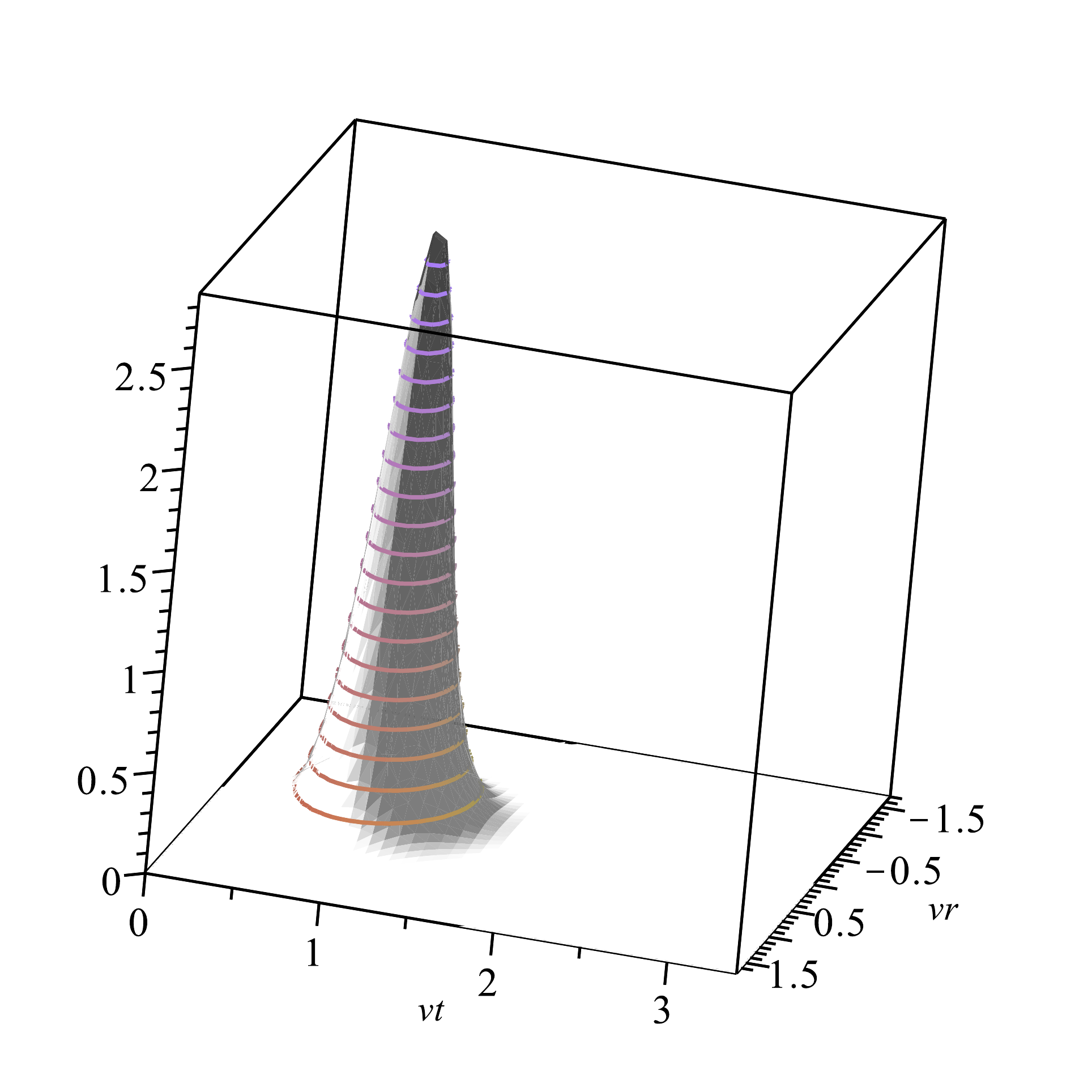},\includegraphics[scale=0.3]{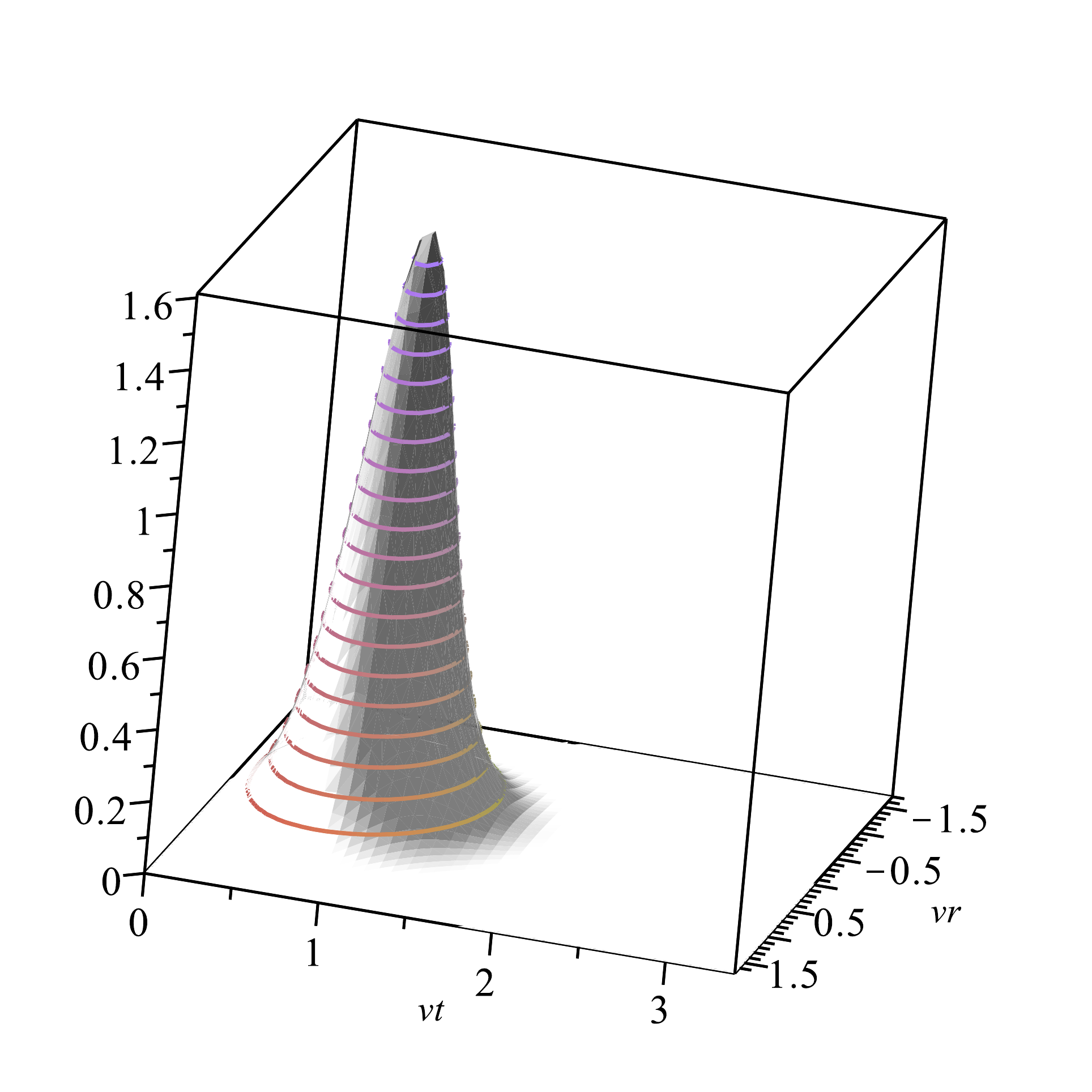},\includegraphics[scale=0.3]{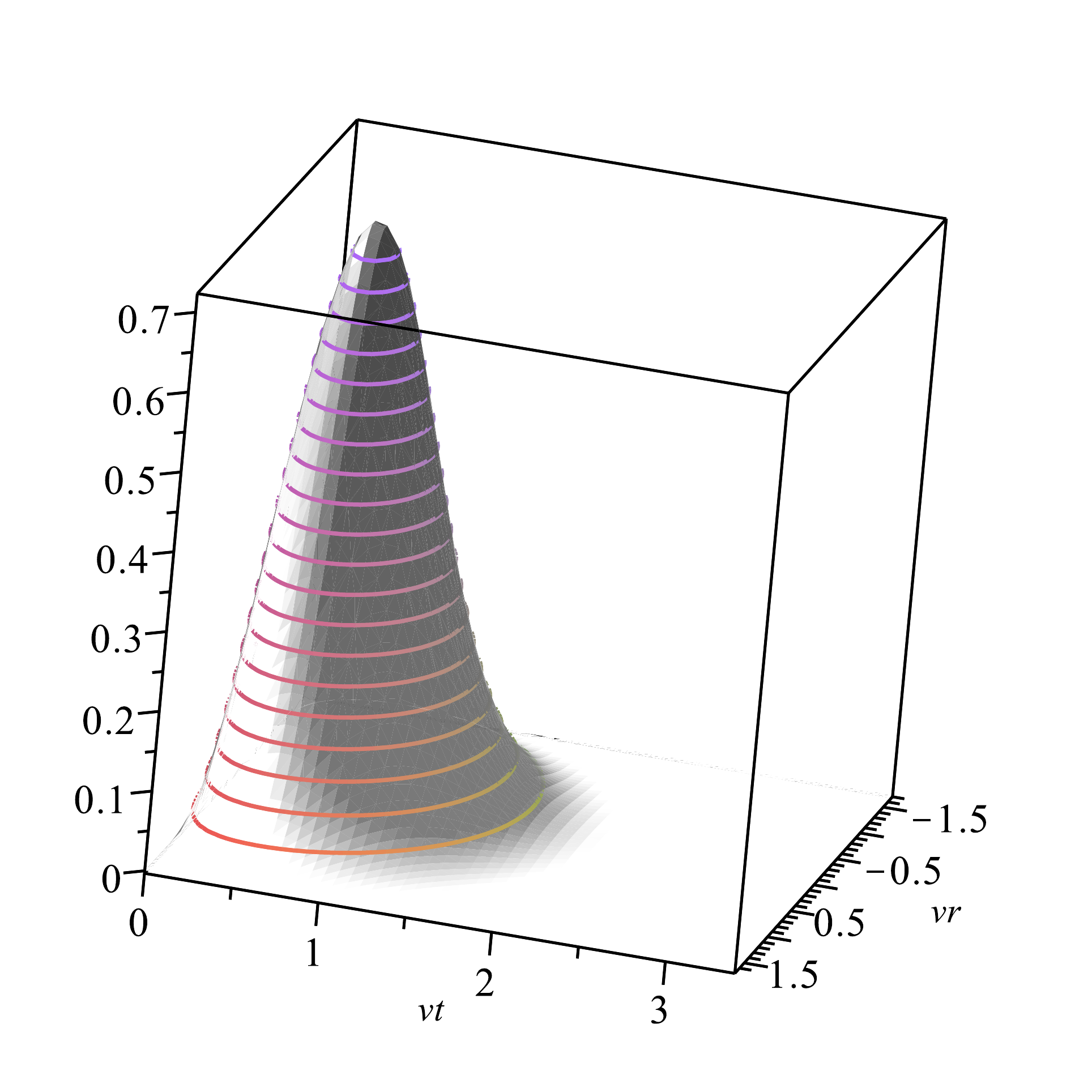},},
\centerline{\includegraphics[scale=0.3]{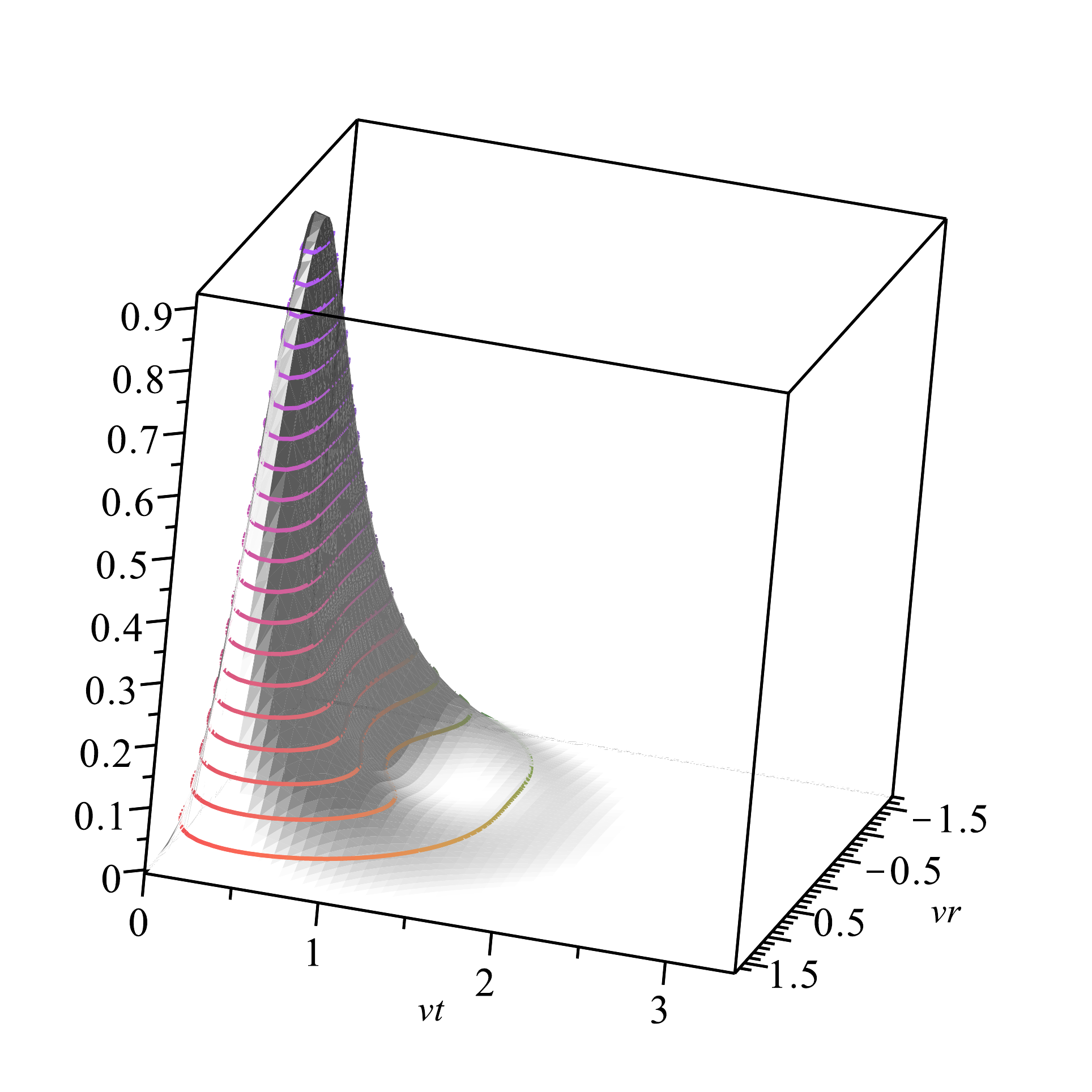},\includegraphics[scale=0.3]{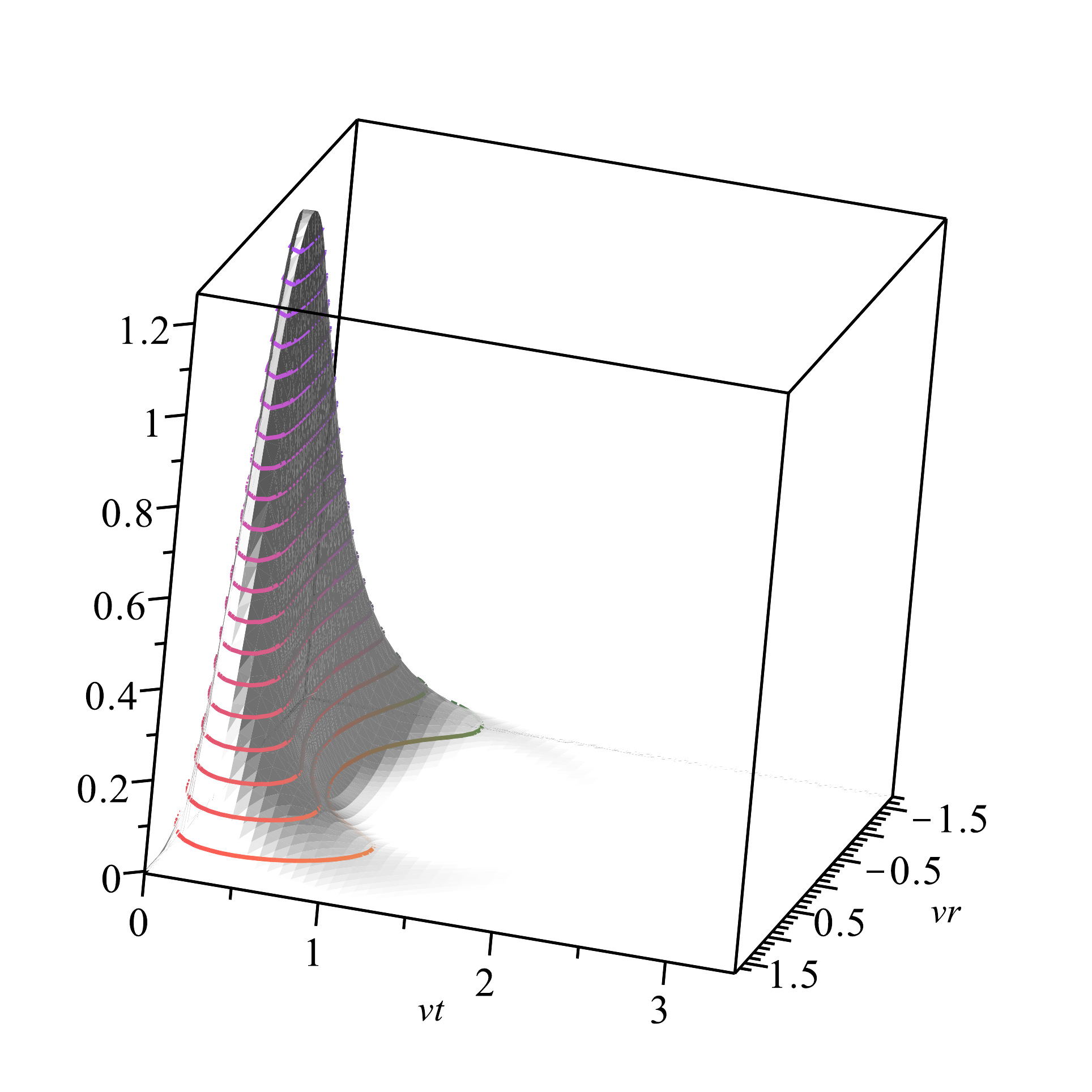}}
\caption{Distribution function $f(v_r,v_t)$ from top-left to bottom-right, with $a=-2,1,0,+1,+2$ 
from tangentially to radially anisotropic velocity distributions.
 }
\label{fig:DF3D}
\end{center}
\end{figure}

Table\,\ref{table:moments1} gives the velocity dispersions for some values of the $a$ parameter when $v_0=1$.
The velocity distribution is  isotropic for $a=0$, and with a radial or tangential bias depending on whether $a>0$ or $a<0$ 
 The  cases $a=-\infty$ or $a=+\infty$ correspond to distributions with purely tangential or purely radial orbits, respectively.
The tangential velocity has two components, according to $\theta$ and $\phi$. $|v_t|$ refers to the tangential velocity modulus, while 
$ \sigma_t / \sqrt{2} $ refers to the dispersion of a single tangential component.
Figure\,\ref{fig:DF3D}  represents the corresponding $f(v_r,v_t) $ distributions.

\begin{table}[htp]
\caption{Velocity moments for Eq. \ref{equ:dfSph} distribution function with $v_0=1$ and $n=+3$.}
\begin{center}
\begin{tabular}{|c|c|c|c|c|c|c|c|c|c|c|c|c|c|}
\hline
$a$ & $\sigma_r$ & $ \frac{\sigma_t}{  \sqrt{2}} $ &  $ \frac{\sigma_r}{ \sigma_t  \sqrt{2}} $ & $<|v_t|>$ & $ \sigma(|v_t|)$ \\
\hline
-10 & 0.13 & 0.70 & 0.19 & 0.99 & 0.09 \\
-2 & 0.30 & 0.67 & 0.45 & 0.93 & 0.21\\
-1 & 0.41 & 0.64 & 0.64 & 0.87 & 0.29 \\
0 & 0.577 & 0577 & 1 & 0.724 & 0.378 \\
+1 & 0.72 & 0.49 & 1.49 & 0.55 & 0.41 \\
+2 & 0.80 & 0.42 & 1.91 & 0.44 & 0.40 \\
+10 & 0.92 & 0.28 & 3.35 & 0.22 & 0.32 \\
\hline
\end{tabular}
\end{center}
\label{table:moments1}
\end{table}%

\subsubsection{$f(E,L_z,L_\perp$) }

 Still within the framework of a spherical and logarithmic potential, the previous distribution function is modified in order to have a triaxial velocity ellipsoid, with the three velocity dispersions $\sigma_r$, $\sigma_\phi,$ and $\sigma_\theta,$ different two-by-two. To do this, we introduced a dependency on the angular momentum $L_z$ and we wrote: 
\begin{equation}
f = \exp \left\{- n\left[E + \frac{1}{r_E^2} \left( b\, L_z^2 + c\, (L^2-L_z^2) \right) \right] \right\} 
\label{equ:dfSph2}.
\end{equation}
As a result, the associated density distribution is no longer necessarily spherical, but may also be oblate or prolate with the $z$ axis as the axis of rotational symmetry.
The density distribution remains proportional to $r^{-n}$, but is now dependent on the angle $\theta$ (the angle between the vertical axis $z$ and the vector radius $\vec{r}$, and $\phi$ is the azimuth angle). The velocity distribution does not depend on the distance to the centre $r$ but only on  the angle $\theta$.

\begin{figure}[!htbp]
\begin{center}
\centerline{\includegraphics[scale=0.3]{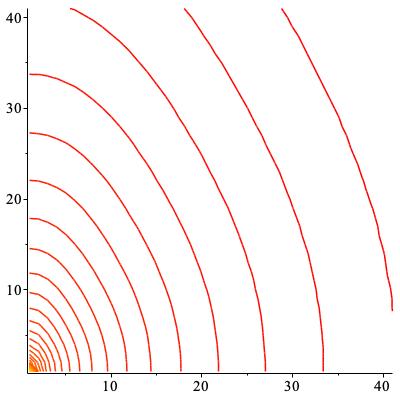},\includegraphics[scale=0.3]{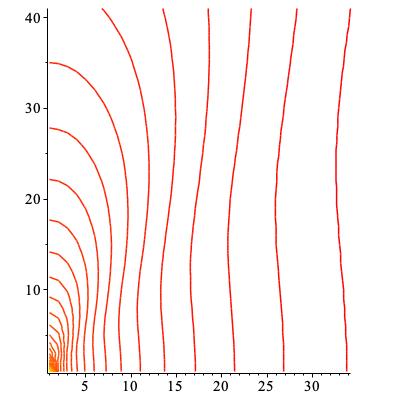},\includegraphics[scale=0.3]{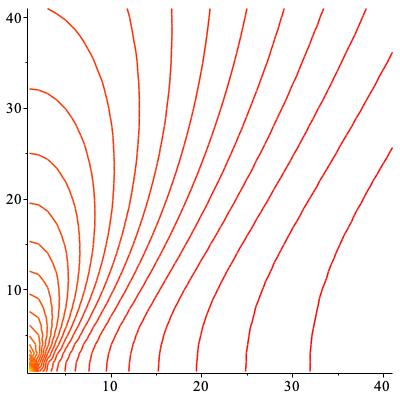}}
\centerline{\includegraphics[scale=0.3]{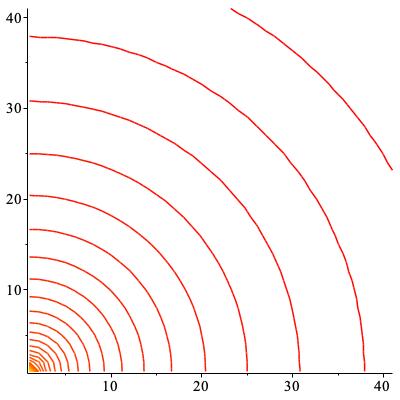},\includegraphics[scale=0.3]{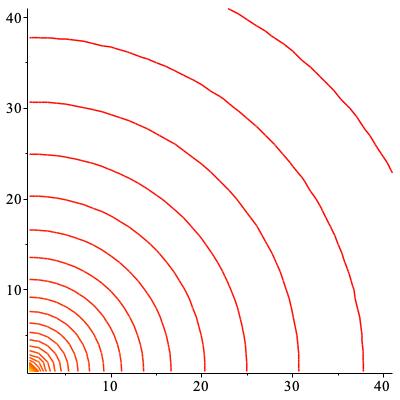},\includegraphics[scale=0.3]{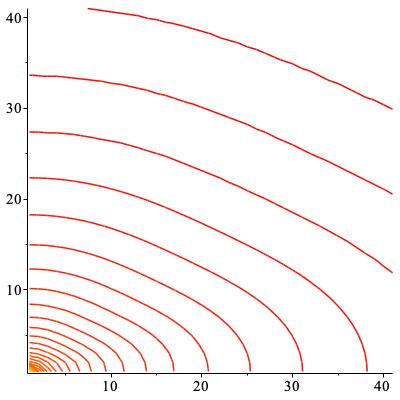}}
\centerline{\includegraphics[scale=0.3]{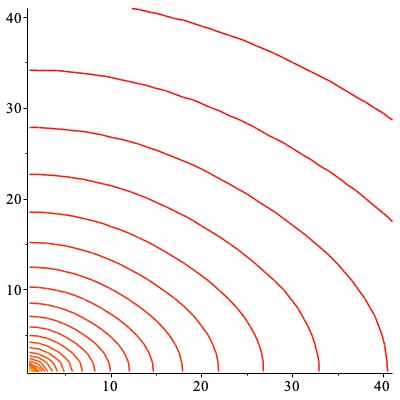},\includegraphics[scale=0.3]{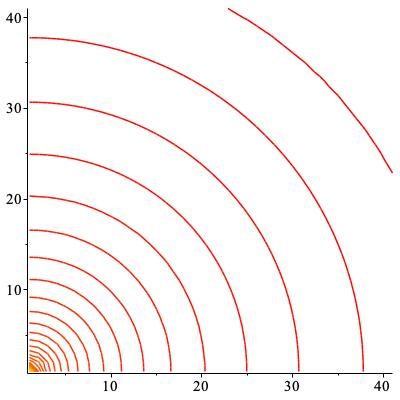},\includegraphics[scale=0.3]{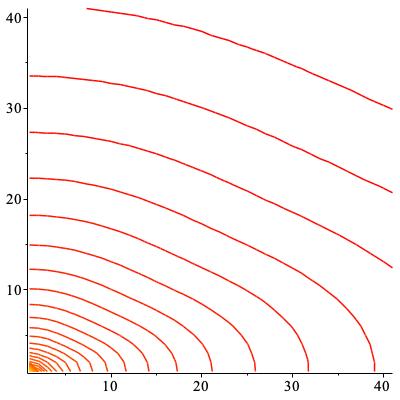}}
\caption{
Density iso-contours (logarithmically equally spaced) for distribution function  (Eq.\,\ref{equ:dfSph2}) for $b$ and $c$ values of Table\,\ref{table:moments2} (raw and column corresponding).
}
\label{fig:iso1}
\end{center}
\end{figure}

The parameters $b$ and $c$ allow modification of the values of the velocity dispersion ratios along the three main axes of the distribution $f(v_r,v_\phi,v_\theta)$.
For $b=c,$ the distribution function has a spherical density (see Eq.\,\ref{equ:dfSph2}). In these cases, the velocity dispersions $\sigma_\phi$ and $\sigma_\theta$ are equal and do not depend on $r$, $\phi,$ or $\theta$ (Tables\,\ref{table:moments2} and \ref{table:moments3} in the second row). 
For the three special cases $b=0$, $c=0$, or $b=c$, in the $z=0$ plane, two of the three components of the velocity dispersions are equal (see  Tables\,\ref{table:moments2} and \ref{table:moments3}). The corresponding densities are shown in Figure\,\ref{fig:iso1}  (also the first and third columns).
It should be noted that if $b<c$, in the $z=0$ plane the vertical velocity dispersion $ \sigma_\theta$ being smaller than in the spherical case, the density distribution is oblate. And conversely for $b>c$,  the density distribution is prolate or less oblate.
The corresponding values of the  velocity dispersion along the three main axes of the velocity distributions are given in Table\,\ref{table:moments3}.
More generally, this distribution function, (see Eq.\,\ref{equ:dfSph2}) , allows us to model the velocity  dispersions with  $\sigma_\phi/\sigma_r$ and $\sigma_\theta/\sigma_r$ ratios arbitrarily chosen within the $z=0$ plane. 
Outside the $z=0$ plane towards the poles $\theta =0$ or $\pi/2$, the velocity distributions along the three main axes vary, and, for symmetry reasons, the dispersions $\sigma_\theta$ and $\sigma_\phi$ become equal to the poles (Table\,\ref{table:moments4}). \\\

\begin{table}[htp]
\caption{$b$ and $c$ co-efficients for  models with density plotted in Figure\,\ref{fig:iso1}.
 Each row and column of the table correspond to those of Figure \,\ref{fig:iso1}, and of Tables \,\ref{table:moments3} and \,\ref{table:moments4}.
}
\begin{center}
\begin{tabular}{|c|c|c|}
\hline
$+4$, $+0$ & +2, $-2$ & $+0$, $-4$ \\
\hline
$+2$, $+2$ & +0, +0 & $-2$, $-2$ \\
\hline
$+0$, $+4$ & $-2$, $+2$ & $-4$, $+0$ \\
\hline
\end{tabular}
\end{center}
\label{table:moments2} 
\end{table}%

\begin{table}[htp]
\caption{$\sigma_r$ $\sigma_\phi$ $\sigma_\theta$ velocity dispersions at $z=0$. 
 Each row and column of the table correspond to those of Figure \,\ref{fig:iso1}, and of Tables \,\ref{table:moments2} and \,\ref{table:moments4}}
\begin{center}
\begin{tabular}{|c|c|c|}
\hline
$0.68 \;\; 0.28  \;\; 0.28$ &  $0.33  \;\; 0.23  \;\; 0.92$ & $0.22  \;\; 0.22  \;\; 0.95$ \\
\hline
$0.80  \;\;  0.42  \;\; 0.42$  & $0.58  \;\;  0.58  \;\;  0.58 $ & $0.30  \;\;  0.67  \;\;  0.67$ \\
\hline
$0.68  \;\;  0.68  \;\;  0.28 $ & $0.33  \;\;  0.92  \;\;  0.23 $ & $0.22  \;\;  0.95  \;\;  0.22 $ \\
\hline
\end{tabular}
\end{center}
\label{table:moments3}
\end{table}%

\begin{table}[htp]
\caption{ $\sigma_r$ $\sigma_\phi$ $\sigma_\theta$ velocity dispersions at $x=y=0$ towards the poles. 
Each row and column of the table correspond to those of Figure \,\ref{fig:iso1}, and of Tables \,\ref{table:moments2} and \,\ref{table:moments3}
}
\begin{center}
\begin{tabular}{|c|c|c|}
\hline
$0.87 \;\; 0.35  \;\; 0.35$ &  $0.80  \;\; 0.42  \;\; 0.42$ & $0.58  \;\; 0.58  \;\; 0.58$ \\
\hline
$0.80  \;\;  0.42  \;\; 0.42$  & $0.58  \;\;  0.58  \;\;  0.58 $ & $0.30  \;\;  0.67  \;\;  0.67$ \\
\hline
$0.58  \;\;  0.58  \;\;  0.58 $ & $0.30  \;\;  0.67  \;\;  0.67 $ & $0.21  \;\;  0.69  \;\;  0.69 $ \\
\hline
\end{tabular}
\end{center}
\label{table:moments4}
\end{table}%

\subsubsection{$f(E,I_2,I_3$) }

Finally, we considered the triaxial logarithmic potential (see Eq.\,\ref{equ:tests-2}) with $m_0=0.3$, $q_y=0.95$, and $q_z=0.85$ and we defined the distribution function (with $n=3$):

\begin{equation}
f = \exp \left[- n \left( E + \frac{1}{r_E^2} ( b\,I_2 + c\,I_3) \right) \right] .
\label{equ:dfSph3}
\end{equation}

For a triaxially symmetrical potential, the quasi-integrals $I_2$ and $I_3$ are  workable generalisations of $L_z^2/2$ and $(L^2-Lz^2)/2$. The figures presented here are made using for $I_2$ and $I_3$ the expressions given by the Eqs.\,\ref{equ:psi} and \ref{equ:xi}.

In the particular case where $b=0$ and $c=0$, the distribution function depends only on energy and is therefore an exact solution of the collisionless Boltzmann equation. When the $b$ or $c$ parameters are non-zero,
the potential is no longer a St\"ackel potential. The integrals $I_2$ and $I_3$ are approximated and the distribution function is therefore not strictly stationary. Moreover, the presence of a non-zero core radius, $m_0=0.3$, means that the potential is no longer strictly scale-free. As a result, the tensor of velocity dispersions varies: it is approximately isotropic in the centre and anisotropic at large radii. The associated density has a core radius of $\sim0.3$ and at larger radii it decreases as $r^{-3}$\\\

\begin{figure}[!htbp]
\begin{center}
\includegraphics[scale=0.30]{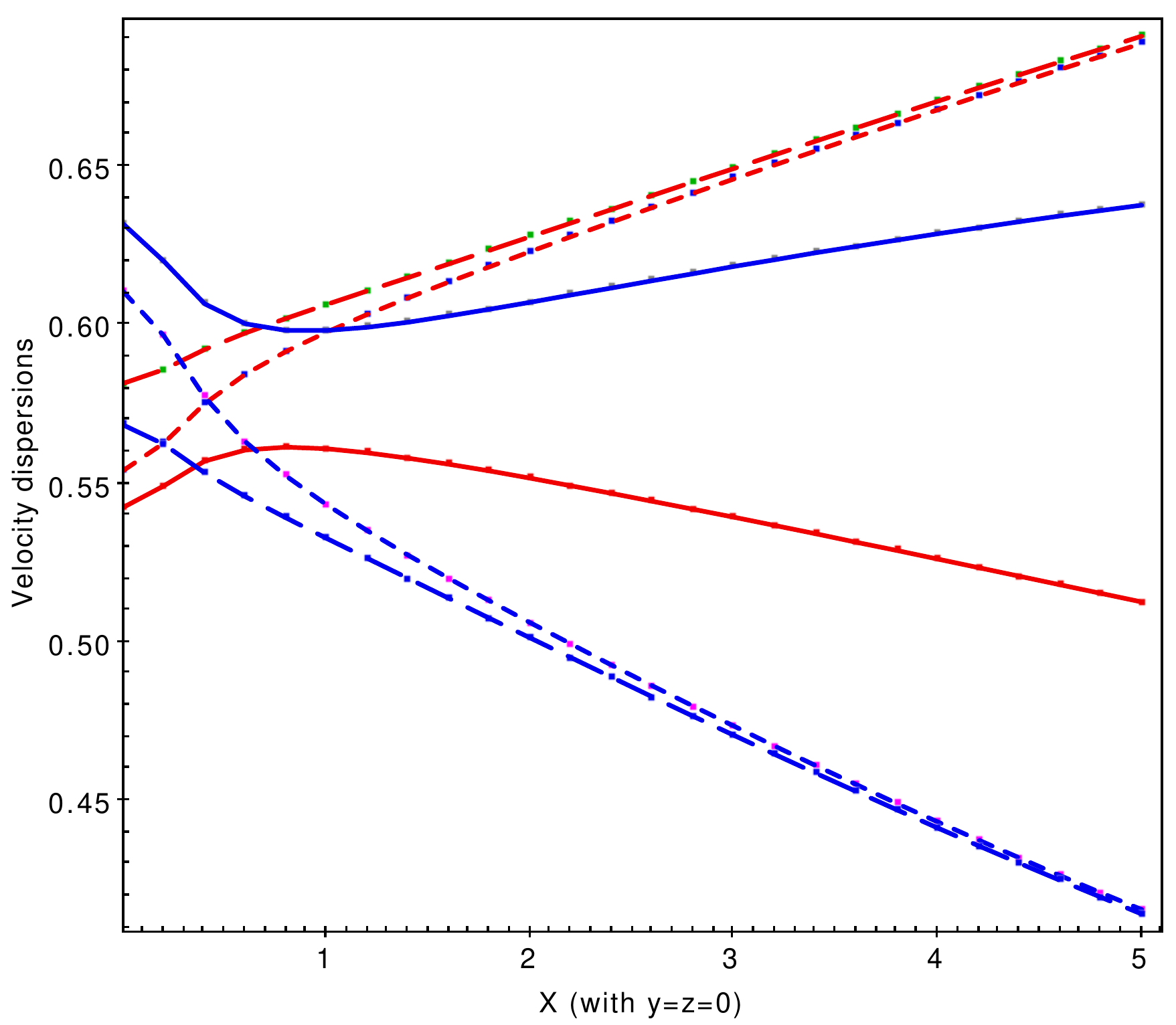}
\caption{
Velocity dispersions versus $x$-distance $(y=z=0)$ to centre for models given by Eq.\,\ref{equ:dfSph3}. 
The blue lines model $(b,c)=(+0.3,+0.3)  $ is dominated by radial obits.
The red lines models $(b,c)=(-0.2,-0.2)$ are dominated by tangential obits.
Continuous lines indicate radial dispersions $\sigma_R$, while short- and long-dashed lines 
indicate tangential dispersions $\sigma_\phi$ and $\sigma_\theta$, respectively.
}
\label{fig:dispersions}
\end{center}
\end{figure}

\begin{figure}[!htbp]
\begin{center}
\includegraphics[scale=0.25]{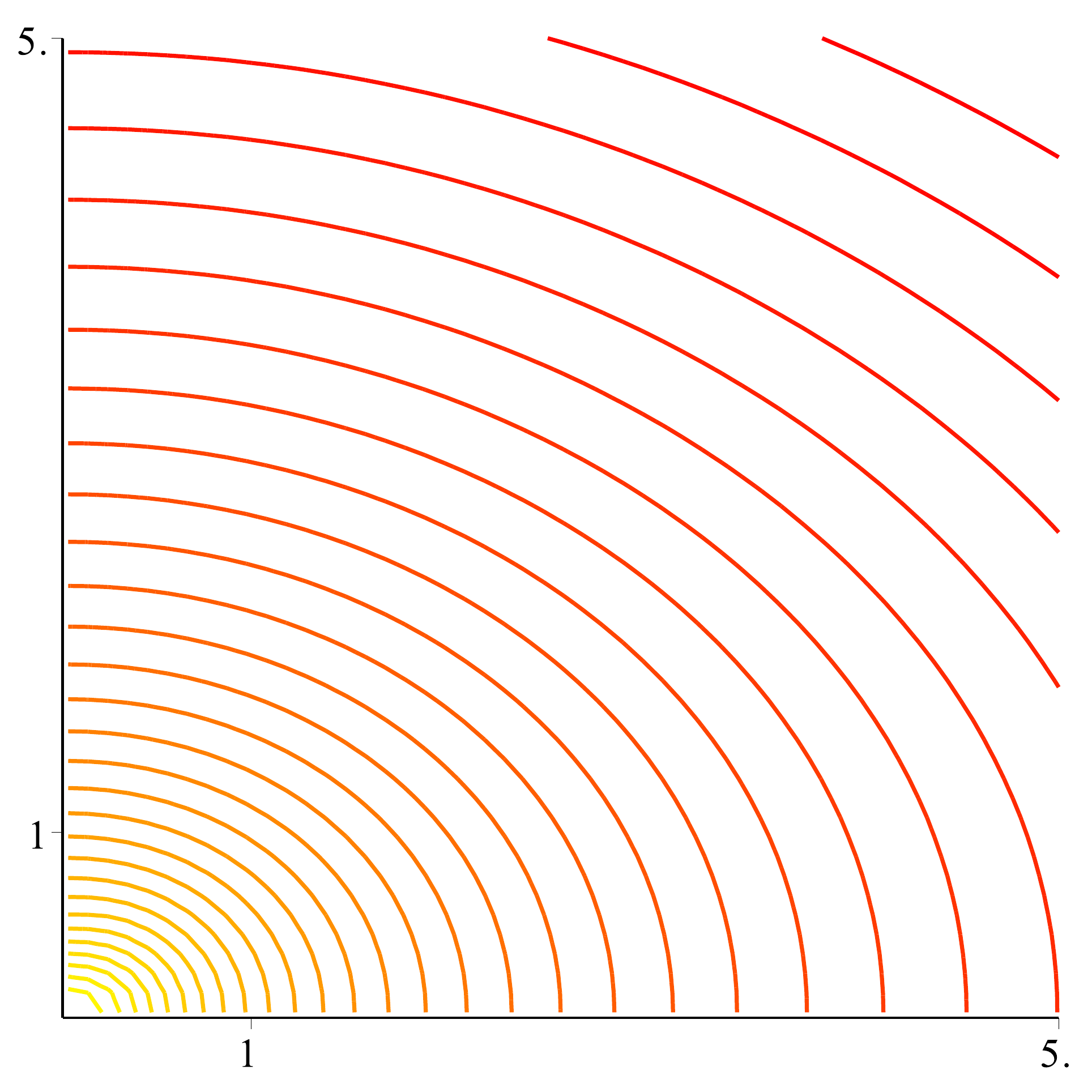}
\caption{
Density for radially biased model $(b=c=+0.3)$.
Shown above are equally spaced iso-contours of the logarithm of the density.
}
\label{fig:iso2}
\end{center}
\end{figure}

The distribution function (see Eq.\,\ref{equ:dfSph3}) is relatively accurate in an extended volume of space and only for small values in modulus of the constants $b$ and $c$. We describe two cases corresponding to oblate distributions of the tracer: $(b,c) = (+0.3,+0.3)$ and $(-0.2,-0.2)$. In both cases, the dispersions are almost isotropic near the centre.   Close to  the galactic plane, $z=0$, the tangential velocity dispersions are similar $\sigma_\phi \simeq \sigma_\theta$.  Figure\,\ref{fig:dispersions}  shows the three components of the velocity dispersions versus the distance $x$ (with $y=z=0)$ to the centre. At large radii, for the first case the distribution function is dominated by radial orbits, in the second case by tangential orbits.
These two examples correspond to triaxial  density distributions of the stellar tracer with axis ratios  $q_z= 0.90$ and $0.77,$ respectively. Figure\,\ref{fig:iso2} shows the iso-contours of the latter model that is dominated with radial orbits.

The degree of accuracy of these distribution functions is evaluated in the following paragraph by applying the Jeans equations to verify that the forces are recovered with sufficient accuracy. 
The main element that limits the accuracy of the distribution function is the modelling of box orbits, for which the quasi-integral orbits are the least well-preserved along the orbits (see also for example the Poincar\'e sections that are less accurate for box orbits than for tube orbits in Fig.\,\ref{fig:f11}).

\subsection{Test with Jeans equations }

The previous distribution function must satisfy the collisionless Boltzmann equation and therefore  the Jeans equations that we rewrote as:

\begin{equation}
 \sum_i \frac{\partial \, <v_i \, v_j> }{\partial \, x_i} +\sum_i <v_i \, v_j> \frac{\partial \, \log \nu }{\partial \, x_i} + \frac{\partial \, \Phi}{\partial \, x_j} = 0
\quad \quad (j=1,2,3) \, . 
\label{equ:jeans}
\end{equation}

These equations allow us to calculate the force field associated with the gravitational potential $\Phi$. To do this, it is sufficient to measure the density and velocity tensor of a stellar tracer represented here by the distribution function Eq.\,\ref{equ:dfSph3}. As the associated density decreases rapidly according to a power law, we used the Jeans equations in the form of Eq.\,\ref{equ:jeans} to obtain better numerical accuracy. We applied these equations for the potential $\Phi$ (Eq.\,\ref{equ:tests-2}) with $m_0=0.3$. This allows the accuracy of the distribution function to be tested and the usefulness of the quasi-integral $I_2$ and $I_3$ to be estimated. 

The results obtained can be compared with those established in the case of axisymmetric potentials with approximate integrals \cite{san12} and in particular with the potential of the Besan\c{c}on Galaxy model \citep{bie15}. While for axisymmetric potentials and distribution functions modelling tube orbits, we achieved accuracies of the order of a thousandth or hundredth over a large volume of the model, here we obtained accuracies of the order of a few per cent over a volume, which, although extended, is comparatively more limited. These results are similar to the work of \cite{san14} in many respects. They use the actions as approximate integrals of motion which are also constructed by approximating orbits using triaxial St\"ackel potentials.  They also find a more limited accuracy than in axisymmetric cases. The main reason for this is that the modelling of box orbits passing near the centre is less accurate than that of tube orbits that remain far from the centre. 

\begin{figure}[!htbp]
\begin{center}
\centerline{\includegraphics[scale=0.25]{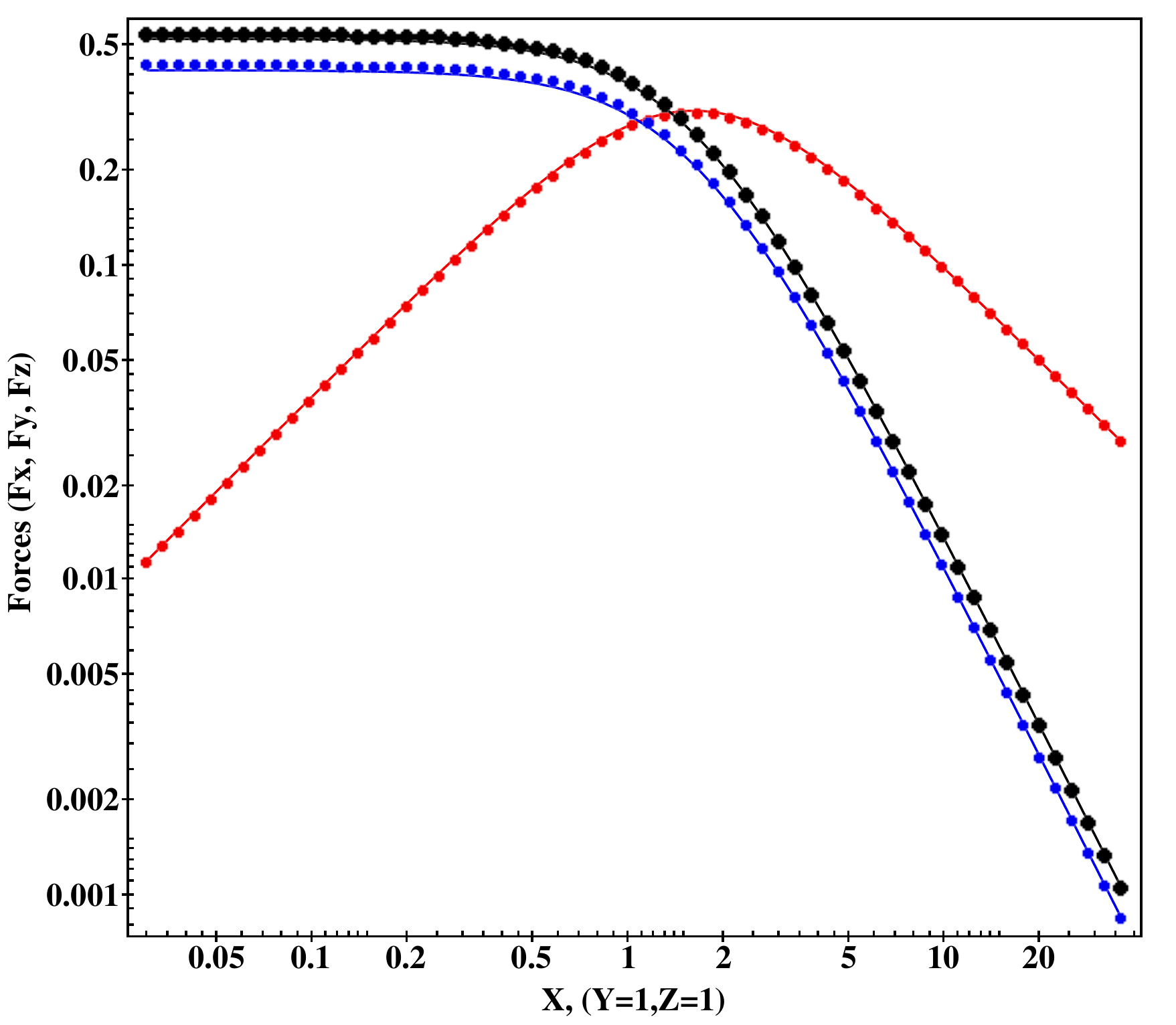},
\includegraphics[scale=0.25]{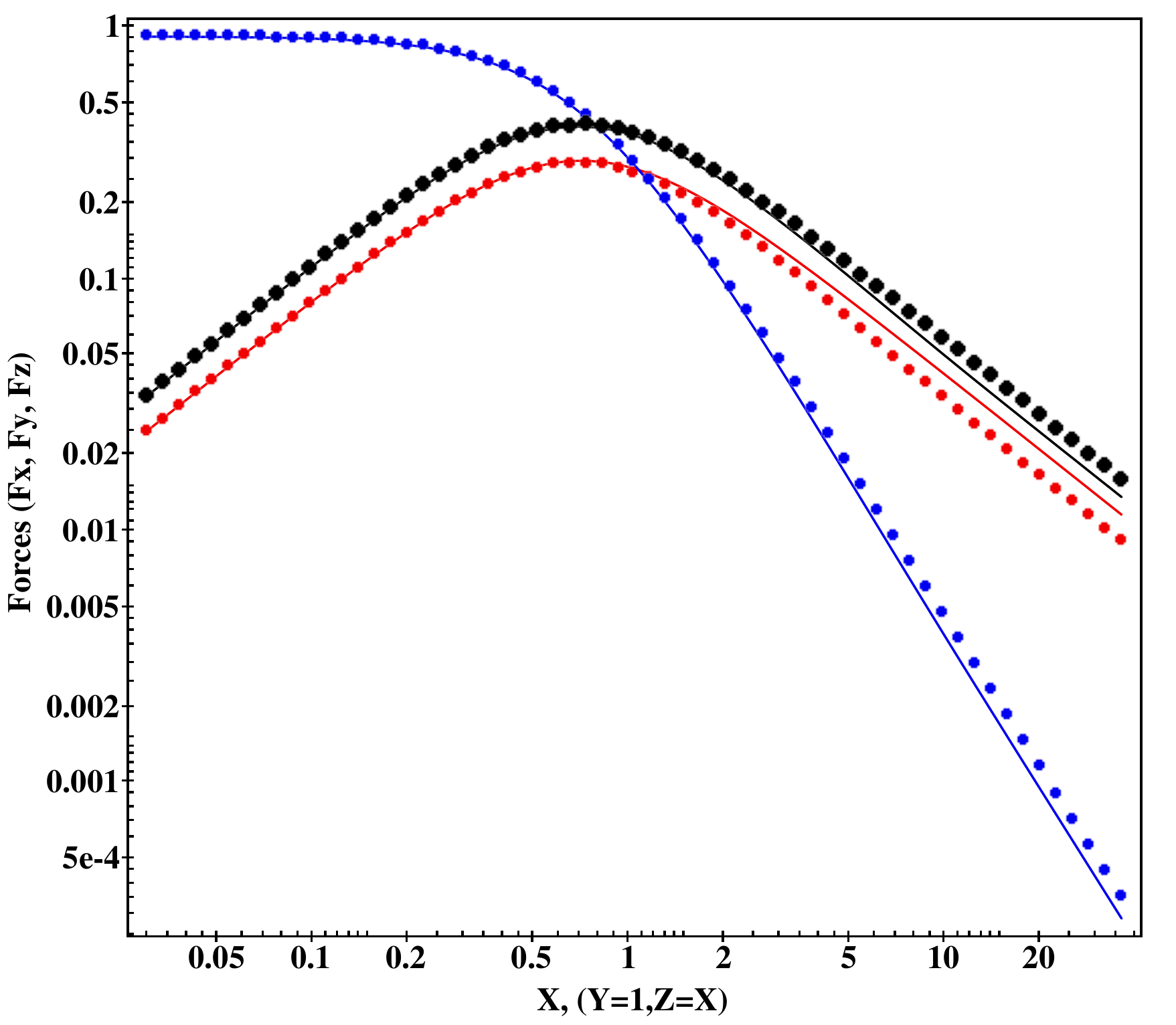},
\includegraphics[scale=0.25]{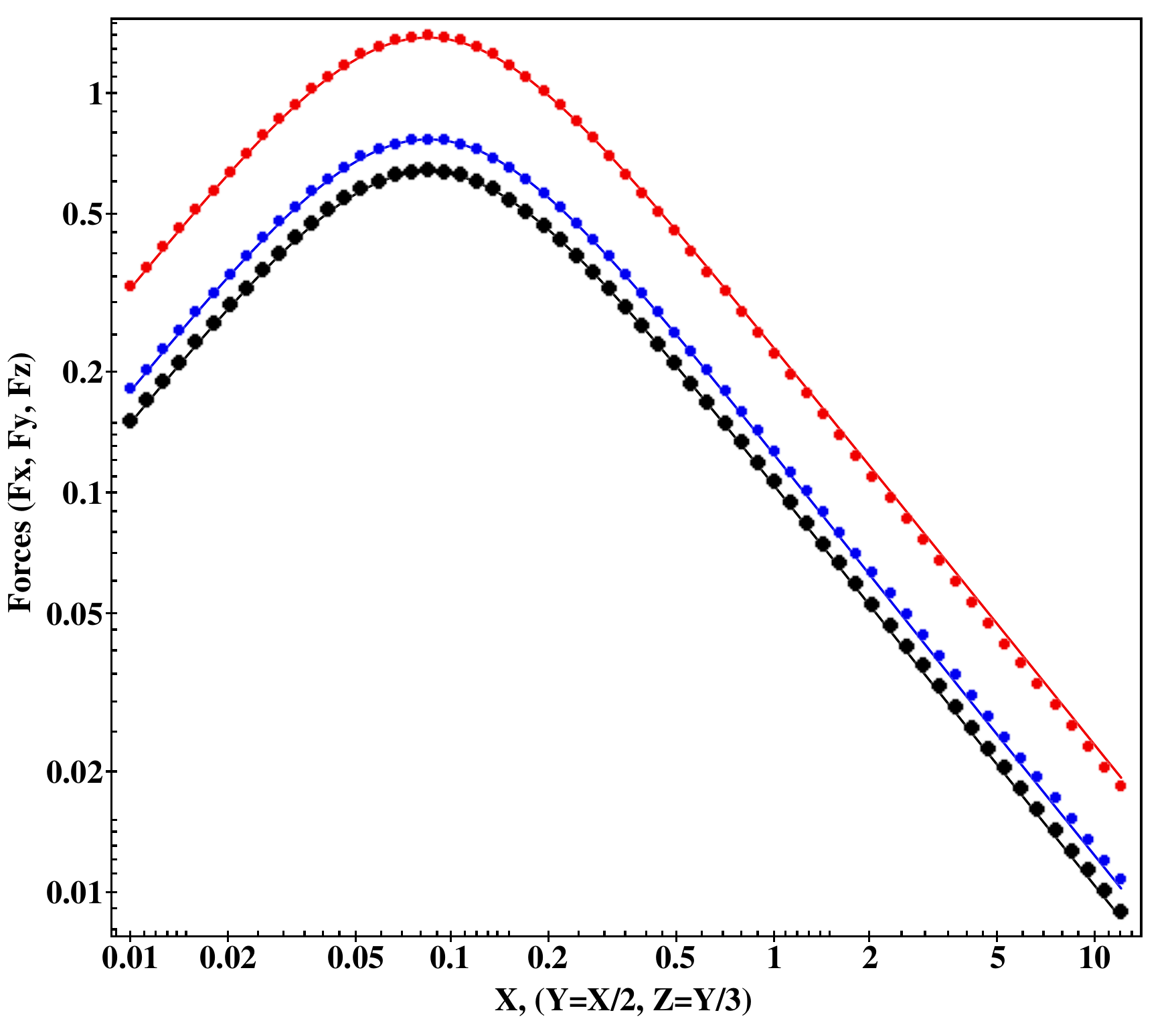}
},
\centerline{\includegraphics[scale=0.25]{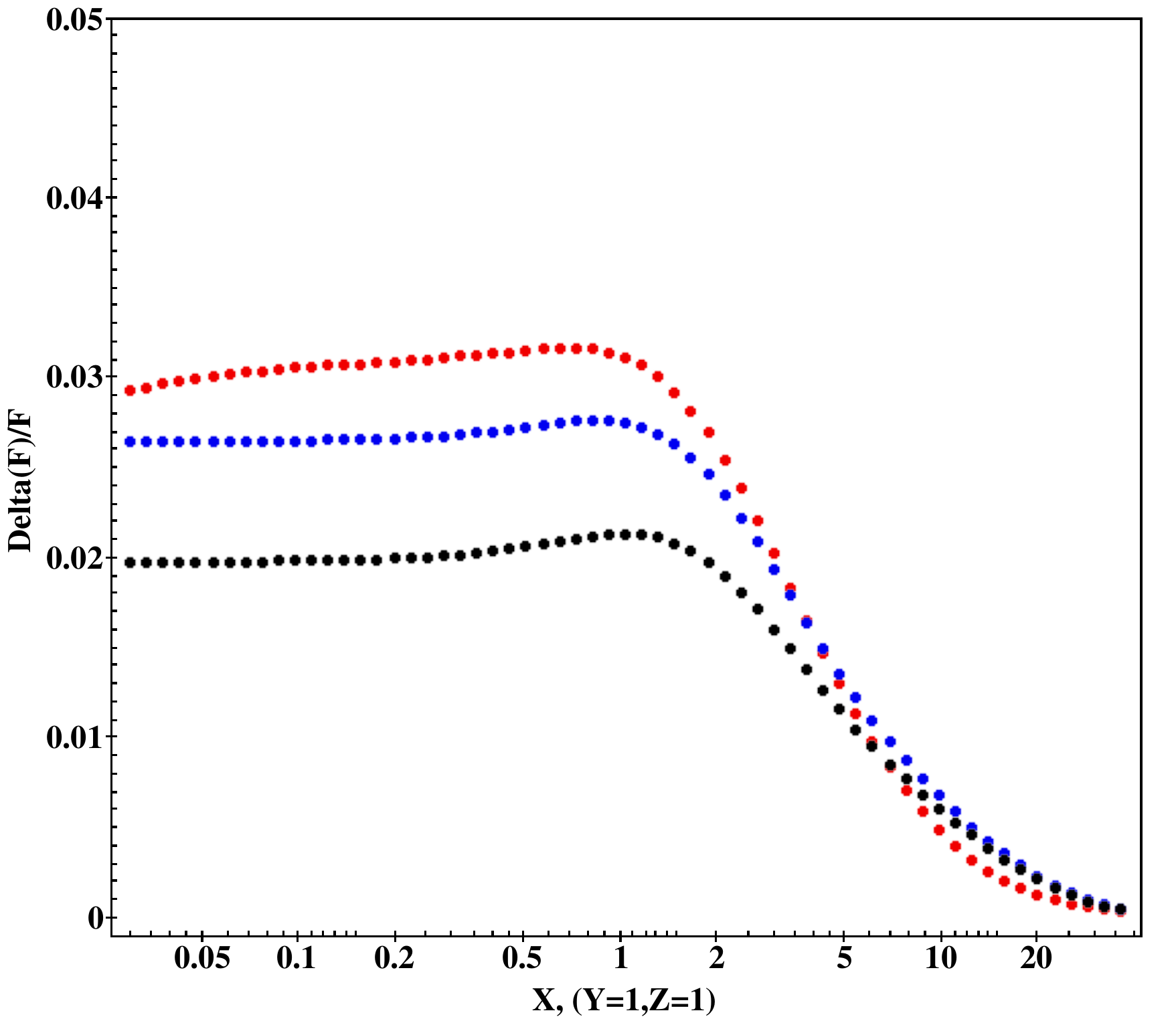},
\includegraphics[scale=0.25]{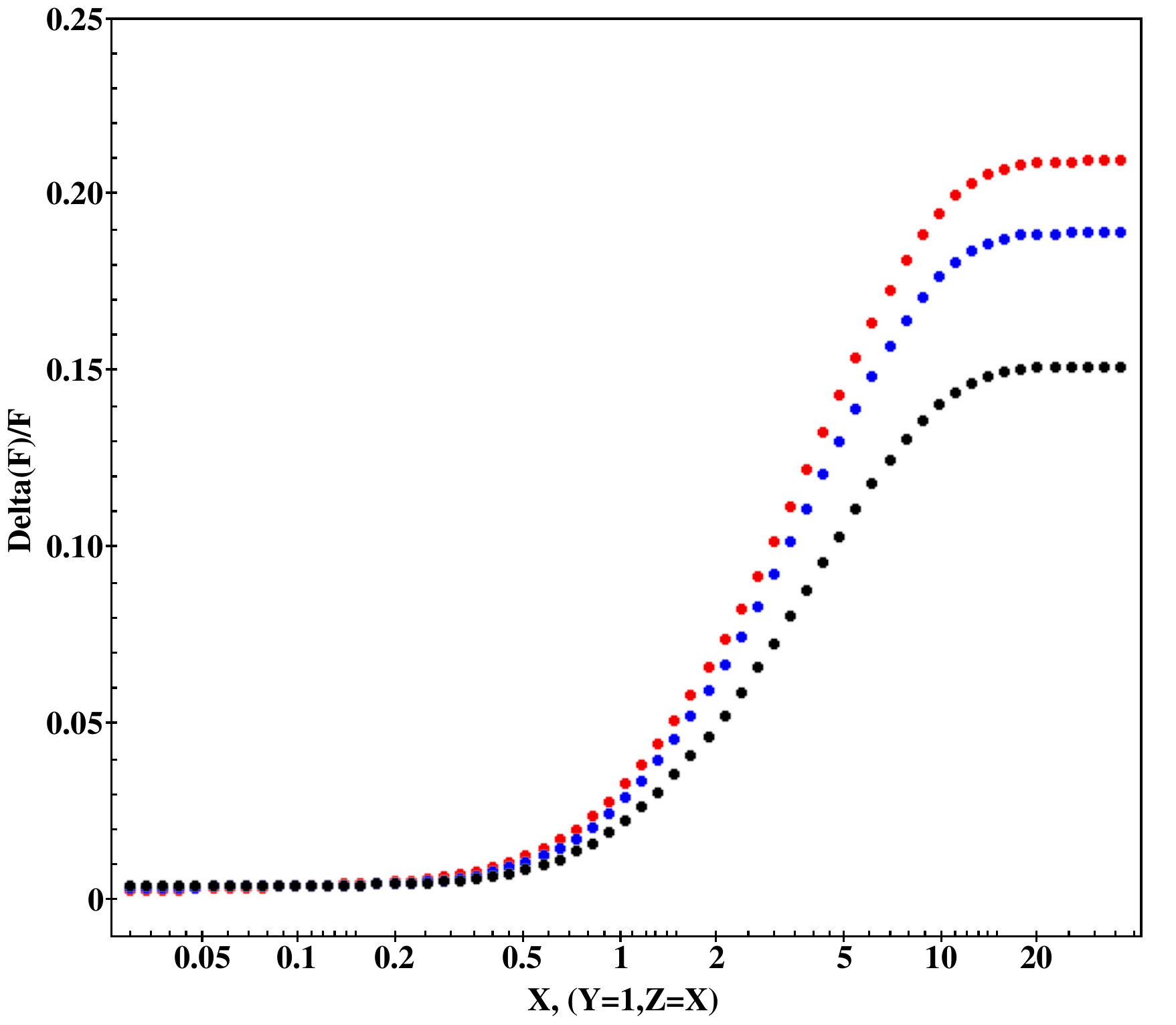},
\includegraphics[scale=0.25]{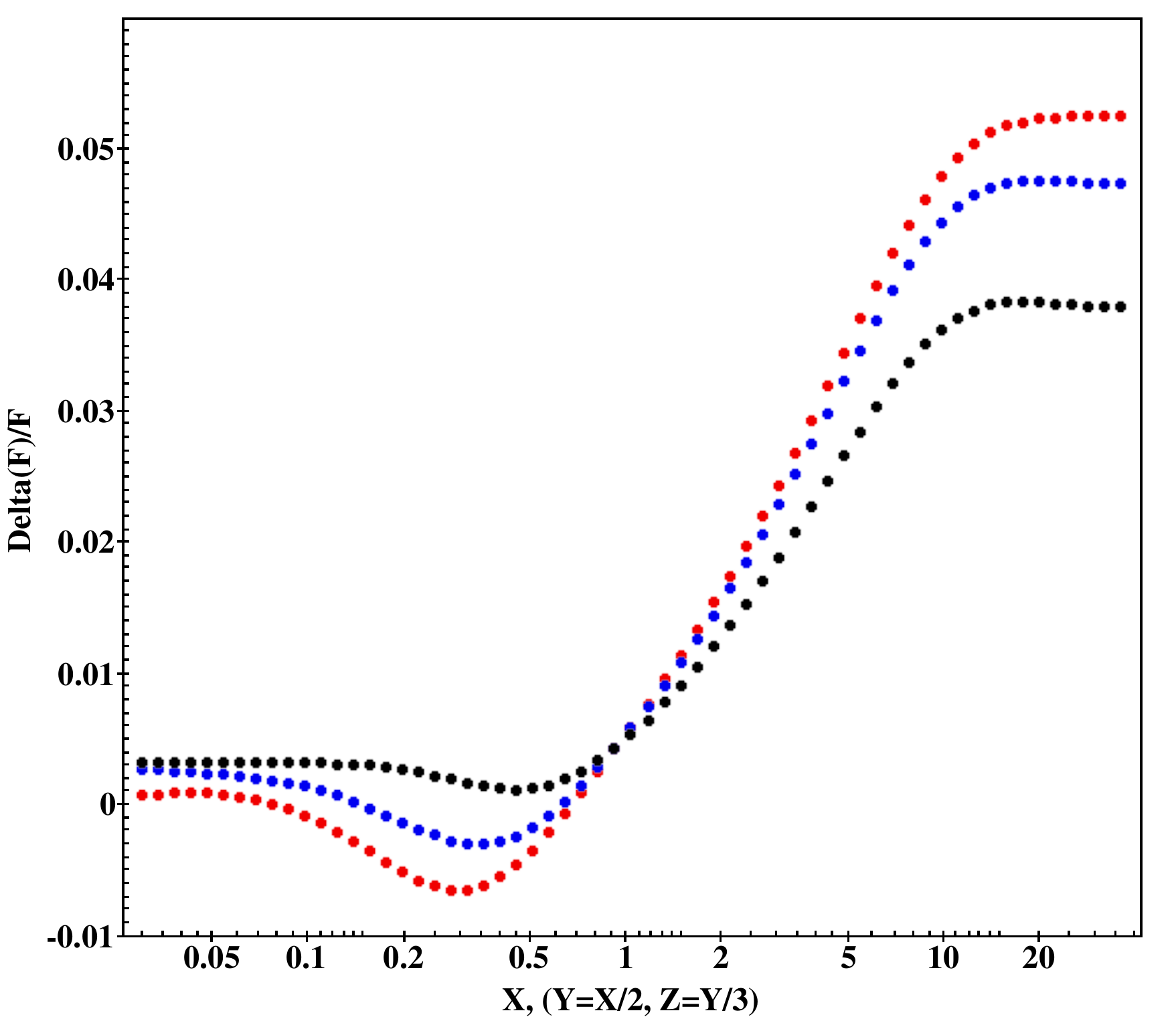}
}
\caption{Accuracy of Jeans equations for 
model with $b=c=-0.2$. The top shows the three forces $F_x,F_y,F_z$ along the three axes $(x,y=1,z=1)$, $ (x,y=1,z=x),$ and  $(x,y=x/2,z=x/3)$. The bottom shows the relative error for the same axes and forces.
}
\label{fig:f1718}
\end{center}
\end{figure}

For the model with predominantly radial velocity dispersions $(b=+0.3,c=+0.3),$  that is, one dominated with box orbits and that also best corresponds to our knowledge of the velocity distribution of galactic halo stars \citep{bla16}, we find the force field with an accuracy of a few per cent within the  volume of radius $r=1$. This is to be compared to the core radius $m_0=0.3$, and up to ten per cent at distances $r=2$ from the centre. The panels of Fig.\,\ref{fig:f1718}  show for the three force components and along three axes,  the differences between the potential gradient $\nabla \Phi,$ and the forces recovered from the Jeans equations. In addition, near the $(x,y)$ plane, if $|z|<1$, the distribution function  allows us to accurately find the force field to within a few per cent and this for $r$ radii up to $\sim 40$.

\begin{figure}[!htbp]
\begin{center}
\centerline{\includegraphics[scale=0.25]{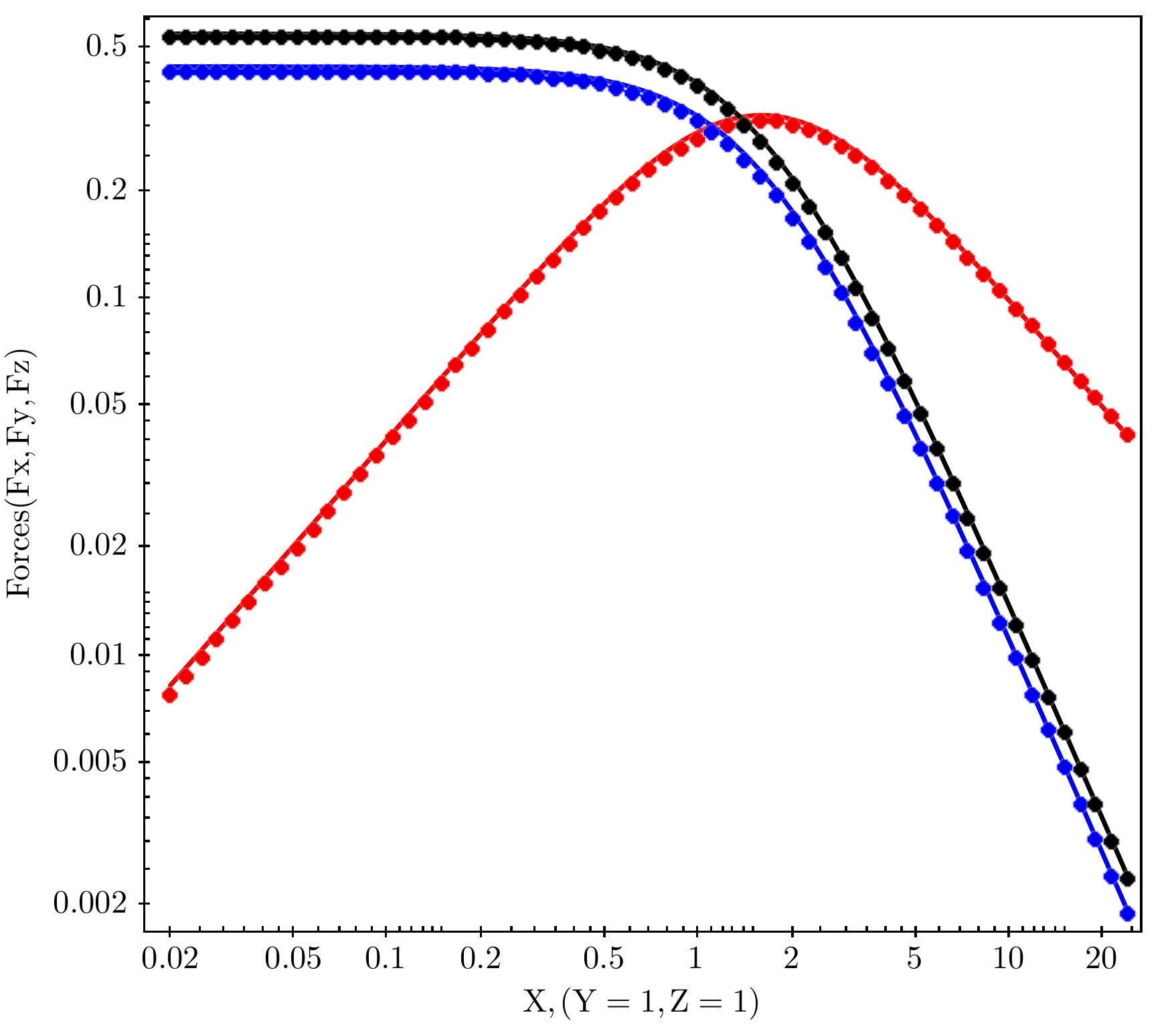},
\includegraphics[scale=0.25]{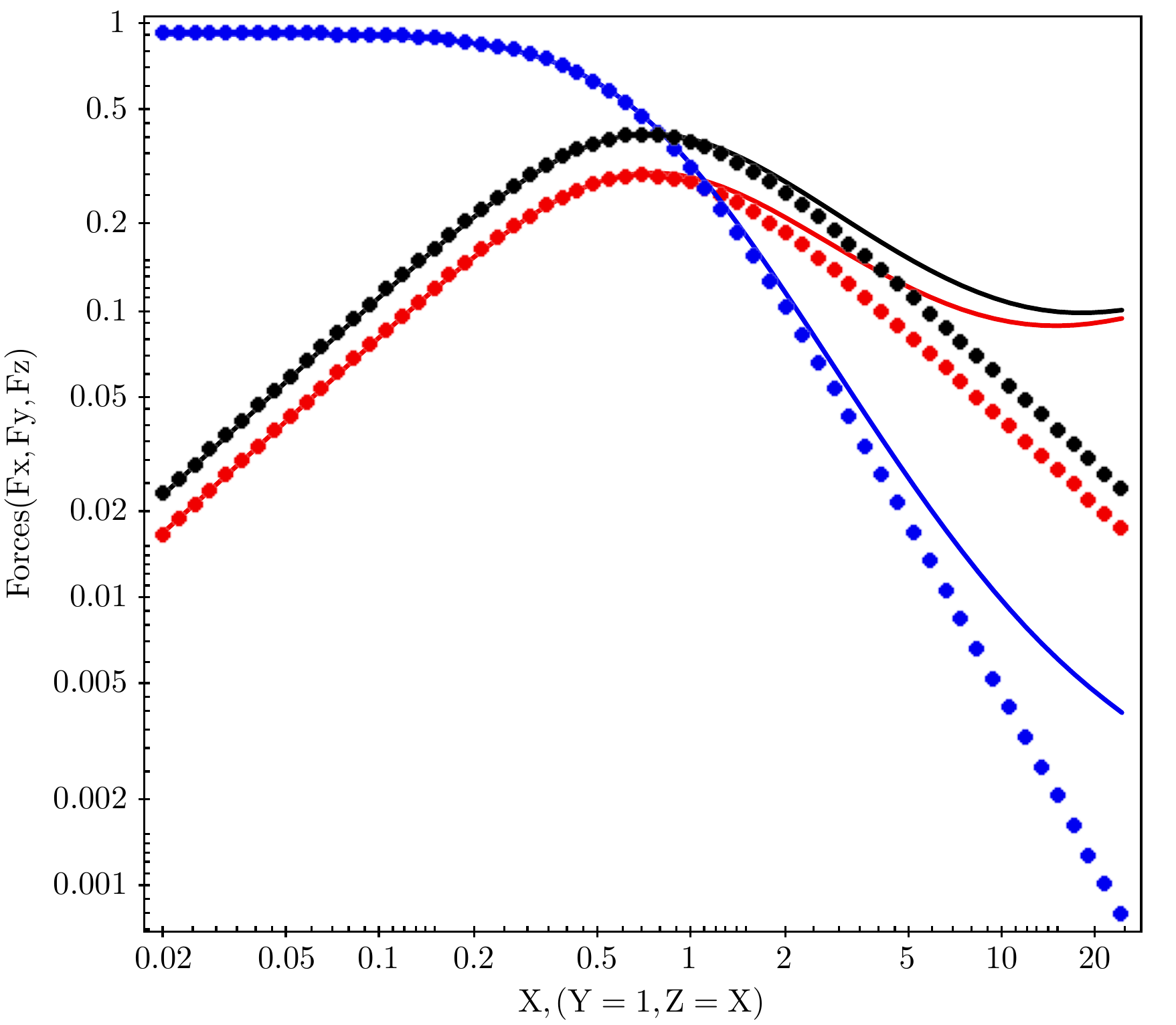},
\includegraphics[scale=0.25]{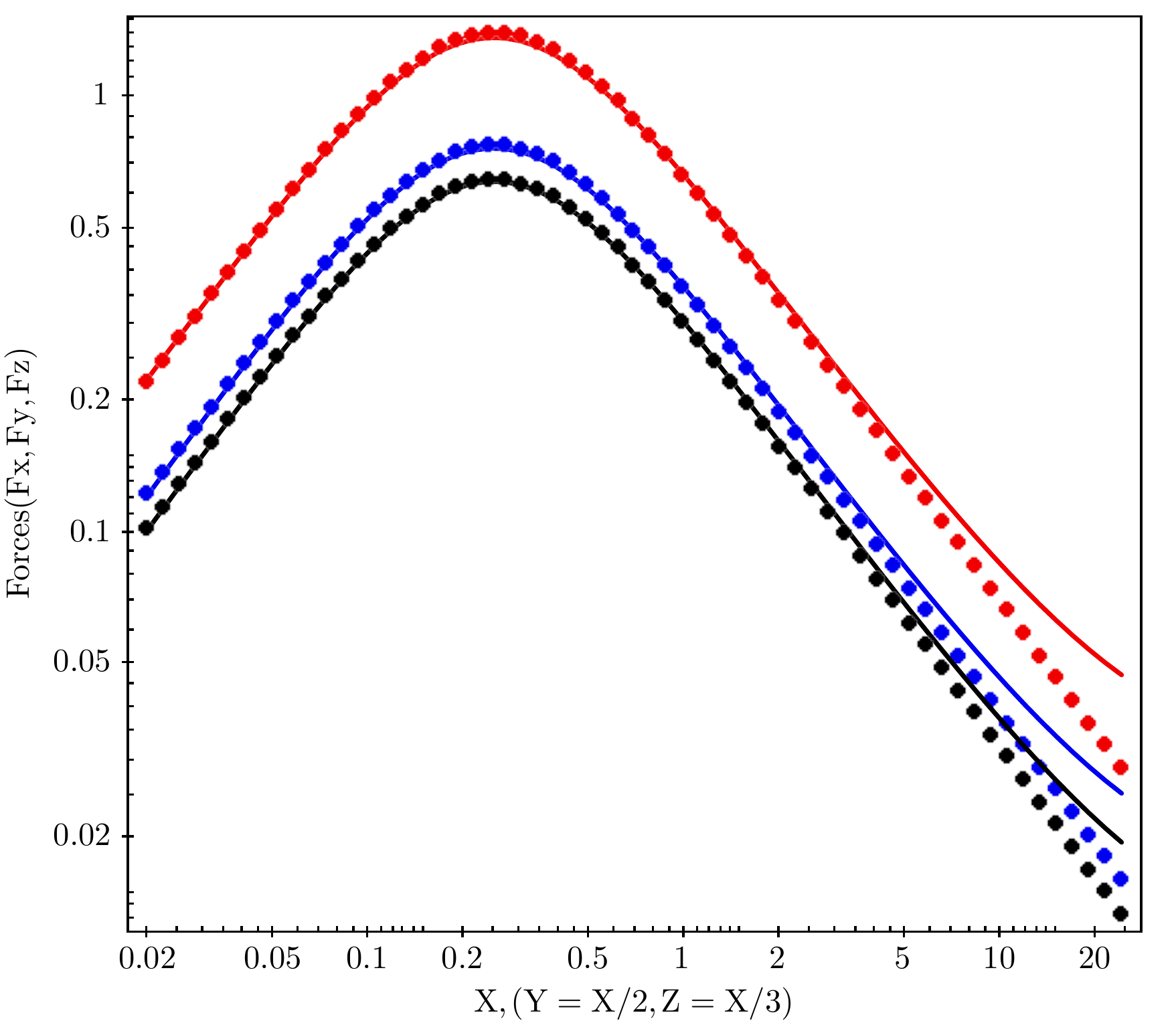}
},
\centerline{\includegraphics[scale=0.25]{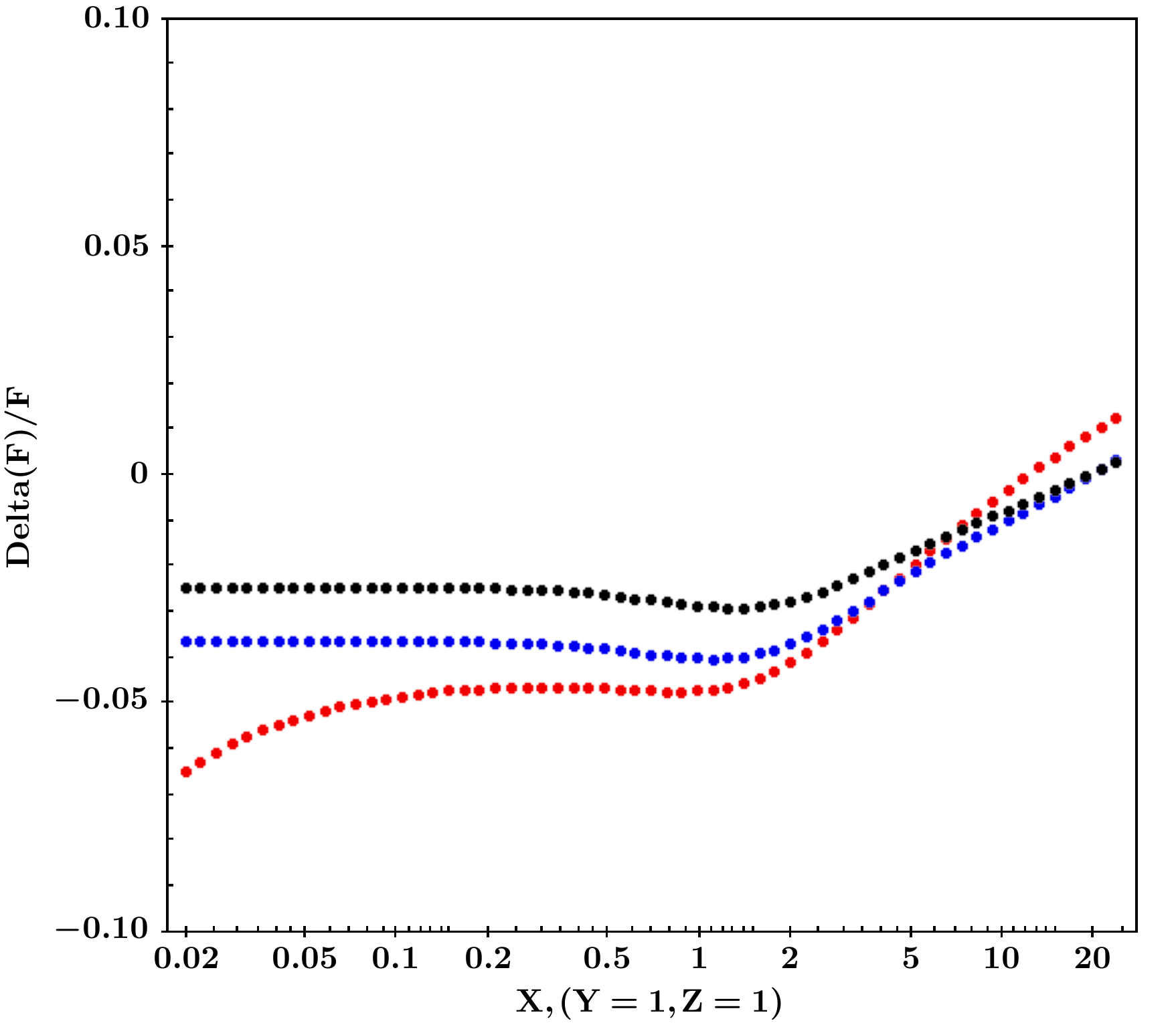},
\includegraphics[scale=0.25]{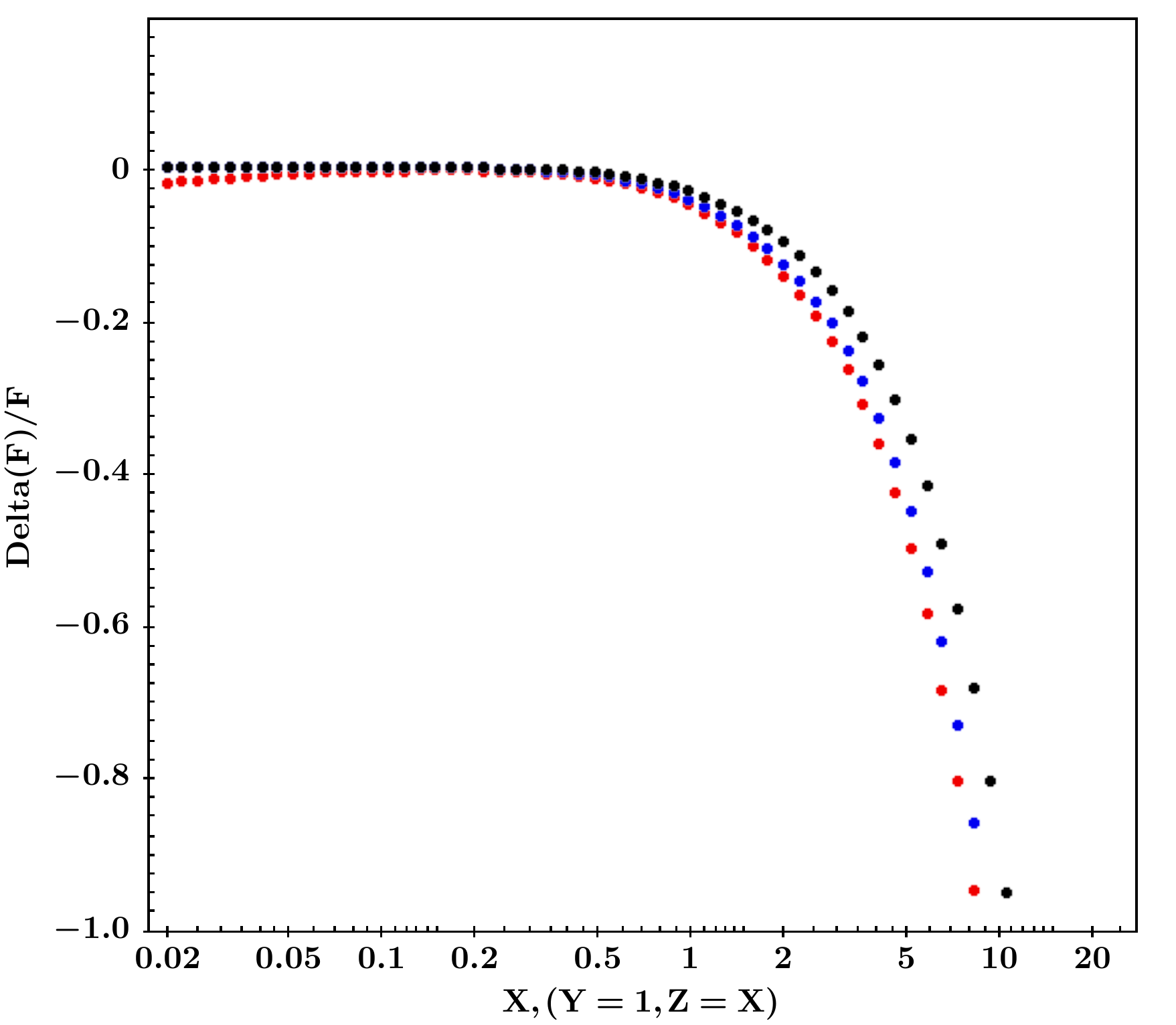},
\includegraphics[scale=0.25]{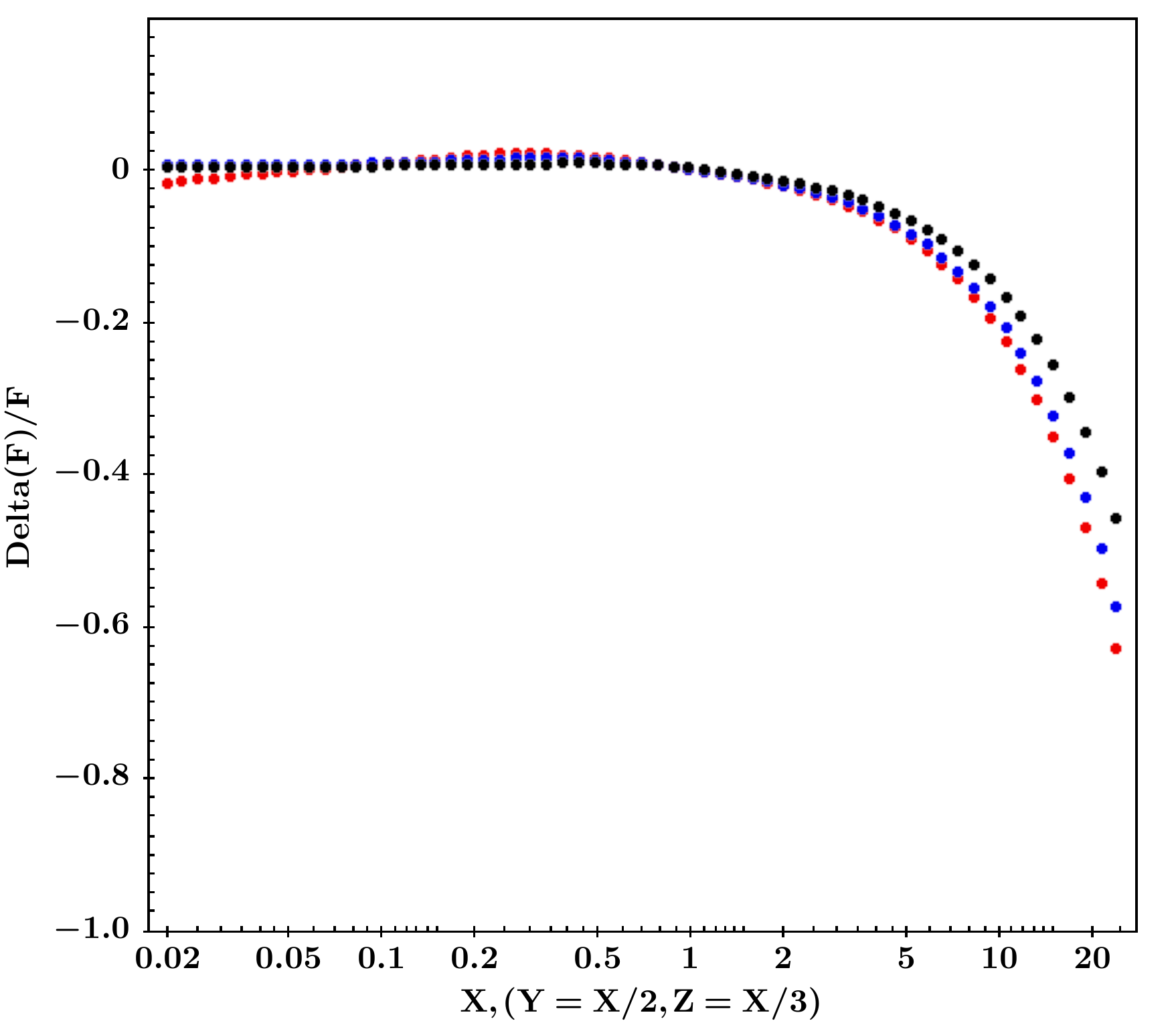}
}
\caption{Accuracy of Jeans equations for 
model with $b=c=+0.3$. The top shows the three forces $F_x,F_y,F_z$ along the three axes $(x,y=1,z=1)$, $ (x,y=1,z=x),$ and  $(x,y=x/2,z=x/3)$. The bottom shows the relative error for the same axes and forces.}
\label{fig:f2021}
\end{center}
\end{figure}

The model with the dominant tangential velocity dispersion $(b=-0.2,c=-0.2)$ allows us to find the force field on a much larger volume going at least to $r=50$. Figure\,\ref{fig:f2021} shows that the relative force errors reach a plateau at $r=20$. Since the distribution function is dominated by tangential orbits (essentially tube or centrophobic orbits) whose quasi-integral orbits are better preserved, the Jeans equations are therefore satisfied with greater precision. \\

\begin{figure}[!htbp]
\begin{center}
\includegraphics[scale=0.25]{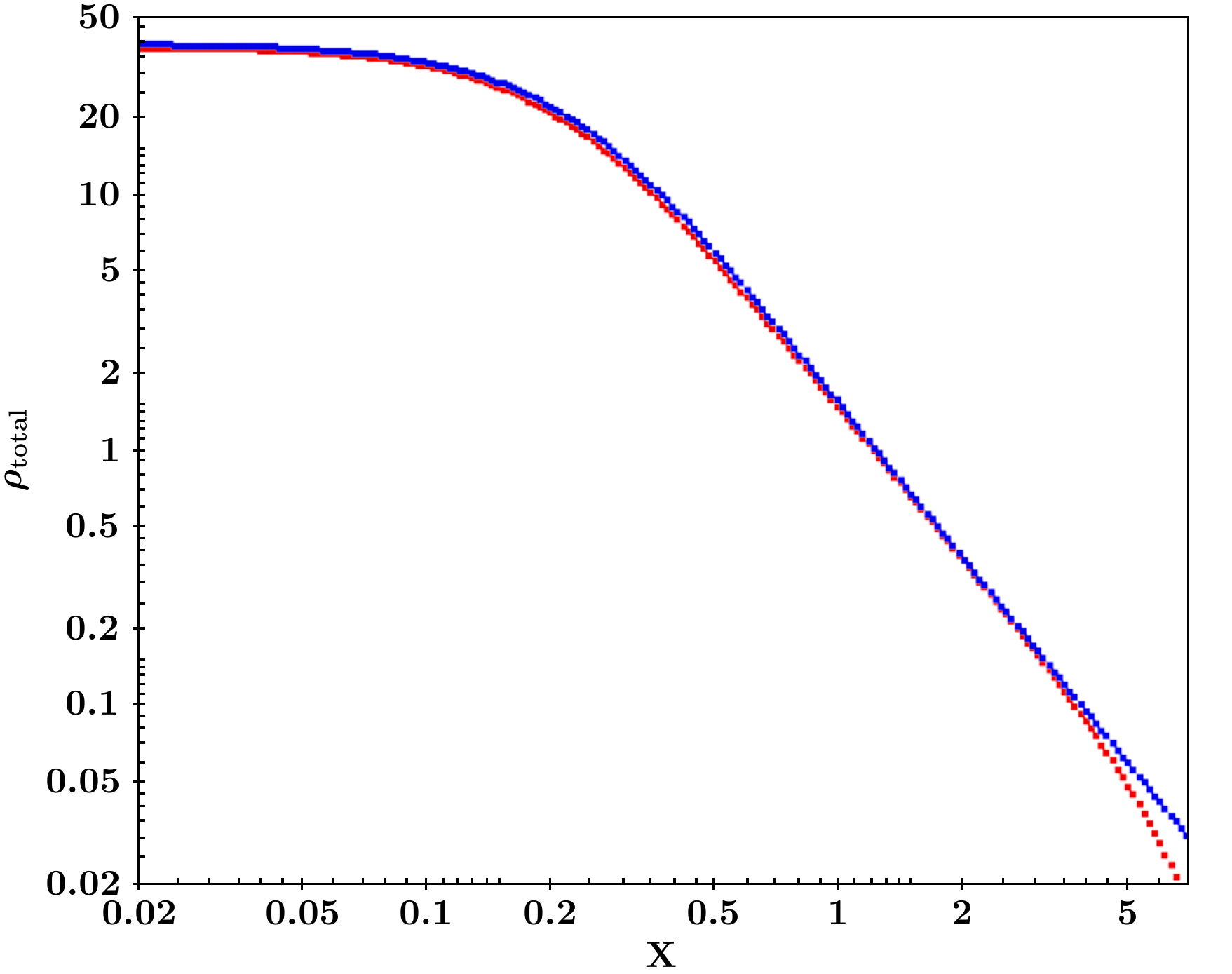}
\includegraphics[scale=0.25]{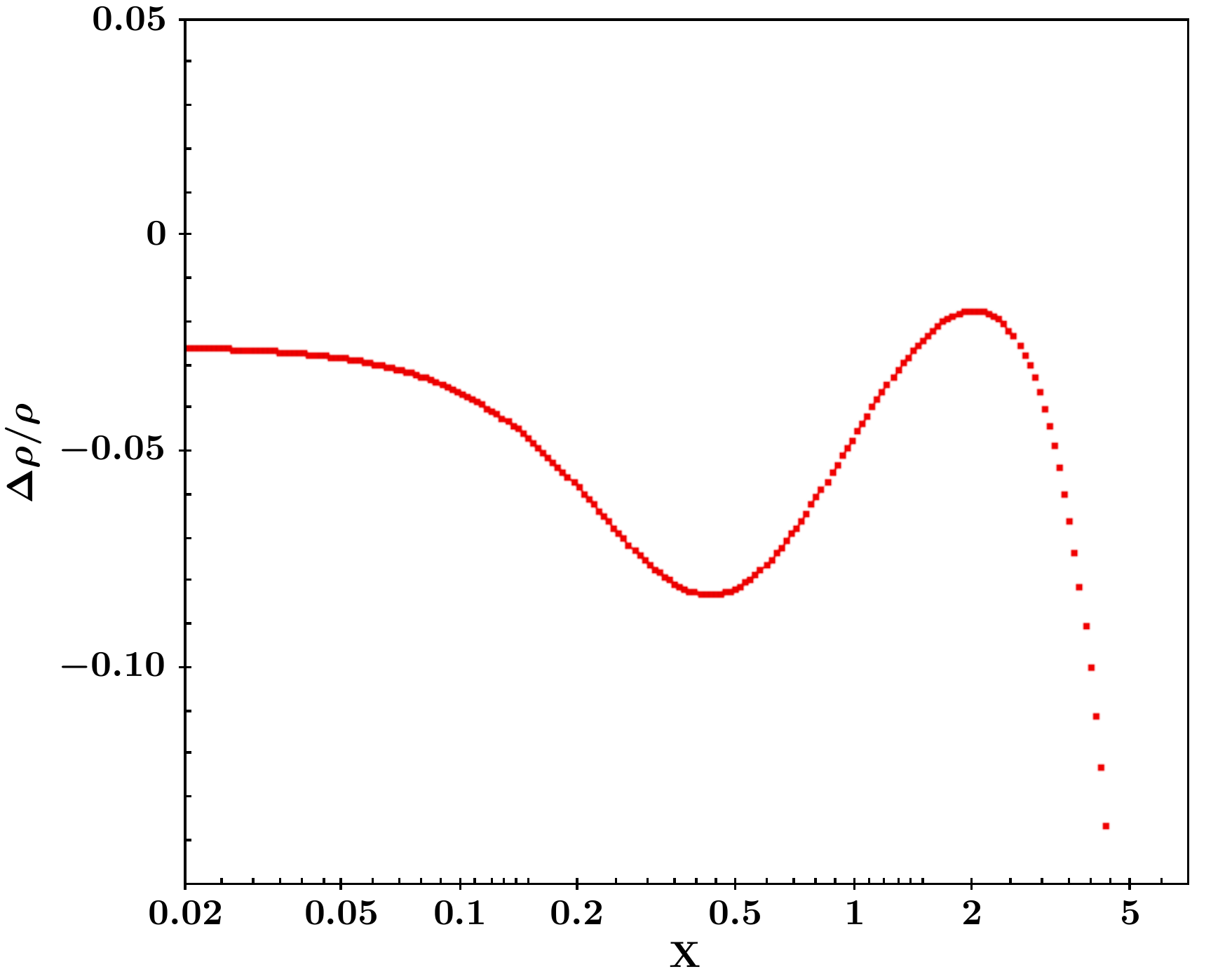}
\caption{
Accuracy of the Jeans equation for model  $b=c=-0.2$ to recover total mass density $\rho_{total}$ from Poisson equation. On the left is shown $\rho_{total}$ along the axis $(x,y-0,z=0)$. On the right is shown the relative error on $\rho_{total}$.
}
\label{fig:f19}
\end{center}
\end{figure}

\begin{figure}[!htbp]
\begin{center}
\includegraphics[scale=0.25]{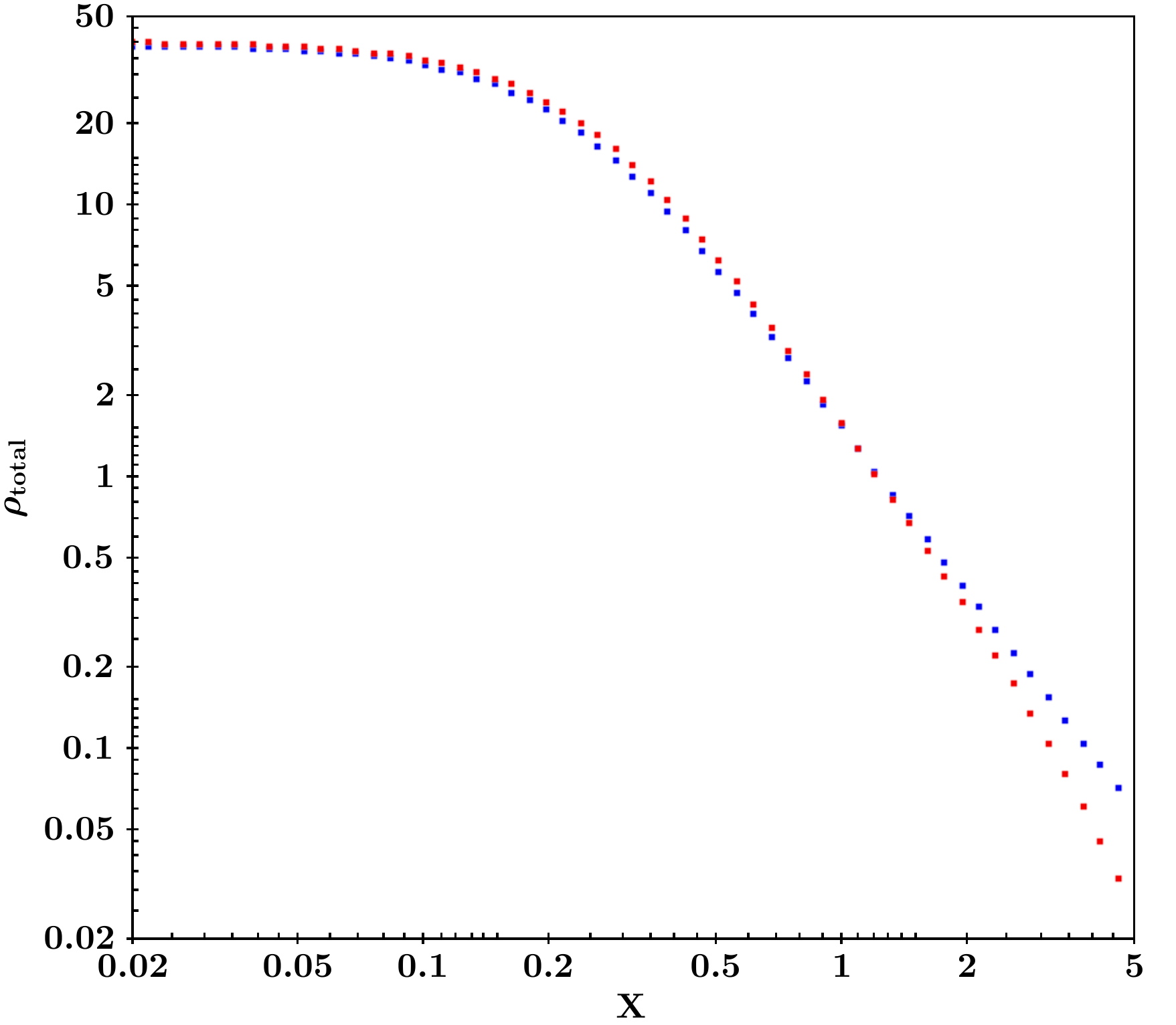}
\includegraphics[scale=0.25]{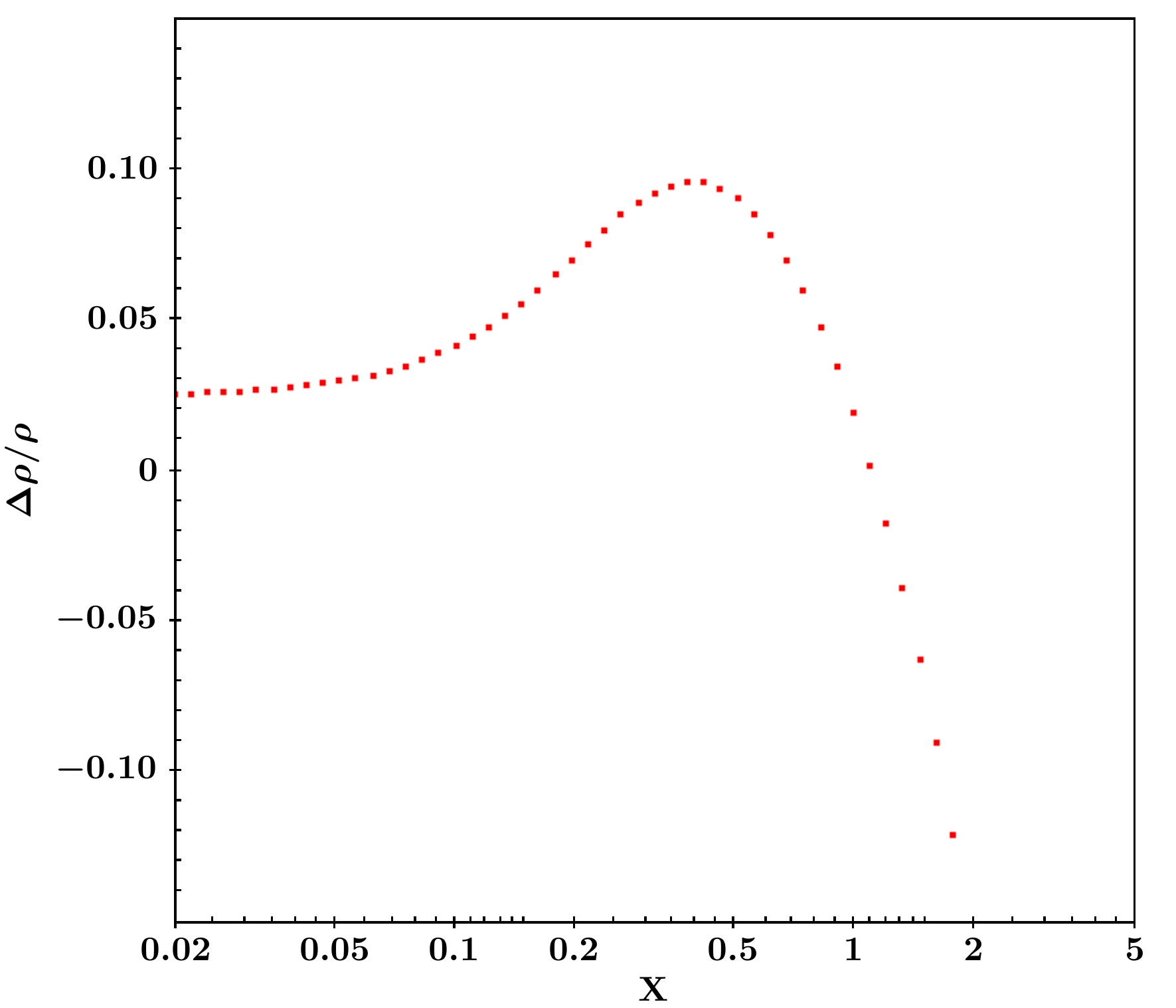}
\caption{Accuracy of the Jeans equation for model  $b=c=+0.3$ to recover total mass density $\rho_{total}$ from Poisson equation. On the left is shown $\rho_{total}$ along the axis $(x,y-0,z=0)$. On the right is shown the relative error on $\rho_{total}$.
}
\label{fig:f22}
\end{center}
\end{figure}

A second and complementary test is the comparison of the total mass of the model, $\rho_{total}$, as given by the Poisson equation applied directly to the potential, and that deduced by using the forces obtained by the Jeans equations
(see Figs.\,\ref{fig:f19} and \ref{fig:f22}). 
For the radial, velocity-dominant model and 
for $r<1,$ the density $\rho_{total}$ is found with systematic deviations, varying according to the position with an average of 6\%, and with a dispersion of 5\%.
For $r<2$ the density $\rho_{total}$ is also found with systematic deviations with  a mean systematic of 6\% but with a greater dispersion of 9\% depending on  position.
For the tangential, velocity-dominant model, the results are quite similar, with a relative error on the total mass density that depends on   position and varies from about zero in the centre to 9\% towards r=2. Globally, we 
have better accuracy in finding the total mass density using the distribution function dominated by tangential orbits.
\\

\section{Conclusion}

We  model the orbits and distribution functions of stellar populations in a triaxial potential.
This is done using an approximation of orbits in the frame of  St\"ackel potentials. \citet{ken91}  already noted that an accurate  approximation of a quasi-invariant integral can be achieved by adjusting a St\"ackel potential for each of the orbits of a galactic  potential. 
Here, we show that quasi-integrals can be written explicitly as potential dependent, by generalising  the usual formulation of St\"ackel potential integrals. 
These analytic expressions of integrals can be used to plot sections in phase space and by a simple quadrature allow the derivation quasi-actions.
They can also be used to build distribution functions.
These integrals allow us to model four different types of orbits in a triaxial potential, and  they are better preserved than the angular moment components, which are sometimes used as approximate integrals.
We note that tube orbits are modelled with very high accuracy, while the modelling of box orbits passing near the galactic centre are less accurate.
The use of quasi-integrals  allows us to build
distribution functions for a triaxial potential with a triaxial  velocity  moment tensor. 
The distribution functions, that are dominated by tube orbits (i.e. dispersions of velocities dominated by tangential components), are more precise than those dominated by radial velocities.
We use the Jeans equations and  find the components of the force of the gravity field with good accuracy over an extended volume. On the other hand,  the total mass density generating the potential and the gravity field is recovered with a significantly lower accuracy. It is indeed necessary to calculate second derivatives of the model distribution, which would require an accuracy on the distribution function that is better by one to two orders of magnitude.\\

We have already performed a  similar analysis for  an axisymmetric version of the Besan\c{c}on Galaxy Model \citep{bie15}. By generalising this work to triaxially symmetrical potentials, we improve the simultaneous modelling of four families of tube and box orbits. The latter have a more complex structure and occupy a very large volume in the configuration space and are modelled with less precision than tube orbits.
Our results can be compared to those of \cite{mcm08, san12,san14,pos15} on axisymmetric or triaxially symmetrical potentials.   
Their analyses are also based on the determination of quasi-integrals, which in their case are actions that they used to define the distribution functions. The accuracies they obtain are of the same order of magnitude as ours. It should be noted that in our case the quasi-integrals  are analytic and  easier to calculate. As for the actions, they can be evaluated here by a simple quadrature using the expressions of quasi-integrals  employed, for example, to plot the different sections of the phase space (Fig.\,\ref{fig:f11}).
Our method and the  St\"ackel fudge \citep{san14} are based  on the same principles and  are less accurate compared to the torus fitting 
 \citep[\citet{mcg90}, and see][for a review of different methods]{san16}.
These methods are included  in  numerical tool boxes proposed by \citet{bov15}  and by \citet{vas19}. It is then necessary to test their accuracy for each new potential and for each distribution function considered. Their accuracy remains limited by the accuracy of the modelled orbits, especially the box orbits. We have shown through an example that if the modelling of integrals and distribution functions made it possible to recover the forces of the gravity field with an accuracy of a few per cent over a large volume of space, the same is not true when it comes to measuring the total mass density of the model, meaning the one that generates the potential.
\\

The advantage of the method presented here is its  numerical speed.
However, more precise methods such as torus fitting are necessary if high precision is desired to accurately recover the potential and the total mass density of a model.
It should be noted that this last method does not make it possible to simultaneously model all the resonances, and that resonances can be numerous for box orbits, for example, in the case  of a significantly flattened logarithmic potential  \citep{bie13}.
Finally, we note that quasi-integral variations are significant when orbits pass near the Galactic centre, while they remain an order of magnitude  more constant between two passages at the Galactic pericentre. This makes it possible to consider using these integrals for the identification of stellar streams that have a limited extension along an orbit. \\

The work presented here suggests possible improvements.
 Figure\,\ref{fig:f11} presents two sections of the phase space and directly shows the limitation of the quasi-integrals  developed here. Finding more precise integrals can then be reduced to the question of finding a more general analytic expression, depending on a few additional parameters in order to reproduce more exactly the shape of the iso-contours of the sections of the phase space.
Another direction for improvement is the  second family of St\"ackel potentials, which has been studied very little and whose equations of motion are separable in a parabolic coordinate system \citep{oll62,lyn62,tsi12}. These potentials also depend on a generic function and therefore have a certain generality that merits further study.
\\

\begin{acknowledgements}

The author  would like to thank the International Space Science Institute, Berne, Switzerland, for providing financial support and meeting facilities.

\end{acknowledgements}

\bibliographystyle{aa} 
\bibliography{Integr.bbl} 

\begin{thebibliography}{56}
\expandafter\ifx\csname natexlab\endcsname\relax\def\natexlab#1{#1}\fi


\bibitem[Bienaym{\'e} et al.(2015)]{bie15} Bienaym{\'e}, O., Robin, A.~C., \& Famaey, B.\ 2015, \aap, 581, A123 

\bibitem[Bienaym{\'e} \& Traven(2013)]{bie13} Bienaym{\'e}, O., \& Traven, G.\ 2013, \aap, 549, A89 

\bibitem[Binney(2012)]{bin12} Binney, J.\ 2012, \mnras, 426, 1324 

\bibitem[Bland-Hawthorn \& Gerhard(2016)]{bla16} Bland-Hawthorn, J., \& Gerhard, O.\ 2016, \araa, 54, 529 

\bibitem[Bovy(2015)]{bov15} Bovy, J.\ 2015, \apjs, 216, 29 

\bibitem[Contopoulos(1960)]{con60} Contopoulos, G.\ 1960, \zap, 49, 273 

\bibitem[de Zeeuw(1985a)]{dez85a} de Zeeuw, T.\ 1985a, \mnras, 216, 273 

\bibitem[de Zeeuw(1985b)]{dez85b} de Zeeuw, T.\ 1985b, \mnras, 216, 599 

\bibitem[de Zeeuw \& Lynden-Bell(1985)]{dez85} de Zeeuw, P.~T., \& Lynden-Bell, D.\ 1985, \mnras, 215, 713 

\bibitem[Fehlberg(1968)]{fel68} Felberg, E.  NASA TECHNICAL REPORT TR R-287, 1968
 
\bibitem[Helmi et al.(2017)]{hel17} Helmi, A., Veljanoski, J., Breddels, M.~A., Tian, H., \& Sales, L.~V.\ 2017, \aap, 598, A58 

\bibitem[H\'enon \& Heiles(1964)]{hen64} H\'enon, M., \& Heiles, C.\ 1964, \aj, 69, 73 

\bibitem[Hietarinta (1987)]{hie87} Hietarinta, J. 1987, Phys. D Nonlinear Phenomena, 28, 248

\bibitem[Kent \& de Zeeuw(1991)]{ken91} Kent, S.~M., \& de Zeeuw, T.\ 1991, \aj, 102, 1994 

\bibitem[Law \& Majewski(2010)]{law10} Law, D.~R., \& Majewski, S.~R.\ 2010, \apj, 714, 229 

\bibitem[Lynden-Bell(1962)]{lyn62} Lynden-Bell, D.\ 1962, \mnras, 124, 95 

\bibitem[Malhan \& Ibata(2018)]{mal18b} Malhan, K., \& Ibata, R.~A.\ 2018, arXiv:1807.05994 

\bibitem[Malhan et al.(2018)]{mal18a} Malhan, K., Ibata, R.~A., \& Martin, N.~F.\ 2018, \mnras, 481, 3442 

\bibitem[McGill \& Binney(1990)]{mcg90} McGill, C., \& Binney, J.\ 1990, \mnras, 244, 634 

\bibitem[McMillan \& Binney(2008)]{mcm08} McMillan, P.~J., \& Binney, J.~J.\ 2008, \mnras, 390, 429 

\bibitem[Merritt(1985)]{mer85} Merritt, D.\ 1985, \aj, 90, 1913 

\bibitem[Morse \& Feshbach(1953)]{mor53} Morse, P.~M., \& Feshbach, H.\ 1953, Methods of Theoritical Physics, Ch 5 McGraw Hill, New York 

\bibitem[Ollongren(1962)]{oll62} Ollongren, A.\ 1962, \bain, 16, 241 

\bibitem[Osipkov(1979)]{osi79} Osipkov, L.~P.\ 1979, Soviet Astronomy Letters, 5, 42 

\bibitem[Posti et al.(2015)]{pos15} Posti, L., Binney, J., Nipoti, C., \& Ciotti, L.\ 2015, \mnras, 447, 3060 

\bibitem[Posti \& Helmi(2019)]{pos19} Posti, L., \& Helmi, A.\ 2019, \aap, 621, A56 

\bibitem[Sanders(2012)]{san12} Sanders, J.\ 2012, \mnras, 426, 128 

\bibitem[Sanders \& Binney(2014)]{san14} Sanders, J.~L., \& Binney, J.\ 2014, \mnras, 441, 3284 

\bibitem[Sanders \& Binney(2015)]{san15} Sanders, J.~L., \& Binney, J.\ 2015, \mnras, 447, 2479 

\bibitem[Sanders \& Binney(2016)]{san16} Sanders, J.~L., \& Binney, J.\ 2016, \mnras, 457, 2107 

\bibitem[Shu(1969)]{shu69} Shu, F.~H.\ 1969, \apj, 158, 505 

\bibitem[Tsiganov(2012)]{tsi12} Tsiganov, A.~V.\ 2012, SIGMA, 8, 031 

\bibitem[Vasiliev(2019)]{vas19} Vasiliev, E.\ 2019, \mnras, 482, 1525 

\bibitem[Whittaker \& Watson(1902)]{whi1902} Whittaker, E.~T., \& Watson, G. N.,\ 1902, A course in Modern Analysis, Cambridge University Press. 



\end{thebibliography}
\appendix

\section{ $(x,y,z)$ $(\lambda,\mu,\nu)$ transformation}

St\"ackel potentials are defined using ellipsoidal coordinates.
The   $(\lambda,\mu,\nu)$ coordinates  define confocal ellipsoids and hyperboloids and are roots of the equation

\begin{equation}
\frac{x^2}{\tau+\alpha}+\frac{y^2}{\tau+\beta}+\frac{z^2}{\tau+\gamma} =1
\end{equation}

with $\alpha$, $\beta$, and $\gamma,  $ which are three fixed numbers with
$-\gamma \le \nu \le -\beta \le \mu \le -\alpha \le \lambda $ .
\\

The expressions for  $x$, $y,$ and $z$ Cartesian coordinates are
\begin{eqnarray}
x^2 & = & \frac{(\lambda+\alpha)(\mu+\alpha)(\nu+\alpha)}{(\alpha-\beta)(\alpha-\gamma)}, \nonumber \\ 
y^2 & = & \frac{(\lambda+\beta)(\mu+\beta)(\nu+\beta)}{(\beta-\gamma)(\beta-\alpha)}, \\
z^2 & = & \frac{(\lambda+\gamma)(\mu+\gamma)(\nu+\gamma)}{(\gamma-\alpha)(\gamma-\beta)}. \nonumber
\end{eqnarray}

The expressions for $\lambda$, $\mu,$ and $\nu$ coordinates are obtained from the relations
\begin{eqnarray}
-b & = & \lambda+\mu+\nu = -\alpha-\beta-\gamma +x^2+y^2+z^2 ,\nonumber \\
c & = & \lambda\mu+\mu\nu+\nu\lambda = \alpha\beta+\beta\gamma+\gamma\alpha -(\beta+\gamma)x^2-(\gamma+\alpha)y^2-(\alpha+\beta)z^2, \\
-d & = & \lambda\mu\nu= -\alpha\beta\gamma +(\beta\gamma)x^2+(\gamma\alpha)y^2+(\alpha\beta)z^2
\nonumber
\end{eqnarray}
and they are the three solutions of the third-degree polynomial equation
\begin{eqnarray}
X^3+b X^2 +cX +d=0
\end{eqnarray}
 or
\begin{eqnarray}
Y^3+pY+q=
Y^3+ \left(-\frac{1}{3}b^2+c\right) Y +\left(\frac{2}{27}b^3+d-\frac{1}{3}cb\right)=0
\end{eqnarray}
with $X=Y-b/3.$ The equation has three solutions given that $4p^3+27q^2 < 0$.

\noindent The solutions are 
$Y_1  =  r  \sin \frac{\varphi}{3}$, 
$Y_2  =  r  \sin \left( \frac{\varphi}{3}+\frac{2\pi}{3 }\right)$,
$Y_3  =  r  \sin \left( \frac{\varphi}{3}+\frac{4\pi}{3 }\right)$, 
with $r=2\sqrt{-\frac{p}{3}}$ and $\sin \varphi = -\frac{3q}{pr}$.\\

\noindent {For spherical  potentials, we have}

\noindent $\alpha=\beta=\gamma$ and  $r^2=x^2+y^2+z^2=\lambda+\alpha$.
\\

\noindent {For prolate potentials,  $(\lambda,\mu)$ define oblate spheroidal coordinates and

\noindent $\beta=\gamma$.  The $x$ axis is the axis of axisymmetry and we have
$\tilde{z}^2=y^2+z^2$,

\noindent 
$\lambda+\mu = -\alpha-\gamma +x^2 + \tilde{z}^2$,
and
$\lambda \mu = \alpha \gamma -\gamma x^2 -\alpha \tilde{z}^2$.
\\

\noindent For oblate potentials, $(\lambda,\nu)$ define prolate spheroidal coordinates and

\noindent $\alpha=\beta$. The $z$ axis is the axis of axisymmetry
and we have 
$R^2=x^2+y^2$

\noindent 
$\lambda+\nu = -\alpha-\gamma +R^2 + z^2$,
and
$\lambda \nu = \alpha \gamma -\gamma R^2 -\alpha z^2$.


\end{document}